\documentclass[aps,twocolumn,nofootinbib]{revtex4}
\usepackage{adjustbox}
\usepackage{graphicx} 
\usepackage{caption,subcaption}
\usepackage{amsmath}
\usepackage{amsthm}
\usepackage{framed}

\usepackage{comment}
\usepackage{bbm}
\usepackage[version = 4]{mhchem}
\usepackage{tikz-cd} 
\usepackage{tikz}
\usepackage{amssymb}
\usepackage{mathtools}
\usepackage{amsthm}

\theoremstyle{plain}
\newtheorem{theorem}{Theorem}[section] 
\newtheorem{lemma}[theorem]{Lemma} 
\newtheorem{proposition}[theorem]{Proposition}

\theoremstyle{definition} 
\newtheorem{definition}{Definition}[section]
\newtheorem{example}{Example}[section] 
\newtheorem{notation}{Notation}[section] 

\theoremstyle{remark} 
\newtheorem{remark}{Remark}

\usepackage{physics}
\usepackage{xcolor}
\usepackage{bm}
\usepackage{mathtools}
\usepackage[hidelinks]{hyperref}
\usepackage{algpseudocode}
\usepackage{xcolor}
\usepackage{xfrac}
\usepackage{bm}
\usepackage[T1]{fontenc}
\usepackage{upquote}\newcommand{\vn}{\varnothing}

\DeclarePairedDelimiter\ceil{\lceil}{\rceil}
\DeclarePairedDelimiter\floor{\lfloor}{\rfloor}

\usepackage[utf8]{inputenc}
\usepackage[english]{babel}
\newcommand{\dsum}{\displaystyle\sum}
\usepackage[lined,boxed]{algorithm2e}

\begin{document}

\title{Polyhedral geometry and combinatorics of an autocatalytic ecosystem}

\author{Praful Gagrani$^{1}$, Victor Blanco$^{2, 3}$, Eric Smith$^{4,5,6,7}$, David Baum$^{1,8}$}

\affiliation{
$^1$Wisconsin Institute for Discovery, University of Wisconsin-Madison, Madison, WI 53715, USA\\
$^2$Institute of Mathematics (IMAG), Universidad de Granada, Spain\\
$^3$ Department of Quant. Methods for Economics \& Business, Universidad de Granada, Spain\\
$^4$ Earth-Life Science Institute, Tokyo Institute of Technology, Tokyo 152-8550, Japan\\
$^5$ Santa Fe Institute, 1399 Hyde Park Road, Santa Fe, NM 87501, USA\\
$^6$ Ronin Institute, 127 Haddon Place, Montclair, NJ 07043, USA\\
$^7$ Department of Biology, Georgia Institute of Technology, Atlanta, GA 30332, USA\\
$^8$ Department of Botany, University of Wisconsin-Madison, Madison,
WI 53706, USA}
\date{\today}
\begin{abstract}
Developing a mathematical understanding of autocatalysis in reaction networks has both theoretical and practical implications. We review definitions of autocatalytic networks and prove some properties for minimal autocatalytic subnetworks (MASs). We show that it is possible to classify MASs in equivalence classes, and develop mathematical results about their behavior. We also provide linear-programming algorithms to exhaustively enumerate them and a scheme to visualize their polyhedral geometry and combinatorics. We then define \textit{cluster chemical reaction networks}, a framework for coarse-graining real chemical reactions with positive integer conservation laws. We find that the size of the list of minimal autocatalytic subnetworks in a maximally connected cluster chemical reaction network with one conservation law grows exponentially in the number of species. We end our discussion with open questions concerning an ecosystem of autocatalytic subnetworks and multidisciplinary opportunities for future investigation.
\end{abstract}
\keywords{Chemical reaction networks}
\maketitle
\begingroup
  \hypersetup{hidelinks}
  \tableofcontents
\endgroup

\section{Introduction}
\begin{figure}
    \centering
    \includegraphics[width = \linewidth]{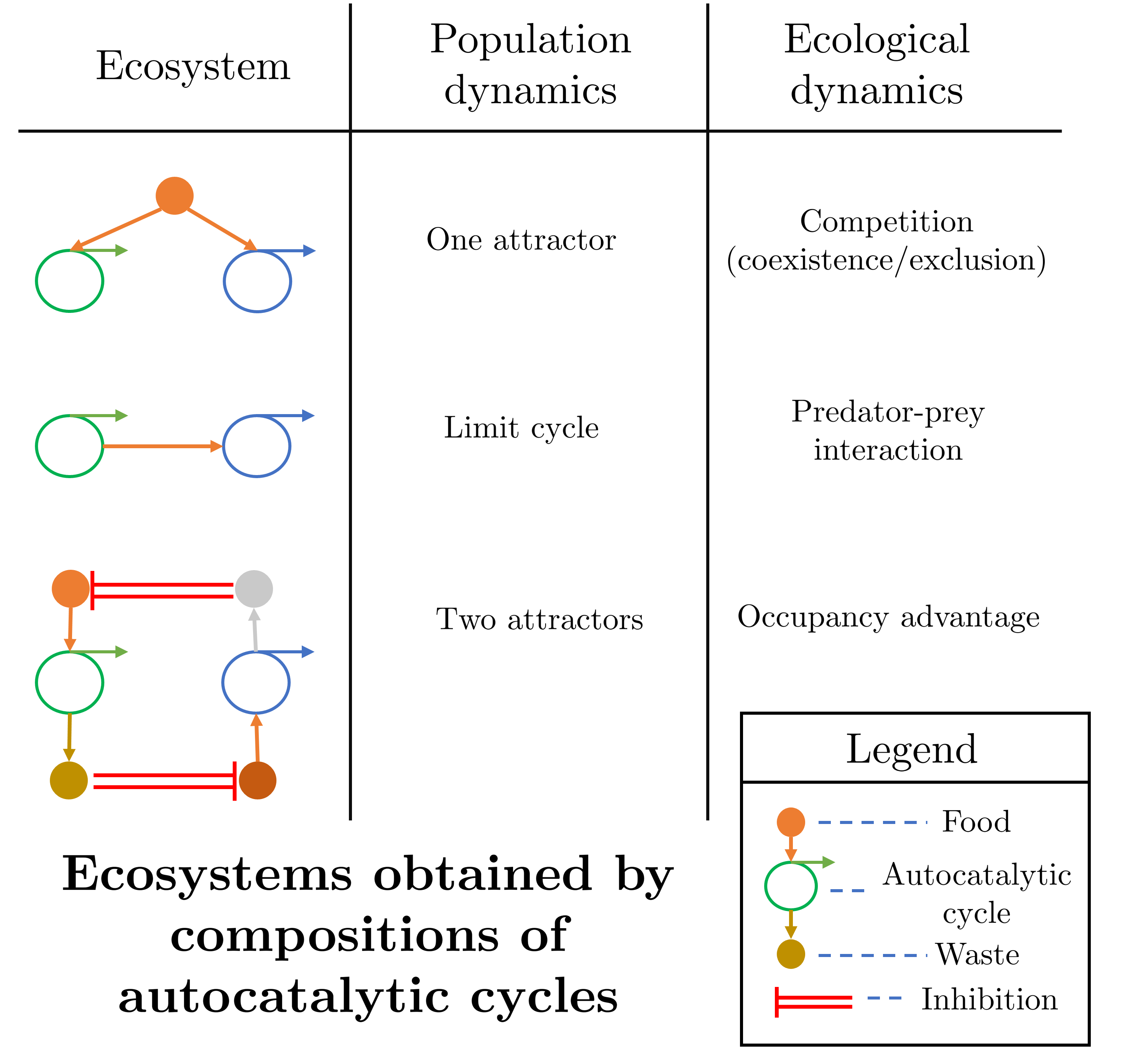}
    \caption{Ecosystems obtained by composing autocatalytic cycles. Each autocatalytic cycle consumes food and produce waste species. Two cycles can be composed in several ways, for example by sharing food (line one), the member species of one could be the food of other (line two), or the waste of one can deplete the food of another (line three). For a detailed account of such interactions and their ecological counterparts, see \cite{peng2020ecological}. }
    \label{fig:Peng_summary}
\end{figure}

Chemical reaction network (CRN) theory offers a versatile mathematical framework in which to model complex systems, ranging from biochemistry and game theory \cite{veloz2014reaction}, to the origins of life \cite{smith2016origin}. The usefulness of CRNs in modelling these phenomena stems from its ability to exhibit a wide range of nonlinear dynamics and it is widely recognized that \textit{autocatalysis} can be seen as a basis for many of them \cite{epstein1998introduction,schuster2019special}. Broadly, autocatalysis is framed as the ability of a given chemical species to make more copies of itself or otherwise promote its own production. Thus, from the perspective of kinetics, one would expect autocatalysts to be able to show super-linear growth, which is indeed possible \cite{hordijk2004detecting,andersen2021defining}. For some examples of nonlinear kinetics obtained by composing two autocatalytic cycles and their relevance to ecological dynamics, see Fig.\ \ref{fig:Peng_summary}.

A comprehensive understanding of autocatalysis in CRNs has great practical values, but its full mathematical treatment remains to be developed. In \cite{andersen2021defining}, Anderson et al.\ conjectured that it is impossible for a CRN to show a temporary speed-up of the reaction before settling down to reach an equilibrium if, at least, it is not \textit{formally} autocatalytic. Recently, in \cite{vassena2023unstable}, Vassena et al.\ show that this is not the case and there can be systems that show superlinear growth that do not exhibit autocatalysis. Also, it is suggested in \cite{deshpande2014autocatalysis} that understanding the interactions between autocatalytic and drainable subnetworks might provide a basis for obtaining a proof for the long-standing \textit{persistence conjecture} \cite{feinberg1989necessary}. 

Andersen et al.\ also defined a notion of \textit{exclusive} autocatalysis for one species, which we generalize for multiple species.
In \cite{blokhuis2020universal}, Blokhuis et al.\ defined a notion of autocatalysis, more restrictive than exclusive autocatalysis, and showed that, within reversible and non-ambiguous CRNS, it has only five types of minimal forms. They also found that autocatalysis is more abundant than previously thought, and simple reaction networks with a few reactions can contain very many cores in them. 

In this work, we focus our attention on the topological notion of autocatalysis, one that can be ascertained purely from the reactions in a CRN. A CRN is autocatalytic in a set of core (autocatalytic) species if there exists a flow on the network such that all core species are necessarily being consumed while being produced at higher rate. We term the subnetwork that consists of all the reactions that have a strictly positive contribution to such a flow as an autocatalytic subnetwork. By imposing further constraints on the subnetworks, we resolve three nested categories of autocatalysis: formal, exclusive, and stoichiometric autocatalysis. We show that stoichiometric autocatalysis is similar to the autocatalysis of Blokhuis et al. (see Remark \ref{remark:blokhuis_comp}), and rederive some of their results for the broader class of exclusive autocatalysis. In particular, we prove that minimal exclusively autocatalytic subnetworks have the property that along with all core species being produced at a higher rate than they are consumed, at least one core species is produced and consumed in each reaction of the subnetwork. We then investigate the polyhedral geometry of these minimal autocatalytic subnetworks (MASs), provide algorithms for exhaustively enumerating them, and propose visualization schemes for understanding their combinatorics. 

We employ the techniques developed in this work on a framework, which we term the \textit{Cluster CRN} (CCRN). A brief description of the motivation and formalism of the CCRN framework is as follows.
Molecules are composed of atoms and atomic conservation is a fundamental law of chemistry. While molecular structure plays a critical role in determining what atomic interchanges are possible in particular environments (and their rate constants), focusing on just atomic composition provides a simple abstraction that, at some level, must resemble real chemistry. In the most basic version with a single conserved quantity (such as a monomer or an atom), the list of all potential reactions is dictated only by arithmetic. By adding other dimensions it is theoretically possible to represent a lot of real chemistry. Adding dimensions can allow for polymers composed of multiple monomer types or for molecular formulae composed of more than one element, where  the \textit{dimension} is the number of monomer types or elements, respectively. Other dimensions can also be used to represent structural variation within a molecular formula or spatial variation, although doing so requires adding constraints besides arithmetic on allowed reaction rules. We thus posit that the CCRN framework is a very flexible coarse-graining of chemistry. Formally, a CCRN is a CRN with positive integer valued conservation laws. The investigations undertaken in this work show that even the simplest version CCRN (in one dimension) can reveal important principles regarding the abundance of MASs and their interactions. 

The layout of the paper is as follows. In Sec.\ \ref{sec:CRN_AC}, we briefly review chemical reaction networks (CRNs) and present a hierarchy of autocatalysis in their graph-theoretic and linear-algebraic manifestations. In Sec.\ \ref{sec:MASs}, we define MASs, prove their properties, organize them in equivalence classes, and explore some geometric quantities that are canonically defined for them. In Sec.\ \ref{sec:organization}, we consider CRNs with multiple MASs, which we term as an \textit{autocatalytic ecosystem}. Given such a CRN, we provide a linear-programming algorithm to exhaustively enumerate the MASs and introduce a visualization scheme in which to represent their polyhedral geometry their combinatorics. In Sec.\ \ref{sec:applications}, we present the cluster chemical reaction network (CCRN) framework and argue that it provides a natural coarse-graining of realistic chemical and biochemical reaction networks. We then explore the combinatorics of MASs within the CCRN framework and provide a worked example to demonstrate the geometry of autocatalysis in the stoichiometric subspace of species concentration. Using elementary number theoretic considerations, we also prove a combinatorial result about the list of MASs. Multiple examples from the CCRN framework are also included as examples at several points in this work. We conclude, in Sec.\ \ref{sec:discussion}, with an overview of our contributions and possible avenues for future investigation. 


\section{Chemical reaction networks and autocatalytic subnetworks}
\label{sec:CRN_AC}
The mathematical theory of chemical reaction networks (CRNs), pioneered by Horn, Jackson, and Feinberg \cite{horn1972general} has ubiquitous applications in understanding nature \cite{feinberg2019foundations,yu2018mathematical}. In particular, for diverse biological and chemical applications, the formalism allows for a notion of self-replication or \textit{autocatalysis} where certain species cause an increase in the population of these same species. Consequently, there is value in rigorously understanding the organization of CRNs containing a collection of autocatalytic subnetworks, which we term as an \textit{autocatalytic ecosystem}. 

The concept of autocatalysis, dating back to the late 1800s \cite{ostwald1890autokatalyse,peng2022wilhelm}, has several distinct formulations in CRN theory. In this section, we explain the correspondence between graph-theoretic and linear-algebraic formulations of CRNs, and employ it to present a nested hierarchy of three types of autocatalysis: formal, exclusive, and stoichiometric. 
In particular, we provide conditions on a subset of reactions (subnetwork) of a CRN for them to be termed autocatalytic for some subset of autocatalytic species. 
Note that, for a subnetwork to be autocatalytic, we require that all the reactions participate with a strictly positive flux. 
So, while a CRN can be said to be autocatalytic if it contains an autocatalytic subnetwork, in our formulation, a subnetwork might loose its autocatalytic property if more reactions are added. 
This choice of definitions is made to simplify analysis (see Remark \ref{remark:blokhuis_comp}) and leads the way to an investigation of minimal autocatalytic subnetworks in the next section.

\subsection{Preliminaries}
\begin{notation}
    \textbf{0} denotes a vector of zeros.
\end{notation}
\begin{notation}
    For a vector $\textbf{v} = [v_1, v_2, \ldots, v_n]^T$ in $\mathbb{R}^n$:
    \begin{enumerate}
        \item We call $\textbf{v}$ positive and write $\textbf{v} \gg \textbf{0}$ if $v_i>0$ for all $i \in [1,n]$.
        \item We call $\textbf{v}$ non-negative and write $\textbf{v} \geq \textbf{0}$ if $v_i \geq 0$ for all $i \in [1,n]$.
        \item We call $\textbf{v}$ semi-positive and write $\textbf{v} > \textbf{0}$ if $v_i \geq 0$ and $\textbf{v} \neq \textbf{0}$.
    \end{enumerate}
\end{notation}

\begin{notation}
    For vectors $\textbf{v}$ and $\textbf{w}$ in $\mathbb{R}^n$, $\textbf{v} \gg \textbf{w} $, $\textbf{v} \geq \textbf{w}$, and $\textbf{v} > \textbf{w}$ mean $\textbf{v}-\textbf{w} \gg \textbf{0}$, $\textbf{v} -\textbf{ w} \geq \textbf{0}$, and $\textbf{v}-\textbf{w}> \textbf{0}$, respectively. 
\end{notation}
\begin{definition}
    For all vectors $\textbf{v}_1, \textbf{v}_2, \ldots, \textbf{v}_k \in \mathbb{R}^n$:
    \begin{enumerate}
        \item Their \textit{convex polyhedral cone} is denoted by
        \[\textrm{Cone}\left(\{\textbf{v}_1, \ldots, \textbf{v}_k\} \right) := \left\{ \sum_{i=1}^k a_i \textbf{v}_i \bigg| a_1,\ldots a_k  \in \mathbb{R}_{\geq 0}\right\}.\]
    \item Their \textit{span} is denoted by
        \[\textrm{Span}\left(\{\textbf{v}_1, \ldots, \textbf{v}_k\} \right) := \left\{ \sum_{i=1}^k a_i \textbf{v}_i \bigg| a_1,\ldots a_k  \in \mathbb{R}\right\}.\]
    \end{enumerate}
\end{definition}

\begin{notation}
    For a matrix $\mathbb{M}$, we denote: 
    \begin{enumerate}
        \item the set rows of $\mathbb{M}$ as $\textrm{rows}(\mathbb{M})$.
        \item the set of columns of $\mathbb{M}$ as $\textrm{cols}(\mathbb{M})$.
        \item the $i^\text{th}$ row as $(\mathbb{M})_i$.
        \item the $j^\text{th}$ column as $(\mathbb{M})^j$.
        \item the entry in the $i^{\text{th}}$ row and $j^\text{th}$ column as $(\mathbb{M})_i^j$.
    \end{enumerate}
\end{notation}

\begin{notation}[Restriction]
    Let $\mathbb{M}$ denote a $m \times n$ matrix with $m$ rows and $n$ columns. Let $M \subseteq \{1,\ldots,m\} $ and $N \subseteq \{1,\ldots,n\}$ be a subset of rows and columns, respectively. Then the \textit{restriction} of $\mathbb{M}$ to the rows of $M$ and columns of $N$ is denoted as $\mathbb{M}_M$ and $\mathbb{M}^N$, respectively. The simultaneous restriction of $\mathbb{M}$ to both sets $M$ and $N$ is denoted as
    \[\mathbb{M}\big|_{(M,N)} := (\mathbb{M})_M^N.\]
\end{notation}

\begin{notation}[Support]
    \label{not:support}
    The \textit{support} of a vector $\textbf{v}$, denoted $\text{supp}(\textbf{v})$, is the set of its non-zero coordinates. 
\end{notation}

\subsection{Chemical reaction networks}
\label{sec:hypergraphs}

\begin{definition}
\label{def:CRN_GT}
    A chemical reaction network (CRN) $\mathcal{G}$ is defined as the triple of a species set, complex set, and reaction set (see \cite{feinberg2019foundations})
    \[\mathcal{G} = (\mathcal{S}, \mathcal{C},\mathcal{R}),\]
    where:
    \begin{enumerate}
    \item the species set $\mathcal{S}$, of size $|\mathcal{S}|$, consists of all the              \textbf{species} that appear in the network,
        \[\mathcal{S} = \{ X_1, \ldots, X_{|\mathcal{S}|}\}.\]
    \item the complex set $\mathcal{C}$ consists of \textbf{complexes} of the network. A complex is a formal non-negative integer linear combination of elements of $\mathcal{S}$. The coefficients of the species in a complex form a vector in $\mathbb{Z}_{\geq0}^{|\mathcal{S}|}$ and denote the \textit{stoichiometry} of the complex. A complex $y \in \mathcal{C}$ will be taken to mean either of the following:
        \[ \sum_{i=1}^{|\mathcal{S}|} y_i X_i \equiv y \in \mathbb{Z}_{\geq0}^{|\mathcal{S}|},\]
    and be denoted by a column vector. 
    \item the reaction set $\mathcal{R} \subseteq \mathcal{C} \times \mathcal{C}$ consists of \textbf{reactions} of the network. 
    A reaction is an ordered pair of complexes, and has the form
        \[ \sum_{i=1}^{|\mathcal{S}|} r^-_i X_i \to \sum_{i=1}^{|\mathcal{S}|} r^+_i X_i, \]
        or, equivalently
        \[ r^- \to r^+.\]
    Here $r^-, r^+ \in \mathcal{C}$ and are referred to as the \textit{input} and \textit{output} complex, respectively.     
    \end{enumerate}
\end{definition}

\begin{remark}
\label{remark:CRN_SR}
    Since the information of the set of complexes $\mathcal{C}$ is implicit in the set of reactions $\mathcal{R}$, Deshpande et al.\ in \cite{deshpande2014autocatalysis} economically define 
    a CRN simply by the pair $\mathcal{G} = (\mathcal{S},\mathcal{R})$. Here the complex set is given by taking the union of all the input and output complexes for all the reactions,
    \[ \mathcal{C} = \bigcup \{r^-,r^+| r^- \to r^+ \in \mathcal{R}\}.\]
\end{remark}

\begin{remark}
    Isomorphically, a CRN $\mathcal{G} = (\mathcal{S}, \mathcal{C}, \mathcal{R})$ can be represented as a directed multi-hypergraph \cite{andersen2019, smith2017flows} whose hyper-vertices are multisets of vertices according to the dictionary: 
    \begin{center}\begin{tabular}{c|c c}
       &CRN&Hypergraph\\\hline
       $\mathcal S$&Species&Vertices\\
       $\mathcal C$&Complexes&Hyper-Vertices\\
       $\mathcal R$&Reactions&Hyper-Edges
    \end{tabular}\end{center} 
\end{remark}

\subsection{Input-output matrix pair and the stoichiometric matrix}

Recall that every reaction is of the form  
\[\textrm{Input complex} \to \textrm{Output complex}.\]
\begin{notation}
We denote a reaction $r \in \mathcal{R}$ as  
\[  r^- \to r^+,\]
where $r^-,r^+$ are column vectors in $\mathbb{Z}_{\geq 0}^{|\mathcal{S}|}$.     
\end{notation}

For any CRN, we can define $\mathcal{S} \times \mathcal{R}$ matrices $\mathbb{S}^-$ and $\mathbb{S}^+$ by collecting all the input and output complexes, respectively, where
\begin{align}
    (\mathbb{S}^-)^r &= r^-,\\
    (\mathbb{S}^+)^r &= r^+.
\label{eq:inp_out_pair}
\end{align}
We refer to $\mathbb{S}^-$ and $\mathbb{S}^+$ as the \textit{input} and \textit{output} matrix, respectively. Observe that $\mathbb{S}^-$ and $\mathbb{S}^+$ have only non-negative entries.

\begin{lemma}
\label{lemma:LA_CRN}
    Each CRN $\mathcal{G}$ has a unique representation as
    \[\mathcal{G} = (\mathcal{S},\mathbb{S}^-,\mathbb{S}^+),\]
    where $\mathcal{S}$ denotes the species set and $(\mathbb{S}^-,\mathbb{S}^+)$ represent the input-output matrix pair.
\end{lemma}
\begin{proof}
Using Remark \ref{remark:CRN_SR}, a CRN $\mathcal{G}$ is given by the pair of species and reactions $\mathcal{S},\mathcal{R}$. By definition, a reaction $r = r^- \to r^+$ is in the set $\mathcal{R}$ if and only if (iff) the $r^\text{th}$ columns of $\mathbb{S}^+$ and $\mathbb{S}^-$ are $r^+$ and $r^-$, respectively. 
\end{proof}

\begin{remark}
    Definition \ref{def:CRN_GT} and the following remarks provide a graph-theoretic (GT) representation of CRNs, and Lemma \ref{lemma:LA_CRN} provides a linear-algebraic (LA) representation of CRNs. 
\end{remark}

\begin{definition}[Stoichiometric matrix]
    The \textit{stoichiometric matrix} $\mathbb{S}$ is a $\mathcal{S} \times \mathcal{R}$ matrix defined as the difference of the output and input matrices,
    \begin{align}
    \mathbb{S} &= \mathbb{S}^+ - \mathbb{S}^-.  \label{eq:stoich_diff_inp_out}
    \end{align}
    Alternatively, for $r \in \mathcal{R}$ and using Eq.\ \ref{eq:inp_out_pair}, each column of the stoichiometric matrix is given by the difference of the output and input vectors of a reaction, 
\begin{align}
(\mathbb{S})^r &= r^+ - r^-. \label{eq:stoich_mat}
\end{align}
\end{definition}

\begin{definition}[Stoichiometric subspace]
\label{def:stoich_sub}
The span of the columns or the image of the stoichiometric matrix is termed as the \textit{stoichiometric subspace}. 
    \[\textrm{Stoichiometric subspace}:= \textrm{Image}(\mathbb{S}).\]
\end{definition}

\subsection{Hypergraph flows and conservation laws}
\label{sec:conservation}
\begin{definition}[Flow]
    For a CRN $\mathcal{G} = (\mathcal{S},\mathcal{R})$, a column vector $\textbf{v} \in \mathbb{R}^{|\mathcal{R}|}_{\geq 0}$ will be said to define a (hypergraph) \textit{flow} on the CRN.   
\end{definition}
A flow on a graph can also be used to create a linear combination of the reactions, also called a composite reaction in \cite{andersen2021defining}. 
While, technically, an abuse of notation, let use use $r$ to index the reaction set $\mathcal{R}$ and also refer to the $r^\text{th}$ reaction. Then, a flow $\textbf{v} = [v_1, \ldots, v_r, \ldots , v_{|\mathcal{R}|}]^T$ on CRN $\mathcal{G}$ would result in the composite reaction
\begin{align*}
    \sum_{r \in \mathcal{R}} v_r  r^- &\to \sum_{r \in \mathcal{R}} v_r r^+. 
\end{align*}

Let vector $\textbf{n} = [n_1, \ldots, n_{|\mathcal{S}|}]^T \in \mathbb{Z}_{\geq 0}^{|\mathcal{S}|}$ denote the population of each species in the system, and $\textbf{v}  \in \mathbb{Z}_{\geq 0}^\mathcal{R}$ denote a flow on $\mathcal{G}$.\footnote{We will use a notation of counts for definiteness and simplicity. A parallel construction goes through with concentrations and flow vectors $\in \mathbb{R}_{\geq0}^{|\mathcal{S}|}$ and can be found in \cite{gagrani2023action}.} 
The amount of species consumed and produced due to this flow are given by $\mathbb{S}^- \cdot \textbf{v}$ and $\mathbb{S}^+ \cdot \textbf{v}$, respectively.
Thus, the change of species population $\Delta \textbf{n}$ resulting from the flow $\textbf{v}$ on the graph $\mathcal{G}$ is given by 
\begin{align*}
    \Delta \textbf{n} &= \mathbb{S} \cdot \textbf{v}.
\end{align*}
Note that, from Definition \ref{def:stoich_sub}, the change in species population under an arbitrary flow on $\mathcal{G}$ lies in the stoichiometric subspace. 
 
The right null space of $\mathbb{S}$ correspond to \textit{null flows} which leave the population of the species unchanged ($\Delta \textbf{n} = 0$). An accounting of the right null space of $\mathbb{S}$ provides a topological classification for CRNs, which we review in Appendix \ref{app:deficiency}. 
\begin{definition}[Null flows]
    A column vector $\textbf{w} \in \mathbb{R}^{|\mathcal{R}|}$ is said to be a null flow on the CRN $\mathcal{G}$ if 
    \[\mathbb{S} \cdot \textbf{w} = \textbf{0}.\]
\end{definition}

The left null space of $\mathbb{S}$, i.e.\ vectors $\textbf{x} \in \mathbb{R}^{|\mathcal{S}|} $ such that $\textbf{x}\cdot\mathbb{S} = \textbf{0}$, correspond to \textit{conservation laws} $$\textbf{x} \cdot \Delta \textbf{n} = 0.$$ Equivalently, the inner product of the conservation law with the population $\textbf{x} \cdot \textbf{n}$ yields a \textit{conserved quantity} that is an invariant for the network under an arbitrary flow. 

\begin{definition}[Conservation law]
    For a CRN $\mathcal{G} = (\mathcal{S},\mathcal{R})$, a column vector $\textbf{x}$ of size $|\mathcal{S}|$ will be said to be a conservation law if
    \[ \textbf{x}^T \cdot \mathbb{S} = \textbf{0}.\]
\end{definition}

    The vector $\textbf{x}$ will be called a \textit{positive conservation law} or \textit{mass-like conservation law} if all its entries are nonnegative. Moreover, if the entries are nonnegative integers, it will be called a \textit{positive integer valued conservation law}.

\begin{figure*}[t]
\centering
\begin{subfigure}{.5\textwidth}
  \centering
  \includegraphics[width=0.8\linewidth]{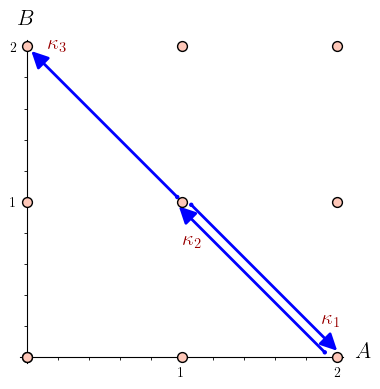}
  
\end{subfigure}%
\begin{subfigure}{.5\textwidth}
  \centering
  \includegraphics[width=.8\linewidth]{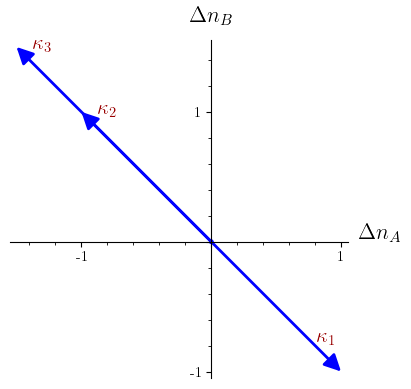}
  
\end{subfigure}
\caption{In the left panel, the Euclidean embedded graph (E-graph, see \cite{craciun2019polynomial}) of the CRN in Example \ref{eg:CRN-1} is shown. The E-graph is obtained by representing the complexes as lattice points in a Euclidean lattice and representing the reactions as edges between them. The flows on the reactions are labelled by $\kappa$ and the change in concentration they induce, $\Delta n = \mathbb{S} \kappa$, is schematically represented in the right panel. }
\label{fig:test_1}
\end{figure*}

\begin{example}
\label{eg:CRN-1}
Consider the CRN $\mathcal{G} = (\mathcal{S},\mathcal{R})$ given as
\begin{align*}
    \ce{A + B <=> 2 A},\\
    \ce{A + B -> 2B}.
\end{align*}
For a graphical representation of this example, see Fig.\ \ref{fig:test_1}. Then the species set $\mathcal{S} = \{ A, B\}$ and the reaction set $\mathcal{R} = \{ A + B \to 2 A , 2A \to A + B ,A + B \to 2B \}$. As can be read from $\mathcal{R}$, the complex set $\mathcal{C} = \{A + B, 2A, 2B \}$, and their corresponding vectors are  
\[y_{A+B}=\begin{bmatrix}
    1\\
    1
\end{bmatrix}, 
y_{2A} = \begin{bmatrix}
    2\\
    0
\end{bmatrix},
y_{2B} = \begin{bmatrix}
    0\\
    2
\end{bmatrix}.
\]
The input and output matrices, $\mathbb{S}^-$ and $\mathbb{S}^+$, respectively, for $\mathcal{G}$ are 
\[ \mathbb{S}^- = \begin{bmatrix}
    1 & 2 & 1\\
    1 & 0 & 1
\end{bmatrix},
\]
and 
\[ \mathbb{S}^+ = \begin{bmatrix}
    2 & 1 & 0\\
    0 & 1 & 2
\end{bmatrix}.
\]
The stoichiometric matrix $\mathbb{S}$ for $\mathcal{G}$ is 
\[ \mathbb{S} = \begin{bmatrix}
    \phantom{-}1 & -1 & -1\\
    -1 & \phantom{-}1 &  \phantom{-}1
\end{bmatrix},
\] with left null space (conservation laws) spanned by $\textbf{x} = [1, 1]$ and right null space (null flows) spanned by the basis $\{ [1, 1, 0 ]^T, [1, 0, 1 ]^T\}$. Notice that the stoichiometric subspace is one-dimensional and the Euclidean inner-product of the spanning vectors with the conservation law is zero. 
\end{example}


\subsection{Formal, exclusive and stoichiometric autocatalysis}
\label{sec:AC-motif}

Consider a CRN $\mathcal{G} = (\mathcal{S},\mathcal{R})$ with its stoichiometric matrix $\mathbb{S}$.

\begin{definition}[Subnetwork] \label{def:subnet} A CRN $\mathcal{G}' = (\mathcal{S}',\mathcal{R}')$ is a subnetwork of $\mathcal{G}$ if:
\begin{enumerate}
    \item The reaction set of $\mathcal{G}'$ is a subset of $\mathcal{G}$,
    \[\mathcal{R}' \subseteq \mathcal{R}.\]
    \item Every species that appears in the subset $\mathcal{R}'$ in $\mathcal{G}$ appears in $\mathcal{S}'$. Equivalently,
    \[X_i\in\mathcal{S'} \text{ iff }r^-\to r^+ \in\mathcal{R'}
\text{ such that } r_i^- > 0 \text{ or } r_i^+ > 0.
\]
\end{enumerate}
\end{definition}
\begin{remark}
The stoichiometric matrix of the subnetwork is obtained from that of the network by keeping only the columns and rows corresponding to the reactions and species participating in the subnetwork, respectively. 
\end{remark}

\begin{definition}[Motif]
\label{def:motif}
     A CRN $\mathcal{G}' = (\mathcal{S}',\mathcal{R}')$ is a motif (or a subhypergraph) of $\mathcal{G}$ if (see \cite{blokhuis2020universal}):
\begin{enumerate}
    \item The reaction set of $\mathcal{G}'$ is a subset of the reaction set of $\mathcal{G}$,
    \[\mathcal{R}' \subseteq \mathcal{R}.\]
    \item The species set of $\mathcal{G}'$ is a subset of the species set of $\mathcal{G}$,
    \[\mathcal{S}' \subseteq \mathcal{S}.\]
\end{enumerate}
\end{definition}
\begin{notation}
The stoichiometric matrix of a motif is obtained from that of the network by keeping only the rows and columns corresponding the species and reactions in the motif, respectively. Henceforth, we denote an arbitrary submatrix of the stoichiometric matrix corresponding to the motif $(\mathcal{S}',\mathcal{R}')$ with an overbar,
    \[\overline{\mathbb{S}} := \mathbb{S}|_{(\mathcal{S}',\mathcal{R}')} =  (\mathbb{S})^{\mathcal{R}'}_{\mathcal{S}'}.\]        
\end{notation}

\subsubsection{Formal autocatalysis}
\begin{definition}
A subnetwork $\mathcal{G}' = (\mathcal{S}',\mathcal{R}')$ of CRN $\mathcal{G}= (\mathcal{S},\mathcal{R})$ is defined to be \textit{formally autocatalytic} in the subset $\mathcal{M}$ of $\mathcal{S}'$ if (see \cite{king1978autocatalysis,andersen2021defining}): 
\begin{enumerate}
    \item There exists a positive real flow ($\gg 0$) on $\mathcal{G}'$ such that the resulting composite reaction is of the form
    \begin{align*}
        \text{(F)} + \textbf{m} \mathcal{M} \to \textbf{n} \mathcal{M} + \text{(W)}, 
    \end{align*}
    where $\textbf{0} \ll \textbf{m} \ll \textbf{n}$. Here $\textbf{m}$ and $\textbf{n}$ are the stoichiometries of the set $\mathcal{M}$ in the input and output of the composite reaction, respectively, and $\textbf{o} \mathcal{M} = \sum_i \textbf{o}_i \mathcal{M}_i$. 
\end{enumerate}
\end{definition}

\begin{remark}
    This definition can be generalized to semi-positive flows, but we will not need that generalization for this work. (Also, see Remark \ref{remark:blokhuis_comp}.)
\end{remark}

\begin{lemma} \label{def:FA_LA}
A subnetwork $\mathcal{G}' = (\mathcal{S}',\mathcal{R}')$ of CRN $\mathcal{G}= (\mathcal{S},\mathbb{S}^-,\mathbb{S}^+)$ is \textit{formally autocatalytic} in the subset $\mathcal{M}$ iff: 
\begin{enumerate}
    \item There exists a flow $\textbf{v} \gg \textbf{0}$ on $\mathcal{G}'$ such that
    \begin{align*}
        \overline{\mathbb{S}} \cdot \textbf{v} = (\mathbb{S})^{\mathcal{R}'}_\mathcal{M} \cdot \textbf{v} \gg 0. 
    \end{align*}
    \item All species in $\mathcal{M}$ are consumed in at least some reaction in $\mathcal{R}'$, or
        \[w > \textbf{0} \text{ for each } w \in \text{rows}((\mathbb{S}^-)_\mathcal{M}^{\mathcal{R}'}).\] 
\end{enumerate}    
\end{lemma}
\begin{proof}
    Let the flow $\textbf{v} \gg \textbf{0}$ on the subnetwork $\mathcal{G}'$ result in a composite reaction of the form 
    \begin{align*}
        \text{(F)} + \textbf{m} \mathcal{M} \to \textbf{n} \mathcal{M} + \text{(W)}. 
    \end{align*}
    Condition $1$ then yields that
    \[ \textbf{n} - \textbf{m} \gg \textbf{0},\]
    and condition $2$ implies that 
    \[\textbf{m} \gg \textbf{0}.\]
    Thus, the conditions stated in the lemma imply that the subnetwork is formally autocatalytic. Conversely, suppose there is a flow $\textbf{v}$ that leads to the composite reaction of the form
    \begin{align*}
        \text{(F)} + \textbf{m} \mathcal{M} \to \textbf{n} \mathcal{M} + \text{(W)}, 
    \end{align*}
    such that $\textbf{0} \ll \textbf{m} \ll \textbf{n}$. Then, 
    \[
    \textbf{m} \gg \textbf{0} \implies \text{rows}((\mathbb{S}^-)_\mathcal{M}) > \textbf{0}, 
    \]
    and 
    \[
    \textbf{n} - \textbf{m} \gg \textbf{0} \implies \mathbb{S}\cdot \textbf{v}|_\mathcal{M} \gg \textbf{0}.
    \]
\end{proof}

\begin{remark}
\label{remark:productivity_implies}
    A matrix $\overline{\mathbb{S}} = \mathbb{S}|_\mathcal{M}$ for which there exists a $\textbf{v} \gg \textbf{0}$ such that 
    \[ \overline{\mathbb{S}} \cdot \textbf{v} \gg \textbf{0} \]
    is also referred to as a \textit{productive} matrix in  \cite[Supplementary Information]{blokhuis2020universal} and is known as a \textit{semi-positive} matrix \cite[Theorem~3.1.13]{johnson_smith_tsatsomeros_2020}. 
    Notice that, productivity implies 
    \[\text{rows}((\mathbb{S}^+)_\mathcal{M}) > \textbf{0}.\]
\end{remark}
\begin{remark}
    The productivity condition can be restated as the condition that the interior of the cone induced by the columns of the stoichiometric matrix intersects the positive orthant in the autocatalytic species, 
    \[ \text{interior(cone(cols(}(\mathbb{S})))) \cap \mathbb{R}_{\geq 0}^{\mathcal{M}} \neq \varnothing. \]
\end{remark}

\begin{remark}
    The pair $(\mathcal{M},\mathcal{R}')$ defines an autocatalytic motif.
\end{remark}
    
\begin{example}
    \label{eg:formal_autocat}
    Consider the reaction network $\mathcal{G}_f = (\{A,B\},\{A \to B, A + B \to  B \})$. Under the flow $\textbf{v} = [1,1]^T$, the resulting composite reaction is 
    \begin{align*}
        2A + B &\to 2B.
    \end{align*}
    Correspondingly, the input and stoichiometric matrix for the CRN are
    \[\mathbb{S}^- = 
    \begin{bmatrix}
    1 & 1 \\
    0 & 1 
\end{bmatrix}
    \]
    and
\[\mathbb{S} = 
    \begin{bmatrix}
    -1 & -1 \\
    \phantom{-}1 & \phantom{-}0 
\end{bmatrix},
    \]
    respectively. It is easy to check that both requirements of Lemma \ref{def:FA_LA} hold true in the species set $\{B\}$. Thus the reaction network is formally autocatalytic for the set $\mathcal{M} = \{B\}$.
\end{example}
\subsubsection{Exclusive autocatalysis}
\begin{definition}[GT]
\label{def:EA_GT}
A formally autocatalytic subnetwork $\mathcal{G}' = (\mathcal{S}',\mathcal{R}')$ of CRN $\mathcal{G}= (\mathcal{S},\mathcal{R})$ is defined to be \textit{exclusively autocatalytic} in the subset $\mathcal{M}$ (see \cite{andersen2021defining,deshpande2014autocatalysis,gopalkrishnan2011catalysis,barenholz2017design}) if: 
\begin{itemize}
    \item Every reaction in $\mathcal{R}'$ consumes at least one species from the set $\mathcal{M}$. This ensures that the flow is inadmissible, or there is no flow, if the population of any species in the set $\mathcal{M}$ is zero. 
\end{itemize}
\end{definition}

\begin{lemma}\label{def:EA_LA}
A formally autocatalytic subnetwork $\mathcal{G}' = (\mathcal{S}',\mathcal{R}')$ of CRN $\mathcal{G}= (\mathcal{S},\mathbb{S}^-,\mathbb{S}^+)$ is \textit{exclusively autocatalytic} in the subset $\mathcal{M}$ if and only if: 
\begin{itemize}
    \item The restriction of the input complexes to the set $\mathcal{M}$ is greater than zero, or
        \[r^- > \textbf{0} \text{ for each } r^- \in \text{cols}((\mathbb{S}^-)_\mathcal{M}^{\mathcal{R}'}).\] 
\end{itemize}
\end{lemma}
\begin{proof}
    It is clear that if the input complex $r^- > \textbf{0}$ for each $r^- \in \text{cols}((\mathbb{S}^-)_\mathcal{M})$, then
    each reaction consumes at least one species from the set $\mathcal{M}$, and vice versa. 
\end{proof}

\begin{remark}
\label{rem:minimality_col}
    If, for a reaction $r^- \to r^+$ in the exclusively autocatalytic subnetwork, the output complex has no species in the set $\mathcal{M}$, then such a reaction can be removed from the subnetwork to yield a smaller exclusively autocatalytic subnetwork. Thus, `smaller' (not necessarily minimal) autocatalytic subnetworks can be obtained by further imposing the condition:
    \[r^+ > \textbf{0} \text{ for each } r^+ \in \text{cols}((\mathbb{S}^+)_\mathcal{M}^{\mathcal{R}'}).\] 
\end{remark}

\begin{remark}
   For a reaction network, a species set $\mathcal{M}$ that satisfies
   \[ \text{cols}((\mathbb{S}^+)_\mathcal{M} > \textbf{0}\]
   and
   \[ \text{cols}((\mathbb{S}^-)_\mathcal{M} > \textbf{0}\]
   is termed as a \textit{siphon} by Deshpande et al.\ in \cite{deshpande2014autocatalysis}. Moreover, if the network is also productive, the set $\mathcal{M}$ is termed as a self-replicating siphon.
\end{remark}

\begin{example}
\label{eg:CRN-exclusive}
One can verify that the CRN $\mathcal{G}$ in Example \ref{eg:CRN-1} is:
\begin{enumerate}
    \item Exclusively autocatalytic in the set $\mathcal{M} = \{A\}$.
    \item Not exclusively autocatalytic in the sets $\mathcal{M} = \{B\}$ or $\{A,B\}$.    
\end{enumerate}
    
Similarly, one can verify that the CRN $\mathcal{G}_f$ in Example \ref{eg:formal_autocat} is not exclusively autocatalytic since the flow proceeds even if the population of $B$ is $0$. 
\end{example}

\subsubsection{Stoichiometric autocatalysis}

\begin{definition}
\label{def:SA_GT}
A subnetwork $\mathcal{G}' = (\mathcal{S}',\mathcal{R}')$ of CRN $\mathcal{G}= (\mathcal{S},\mathcal{R})$ is \textit{stoichiometrically autocatalytic} in the subset $\mathcal{M}$ if: 
\begin{enumerate}
    \item It is exclusively autocatalytic. 
    \item For every reaction $r^- \to r^+ \in \mathcal{R}'$:
    \begin{enumerate}
        \item each of $r^-$ and $r^+$ contain at least one species from the set $\mathcal{M}$.
        \item no species is in both $r^-$ and $r^+$,
        \[ \text{supp}(r^-) \cap \text{supp}(r^+) = \varnothing.\]
    \end{enumerate}
\end{enumerate}
\end{definition}

\begin{lemma}\label{def:SA_LA}
An exclusively autocatalytic subnetwork $\mathcal{G}' = (\mathcal{S}',\mathcal{R}')$ of CRN $\mathcal{G}= (\mathcal{S},\mathbb{S}^-,\mathbb{S}^+)$ is \textit{stoichiometrically autocatalytic} in the subset $\mathcal{M}$ only if: 
\begin{enumerate}
    \item Each column in $\overline{\mathbb{S}} = (\mathbb{S}^+ - \mathbb{S}^-)_\mathcal{M}$ has at least one positive and one negative entry. 
    \item Each row in $\overline{\mathbb{S}}$ has at least one positive and one negative entry.     
\end{enumerate}
\end{lemma}
\begin{proof}
The conditions in Definition \ref{def:SA_GT} implies that every column in the stoichiometric matrix $\mathbb{S}^+-\mathbb{S}^-$ restricted to the autocatalytic set contains a positive and a negative entry. From condition $2$ in Lemma \ref{def:FA_LA} and Remark \ref{remark:productivity_implies}, we know that every row in the input and output matrix pair restricted to the autocatalytic set is semi-positive, respectively. This, along with the condition that every reaction has distinct input and output species, yields that every row in the stoichiometric matrix restricted to the autocatalytic set must contain a positive and a negative entry.   
\end{proof}
\begin{remark}
\label{remark:blokhuis_comp}
    The preceding lemma alludes to the equivalence of our notion of stoichiometric autocatalysis and the autocatalysis of Blokhuis et al.\ in \cite{blokhuis2020universal}. They term the first condition as \textit{autonomy} of the submatrix $\overline{\mathbb{S}}$. The second condition implies that each autocatalytic species is consumed (and produced). However, there is a caveat. The productivity condition for Blokhuis et al.\ is defined over an arbitrary flow vector which can take negative values. In their treatment, they only consider reversible reaction networks where for a reaction with a negative flow, an opposite reaction with a positive flow exists. Since, we do not make such assumptions on our reaction network, we restrict our flows to be strictly in the positive orthant. 
\end{remark}

The condition $\text{cols}((\mathbb{S}^-)_\mathcal{M}) > 0$, ensures that the production of any species \textit{consumes} at least one other species of the motif, disallowing unconditional growth. Notice that this condition is satisfied by exclusive and stoichiometrically autocatalytic subnetworks. However, the condition that the supports of the input and output complexes are disjoint makes the definition of stoichiometric autocatalysis more restrictive than exclusive autocatalysis. A useful feature of stoichiometric autocatalysis is that it can be inferred from the stoichiometric matrix without referring to the underlying CRN, due to which their criterion is termed stoichiometric autocatalysis \cite{peng2022hierarchical}. 

Following the terminology in \cite{deshpande2014autocatalysis}, if there exists a nonnegative flow such that $\overline{\mathbb{S}} \cdot \textbf{v} \ll \textbf{0}$, we would term the motif $(\mathcal{M},\mathcal{R}')$ a \textit{drainable motif}, where the change in concentration of all the species in the subset $\mathcal{M}$ is strictly negative. While we focus our attention on autocatalytic motifs, our results are easily extended to their drainable counterparts. In \cite{deshpande2014autocatalysis}, the authors prove that a network is persistent if it has no drainable motifs and also remark on the importance of autocatalytic and drainable motifs for understanding the dynamics of a network in ecological terms.

\section{Minimal autocatalytic subnetworks}
\label{sec:MASs}

In the previous section, we gave the criteria for ascertaining when a subnetwork is exclusively autocatalytic. However, one autocatalytic subnetwork can have several subnetworks which are also autocatalytic. In Sec. \ref{sec:properties_MAS}, we define and prove some properties of minimal autocatalytic subnetworks (MASs). In Sec.\ \ref{sec:definitions}, we demonstrate that these subnetworks organize within equivalence classes. Finally, in Sec.\ \ref{sec:cones} we investigate how the subnetworks within an equivalence class can be organized by analyzing what we define as their \textit{flow-productive, species-productive} and \textit{partition-productive cones}. 

\subsection{Properties}
\label{sec:properties_MAS}

To define a MAS, we first define an \textbf{autocatalytic core} \cite{blokhuis2020universal}. Recall from Definition \ref{def:motif} that a motif is an arbitrary subhypergraph which is obtained by selecting a subset of the reactions and species in the original CRN. 
\begin{definition}[Autocatalytic core]
    An \textit{autocatalytic core} is an exclusively autocatalytic motif that does not contain a smaller exclusively autocatalytic motif. 
\end{definition}
\begin{remark}
    Note that our definition is a generalization of an \textit{autocatalytic core} as defined by Blokhuis et al.\ as ours is defined for \textit{exclusively}, as opposed to \textit{stoichiometrically}, autocatalytic motifs. 
\end{remark}
\begin{notation}
    We denote the motif of the autocatalytic core $A = (A_C,\mathcal{R}_A)$, its input-output matrix pair as $(\overline{\mathbb{S}^+},\overline{\mathbb{S}^-})$, and term $A_C$ as the \textit{core set}.
\end{notation}

\begin{theorem}
\label{theorem:properties_MAS}
A motif $(A_C,\overline{\mathbb{S}^+},\overline{\mathbb{S}^-})$ is an autocatalytic core only if all the following conditions are satisfied:
\begin{enumerate}
    \item The stoichiometric matrix $\overline{\mathbb{S}} = \overline{\mathbb{S}^+} - \overline{\mathbb{S}^-}$ is productive, that is
    \[ \exists \textbf{v} \gg \textbf{0} \text{ such that } \overline{\mathbb{S}} \cdot \textbf{v} \gg \textbf{0}. \]
    \item Every species in the core set is consumed at least once,
    \[ \text{rows}(\overline{\mathbb{S}^-})>\textbf{0}.\]
    \item Every species in the core set is produced at least once,
    \[ \text{rows}(\overline{\mathbb{S}^+})>\textbf{0}.\]
    \item Every reaction in the motif consumes some species in the core set,
    \[ \text{cols}(\overline{\mathbb{S}^-})>\textbf{0}.\]
    \item Every reaction in the motif produces some species in the core set,
    \[ \text{cols}(\overline{\mathbb{S}^+})>\textbf{0}.\]
\end{enumerate}
\end{theorem}
\begin{proof}
First, notice that every species in the autocatalytic core must be in the autocatalytic set, otherwise it could be removed to yield a smaller autocatalytic motif. Using Lemma \ref{def:EA_LA}, properties $1$, $2$, and $4$ follow. As remarked in \ref{remark:productivity_implies}, property $3$ follows from property $1$. Finally, as remarked in \ref{rem:minimality_col}, if a reaction does not produce species in the autocatalytic set, it could be removed without affecting productivity.  
\end{proof}

\begin{figure*}[t]
\centering
  \includegraphics[width=0.6\linewidth]{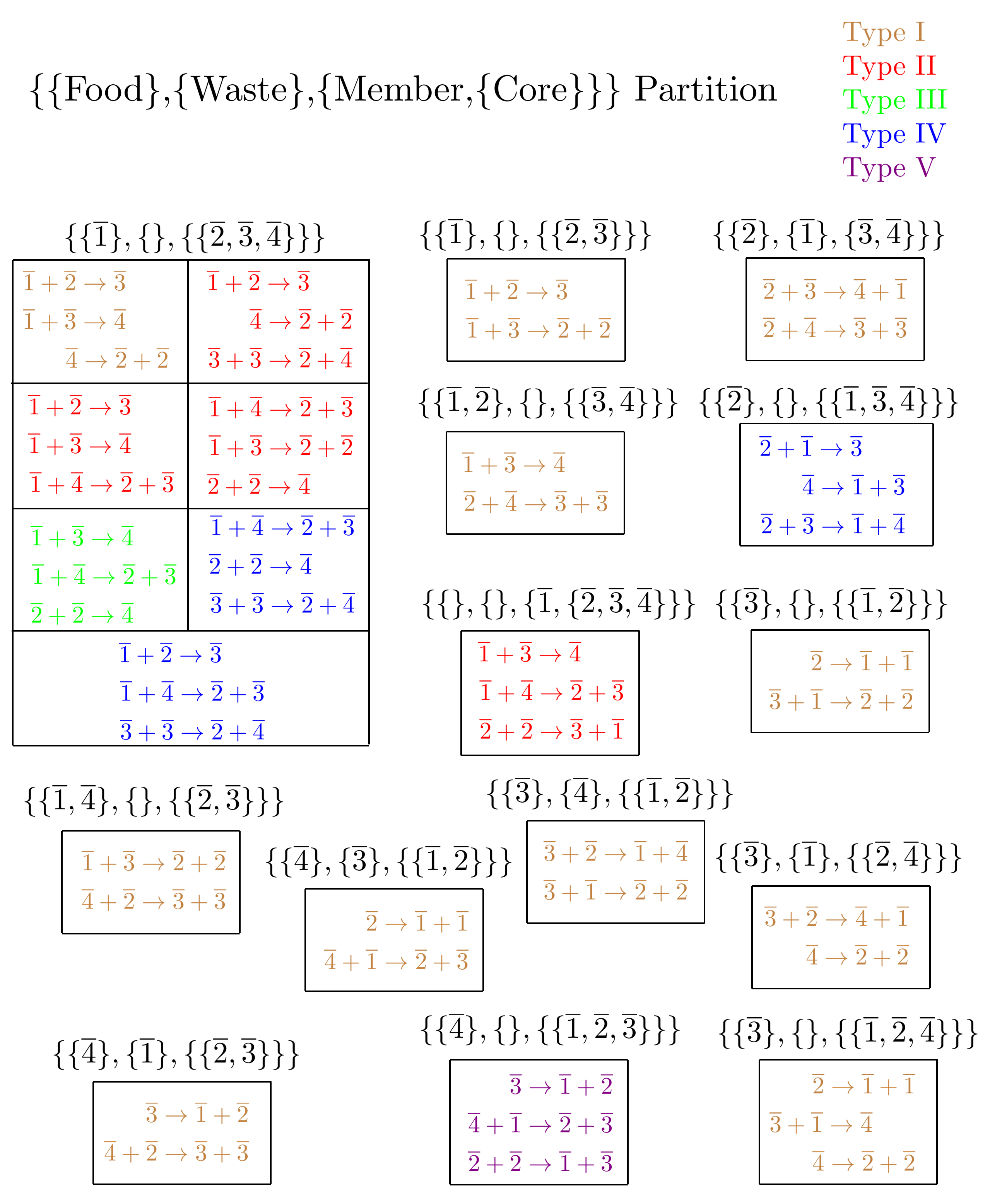}
\caption{All minimal autocatalytic subnetworks (MASs) for the complete 1-constituent CCRN with $L=4$ are shown. The core types are also color coded using the typology of Blokhuis et al.\ in \cite{blokhuis2020universal}. In a CCRN, each species is denoted by a vector of positive integers with an overbar, and each reaction satisfies certain conservation laws. For a definition of a CCRN, see Sec.\ \ref{sec:applications}. The list of reactions can be found in Appendix \ref{sec:MAS_list} Table \ref{t:react4}.}
    \label{fig:cluster_n4}        
\end{figure*}

Furthermore, we can easily extend the proof of Proposition $2$ in the Supplementary Information for \cite{blokhuis2020universal} by Blokhuis et al.\ to show that (exclusively) autocatalytic cores are always square and invertible. 
\begin{lemma}
    Every species in the core set is the only reactant of at least one reaction in the autocatalytic core.  
    \label{lemma:spec_react_core}
\end{lemma}
\begin{proof}
    Let there be a species in the core set that is never the only reactant of any reaction and always occurs as a co-reactant. Since this species is in an autocatalytic core, it must be produced by some reactions. This species occurs either as:
    \begin{enumerate}
        \item an only product of reactions.
        \item a co-product of reactions.        
    \end{enumerate}
    If the species under consideration occurs as an only product of any reaction, then this reaction can be removed to yield a smaller motif which is autocatalytic in the remaining species. If the species only occurs as a co-product of a reaction alongside other species, then the species under consideration can be removed to yield a smaller autocatalytic motif. In either case, minimality of the autocatalytic core is violated, leading to a contradiction.
\end{proof}

Recall from Remark~\ref{remark:productivity_implies} that the stoichiometric matrix of an exclusively autocatalytic motif is semi-positive.
A semi-positive matrix is \textit{minimal semi-positive} \cite[Section~3.5]{johnson_smith_tsatsomeros_2020} if no column can be removed resulting in a semi-positive matrix.

\begin{lemma}
\label{lemma:msp}
    The stoichiometric matrix of an autocatalytic core is a minimal semi-positive matrix.
    \begin{proof}
       Removing a column of the stoichiometric matrix corresponds to removing a reaction from the motif. 
       
       Suppose for the sake of contradiction that removing a column of the stoichiometric matrix of an autocatalytic core results in a motif with semi-positive stoichiometric matrix.
       
       Then removing the species not consumed by any of the remaining reactions results in a smaller motif that also has a semi-positive stoichiometric matrix, and in which every species is consumed by some reaction.
       Moreover, any reaction consuming a species in the original motif still consumes a species in the smaller motif, so the resulting motif is exclusively autocatalytic, contradicting minimality of the core.
    \end{proof}
\end{lemma}
\begin{theorem}
\label{theorem:core_inverse}
The stoichiometric matrix of an autocatalytic core is a square and invertible matrix. In particular, the number of reactions in autocatalytic core equals the number of species in the core set.
\end{theorem}
\begin{proof}
    By Lemma~\ref{lemma:msp}, the stoichiometric matrix of an autocatalytic core is a minimal semi-positive matrix, and so has linearly independent columns whose number is bounded by the number of rows \cite[Theorem~3.5.6]{johnson_smith_tsatsomeros_2020}.
    Lemma \ref{lemma:spec_react_core} shows the number of species, i.e.\ rows, is bounded by the number of reactions, i.e.\ columns. Thus the stoichiometric matrix is square with linearly independent columns, and hence also invertible.
\end{proof}

\begin{remark}
The autocatalytic cores of \cite{blokhuis2020universal} are autocatalytic cores in our sense that are stoichiometrically autocatalytic and satisfy the additional property that no smaller motif becomes autocatalytic after reversing any of its reactions. Blokhuis et al.\ provide a graphical characterization of the stoichiometrically autocatalytic cores and show that they can only be of five types. In Fig.\ \ref{fig:cluster_n4} we have listed all the minimal autocatalytic subnetworks for the complete 1-constituent CCRN (up to a maximum cluster size of $4$) and have color coded them by their types. 
\end{remark}

In a CRN, an autocatalytic core will generally occur embedded in a set of reactions. From Definition \ref{def:subnet}, recall that a subnetwork is obtained from a CRN by selecting a subset of reactions and retaining all the species that participate in the subset. We define \textbf{minimal autocatalytic subnetworks} as the following.
\begin{definition}[MAS]
    A \textit{minimal autocatalytic subnetwork} (MAS) is defined to be the subnetwork with the least number of reactions containing a particular autocatalytic core.   
\end{definition}

\begin{remark}
    For a particular autocatalytic core, a MAS loses its autocatalytic property over all the core species if any reaction is removed. Moreover, if any reaction is added, the subnetwork is not minimal. Thus, a MAS contains exactly the number of reactions as its particular core.  
\end{remark}

\begin{remark}
   \label{rem:2_AC} A MAS can also contain two distinct but overlapping autocatalytic cores. A simple example of such a network is: 
    \[  \mathcal{G} = \{A + B \to C, C \to 2 A + 2 B\}.\]
    Observe that, while $\mathcal{G}$ is a MAS, it contains two autocatalytic cores consisting of autocatalytic species sets $\{A,C\}$ and $\{B,C\}$. 
\end{remark}

\begin{remark}
\label{remark:inflow}
    Observe that, using Theorem \ref{theorem:properties_MAS}, an \textit{inflow reaction} of the type 
    \[ \varnothing \to X_i,\]    
    or an \textit{outflow reaction} of the type 
    \[ X_i \to \varnothing,\]
    can never be in the reaction set of a MAS. We return to this point in Sec.\ \ref{sec:results}.
\end{remark}

\begin{example}
    Consider the CRN $\mathcal{G}$ given by the reaction set
    \[ \mathcal{R} = \{ A \to A + B, B \to A\}.\]
    The input-output matrix pair and the stoichiometric matrix for this CRN are given by:
    \[
    \mathbb{S}^- = 
    \begin{bmatrix}
    1 & 0 \\
    0 & 1 
\end{bmatrix},
\]
\[\mathbb{S}^+ = 
    \begin{bmatrix}
    1 & 1 \\
    1 & 0 
\end{bmatrix},
\]
\[\mathbb{S} = 
    \begin{bmatrix}
    0 & \phantom{-}1 \\
    1 & -1 
\end{bmatrix}.
\]
It can be verified that this CRN is exclusively autocatalytic but \textit{not} stoichiometrically autocatalytic. Note, moreover, consistent with Theorem \ref{theorem:core_inverse}, the stoichiometric matrix is square and invertible.
\end{example}

\begin{example}
\label{eg:CRN-extension}
    Recall Example \ref{eg:CRN-1} with the stoichiometric matrix
\[ \mathbb{S} = \begin{bmatrix}
    \phantom{-}1 & -1 & -1\\
    -1 & \phantom{-}1 &  \phantom{-}1
\end{bmatrix}.
\] 
It can be verified that this CRN is exclusively autocatalytic but not stoichiometrically autocatalytic. To convert this to a stoichiometrically autocatalytic network, we can modify the network by replacing each reaction whose reactant and product complex shares a species with two new reactions with a fictitious distinct intermediate species. This then yields a new network $\mathcal{G}^*$ given by 
\[
\ce{   2 A  <=>  C <=>  A + B  ->  D -> 2 B  },
\]
where $C,D$ are the newly added fictitious species. The stoichiometric matrix for $\mathcal{G}^*$ is given by
\[ \mathbb{S}^* = \begin{bmatrix}
      \phantom{-}2  &    -2        & -1              & \phantom{-}1 & -1         & \phantom{-}0\\
    \phantom{-}0 & \phantom{-}0 &-1              &    \phantom{-}1 & -1         & \phantom{-}2\\
       -1   & \phantom{-}1 & \phantom{-}1   & -1           & \phantom{-}0 &\phantom{-}0\\
       \phantom{-}0 & \phantom{-}0 & \phantom{-}0   &  \phantom{-} 0           & \phantom{-}1 & -1
\end{bmatrix}.
\] 
Notice that the restriction of the stoichiometric matrix to the species set $\mathcal{S}' = \{A,C\}$ and the reactions set $\mathcal{R}' = \{A+B \to C, C\to 2A\}$, denoted by $(\mathbb{S}^*)^{\mathcal{R}'}_{\mathcal{S}'}$, indeed satisfies the properties for an autocatalytic core, and thus $\mathcal{G}^*$ is indeed autocatalytic with core species $\{A,C\}$. A similar construction shows that the network is also autocatalytic in the set $\{B,D\}$. (In the typology of Blokhuis et al.\, both of these cores are of type I.)
\end{example}

\subsection{Food-waste-member-core partition}
\label{sec:definitions}

Let $\mathcal{G} = (\mathcal{S},\mathcal{R})$ be a CRN and $A = (\mathcal{S}_A,\mathcal{R}_A)$ be a MAS autocatalytic in the core species $A_C$. From Theorem \ref{theorem:core_inverse}, we know that the number of core species is the same as the size of $\mathcal{R}_A$. In general, however, there will also be disjoint subsets of $\mathcal{S}_A$ of species that only occur as co-reactants, co-products, and both co-reactants and co-products in $\mathcal{R}_A$. As explained below, given a MAS and an associated autocatalytic core, we can partition the species set into \textit{food-waste-member-core} (FWMC) sets.
\begin{definition}[FWMC]
\label{def:FWMC_def}
    Let the stoichiometric matrix of MAS $A=(\mathcal{S}_A,\mathcal{R}_A)$ with a particular core be denoted by $\overline{\mathbb{S}}$. Denoting the $i^\text{th}$ row of a matrix $\mathbb{M}$ as $(\mathbb{M})_i$, we define:
\begin{enumerate}
    \item the species set corresponding to all rows of $\overline{\mathbb{S}}$ with nonpositive coefficients as the \textit{food set} and denote it by $A_F$,
    \[ X_i \in A_F \text{ iff } (\overline{\mathbb{S}})_i < \textbf{0}.\]
    \item the species set corresponding to all rows of $\overline{\mathbb{S}}$ with nonnegative coefficients as the \textit{waste set} and denote it by $A_W$,
        \[ X_i \in A_W \text{ iff } (\overline{\mathbb{S}})_i > \textbf{0}.\]
    \item the species set corresponding to all rows of $\overline{\mathbb{S}}$ with both positive and negative coefficients as the \textit{member set} and denote it by $A_M$.
    \item the species set corresponding to all rows of $\overline{\mathbb{S}}$ as the \textit{core member set} and denote it by $A_C$. $A_C$ is a subset of $A_M$.    
\end{enumerate}    
\end{definition}

\begin{figure}
    \centering
    \includegraphics[width=\linewidth]{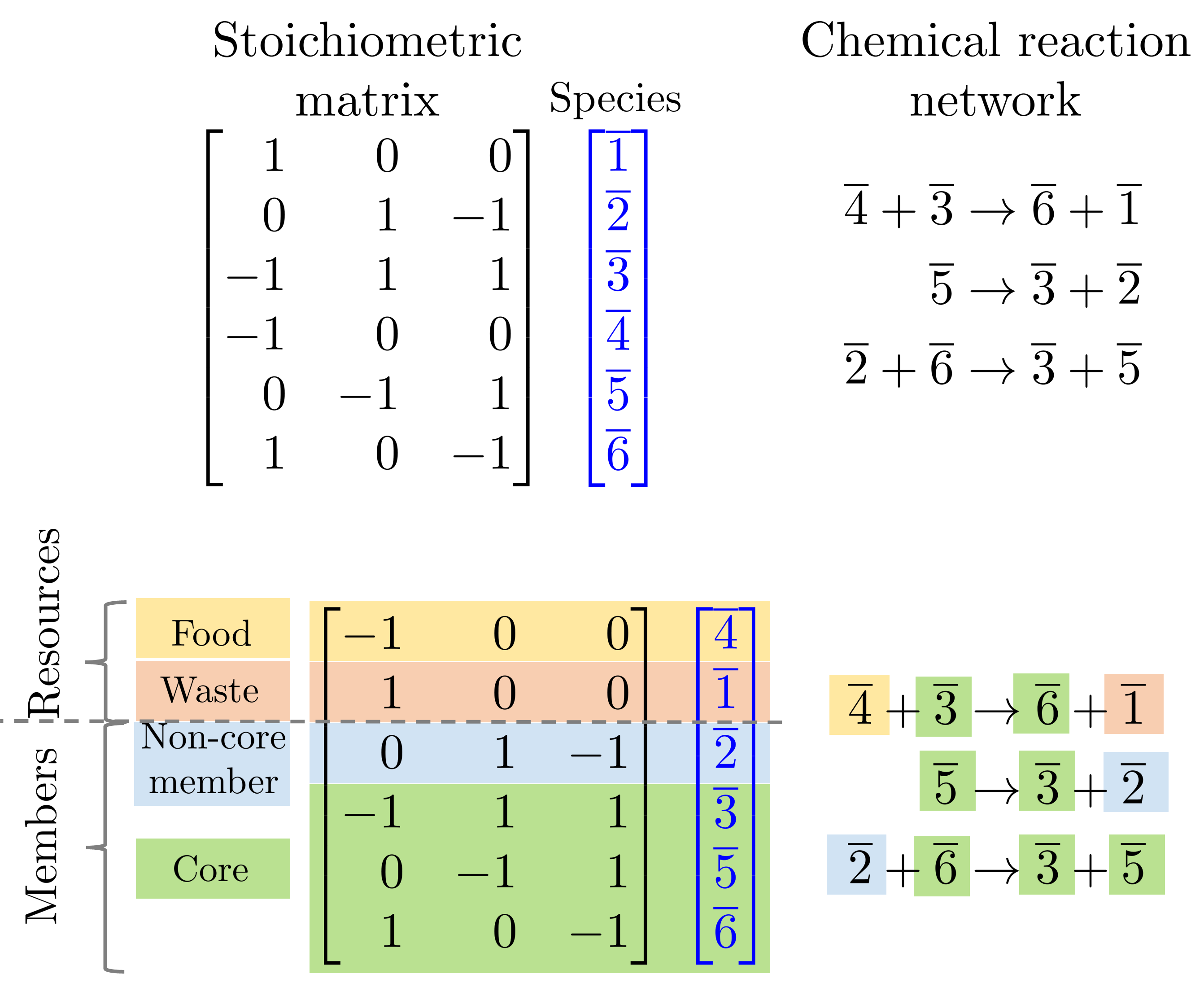}
    \caption{Food-waste-member-core and resource-member partition of a minimal autocatalytic subnetwork in the 1-constituent Cluster CRN (CCRN) with $L=6$. Using the typology of Blokhuis et al.\, this subnetwork has a type two core. }
    \label{fig:FWMC_example}
\end{figure}

\begin{remark}
    A MAS with distinct, but overlapping, autocatalytic cores (for example, see Remark \ref{rem:2_AC}), can have a non-unique FWMC partition.   
\end{remark}
\begin{notation}
\label{not:S_C}
 We will refer to the submatrices of the stoichiometric matrix generated by restricting to the food, waste, member, and core species, as $\overline{\mathbb{S}}_F$, $\overline{\mathbb{S}}_W$, $\overline{\mathbb{S}}_M$, and $\overline{\mathbb{S}}_C$, respectively.     
\end{notation}
\begin{notation}
We denote the exclusion of a subset $K$ from the species set by $A_{/K}$.    
\end{notation}
For example, all species in the species set of an autocatalytic subnetwork $A$ except the core member species will be denoted by $A_{/C}$. Note that the core member set, or simply \textit{core set,} is a subset of the member set, or $A_C \subseteq A_M$. We refer to the species in the member set that are not in the core set as \textit{non-core member species} and denote their set as $A_{M/C}$. For an example of the partitioning process on a reaction network where each set is non-empty, see Fig.\ \ref{fig:FWMC_example}.

We refer to the above partition of the species set for a particular autocatalytic core in a MAS as the \textit{food-waste-member-core} (FWMC) partition of the MAS.
Notice that equality of the FWMC partitions of multiple autocatalytic cores can serve as the basis for an equivalence relation.
\begin{definition}
    Let $a$ and $b$ be two autocatalytic cores embedded in MASs $A$ and $B$. The autocatalytic cores $a$ and $b$ are equivalent if and only if the FWMC partitions of $A$ and $B$ under their associated autocatalytic cores is identical.
\begin{align*}
a \equiv b &\iff \text{FWMC}(A) = \text{FWMC}(B).    
\end{align*}
\end{definition}
\begin{remark}
    In the rest of the work, unless specified otherwise, we will assume that a MAS has a unique autocatalytic core. Under this assumption, each MAS can be uniquely assigned its FWMC partition.
\end{remark}

Our \textit{quadpartite} partition of the species set can be coarse-grained to a bipartite \textit{resource-member} partition where the \textit{resource} set is the union of food and waste sets (and member set is the same as above) (see Fig.\ \ref{fig:FWMC_example}). Our partitioning of the species set by examining the entries of the stoichiometric matrix is cognate with the partitioning done by Avanzini et al.\ in \cite{avanzini2022circuit}, where they partition the set of species into a \textit{resource-member} partition. Also, it is a refinement of, the \textit{food-waste-member} partitioning by Peng et al.\ in \cite{peng2020ecological}, where it is shown that different interactions between cycles can be defined based on which types of species are in common. For example, competition applies when two MASs share a food species or a waste species, mutualism, when the food of one is the waste of another, and predation/parasitism, when a member of one is the food of another (see Fig.\ \ref{fig:Peng_summary}). 

\begin{example}
\label{eg:FWMC}
In the network $\mathcal{G}^*$ from Example \ref{eg:CRN-extension}, the MASs are:
\begin{enumerate}
    \item $A_1 = (\{A,B,C\}, \{A+B \to C, C \to 2 A \}$ with FWMC partition $\{\{B\},\{\},\{\{A,C\}\}\}$.
    \item $A_2 = (\{A,B,D\}, \{A+B \to D, D \to 2 B \}$ with FWMC partition $\{\{A\},\{\},\{\{B,D\}\}\}$.
\end{enumerate}
Notice that both these subnetworks have empty waste and non-core member species sets.

\end{example}

\begin{figure*}[t]
\centering
\begin{subfigure}{.5\textwidth}
  \centering
  \includegraphics[width=0.8\linewidth]{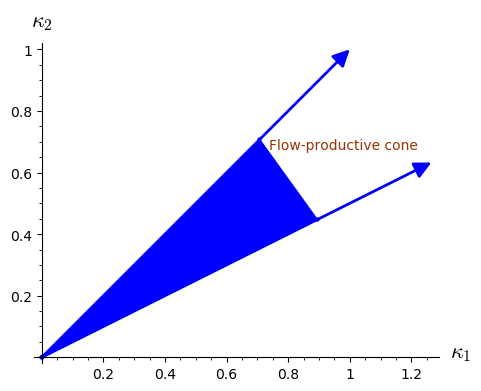}
  %
\end{subfigure}%
\begin{subfigure}{.5\textwidth}
  \centering
  \includegraphics[width=.8\linewidth]{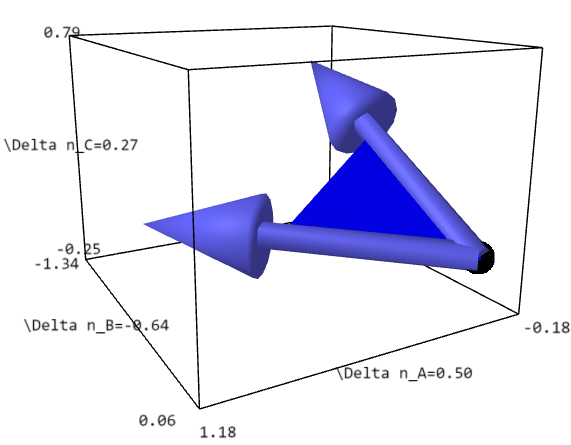}
  %
\end{subfigure}
\caption{The flow and species productive cones for the Example \ref{eg:all-cones} are shown in the left and right panel, respectively. The dimensions of the embedding space of the flow and species productive cones are the number of reactions and species in the minimal autocatalytic subnetwork, respectively. Notice that the food species $B$ is strictly consumed (nonpositive) and the core species $A,C$ are strictly produced (nonnegative) in the species-productive region. For this example, the partition-productive cone is identical to the species-productive cone.  }
\label{fig:test}
\end{figure*}

\subsection{Flow, species and partition productive cones}
\label{sec:cones}
Recall from Notation \ref{not:S_C} and Theorem \ref{theorem:core_inverse} that $\overline{\mathbb{S}}_C$ is invertible. The chemical interpretation of the inverse is that the $k^\text{th}$ column is a flow vector that increases the $k^\text{th}$ species by exactly one unit, making it an elementary mode of production.

\begin{theorem}
    The columns of $\overline{\mathbb{S}}_C^{-1}$ lie in the semi-positive orthant.  
    \label{theorem:inverse_S_C}
\end{theorem}
\begin{proof}
The stoichiometric matrix is a square minimal semi-positive matrix by Lemma~\ref{lemma:msp} and Theorem~\ref{theorem:core_inverse}, so its inverse has non-negative entries by \cite[Corollary~3.5.8]{johnson_smith_tsatsomeros_2020}, and the columns are non-zero non-negative so semi-positive.
\end{proof}
\subparagraph{Flow-productive cone:} 
\begin{definition}
\label{def:flow_productive}
The \textit{flow-productive cone}, denoted by $\mathcal{F}$, is defined to be the cone generated by the elementary modes of production of an autocatalytic core in a MAS,
\begin{align}
    \mathcal{F}(A) &= \text{cone}(\text{cols}(\overline{\mathbb{S}}_C^{-1})).
\end{align}    
\end{definition}
\begin{remark}
    Using Theorem \ref{theorem:inverse_S_C}, the inverse of $\overline{\mathbb{S}}_C$ never contains negative entries. Thus $\mathcal{F}(A)$ lies in the non-negative orthant.
\end{remark}
By definition, an element in the interior of the flow-productive cone, the \textit{flow productive region}, of $A$ corresponds to a flow vector for which $A$ is productive, i.e.\ the core member species of $A$ are strictly produced. 

\subparagraph{Species-productive cone:} 
\begin{definition}
The \textit{species-productive cone}, denoted by $\mathcal{P}$, is defined to be the image of the flow-productive cone of an autocatalytic core in a MAS under its stoichiometric matrix $\overline{\mathbb{S}}$,
\begin{align}
    \mathcal{P}(A) &= \overline{\mathbb{S}} \cdot \mathcal{F}(A) \nonumber \\
    &= \text{cone}(\text{cols}(\overline{\mathbb{S}} \cdot \overline{\mathbb{S}}_C^{-1}).
\end{align}    
\end{definition}
Therefore, each vector in the interior of the species-productive cone for a MAS, which we refer to as the \textit{species-productive region}, describes a change in concentration of the participating species for which the subnetwork is productive in the core member species. 
\begin{remark}
    The \textit{species-consumptive cone} for the \textit{drainable subnetwork} $A^\text{op}$, which is obtained by reversing the edges of the autocatalytic subnetwork $A$, is the opposite of the species-productive cone:
\[\mathcal{P}(A^\text{op})=-\mathcal{P}(A).\]   
\end{remark}

\subparagraph{Partition-productive cone:} Consider a MAS with species set $\mathcal{S}$. Let us denote the FWMC partition of $\mathcal{S}$ as $T = \{T_F, T_W, T_{M/C}, T_C \}$. Let $e_i$ be a column vector of size $\mathcal{S}$ with $1$ at the $i^\text{th}$ position and $0$ elsewhere. We define the set of basis vectors $\mathcal{B}_T$ for the partition $T$ to be
\begin{align}
    \mathcal{B}_T &= \mathcal{B}_T^F \cup \mathcal{B}_T^W \cup \mathcal{B}_T^{M/C} \cup \mathcal{B}_T^C  \label{eq:part_prod_cone}
\end{align}
where 
\begin{align*}
    \mathcal{B}_T^F &= \{ -e_f | f \in T_F \},\\
    \mathcal{B}_T^W &= \{ e_w | w \in T_W \},\\
    \mathcal{B}_T^{M/C} &= \{ -e_m \cup e_m | m \in T_{M/C} \},\\
    \mathcal{B}_T^C &= \{ e_c | c \in T_C \}.\\
\end{align*}
Thus the basis $\mathcal{B}_T$ for a partition $T$ spans the negative half-space of each species in the food set, the positive half-space of each species in the waste and core set, and both negative and positive half-spaces of each species in the non-core member set. Let the reaction network have conservation laws $\{\textbf{x}_i\}$, such that $\textbf{x}_i\cdot \mathbb{S} = 0$. 
\begin{definition}
    The \textit{partition-productive cone} for the partition $T$, denoted by $\mathcal{Q}(T)$, is defined to be the cone generated by the basis vectors restricted to the stoichiometric subspace
\begin{align*}
    \mathcal{Q}(T) &= \text{cone}(\mathcal{B}_T) \perp \{\textbf{x}_i\}.
\end{align*}
\end{definition}
As noted in the previous subsection, belonging to an FWMC partition is an equivalence relation. By definition, the species-productive cones of all equivalent autocatalytic subnetworks belonging to the partition $T$ will lie in the partition-productive cone $\mathcal{Q}(T)$.  

\begin{example}
\label{eg:all-cones}
    In this example, we will identify the flow-productive, species-productive and partition-productive cones for the MAS $A_1$ from Example \ref{eg:FWMC} (shown in Fig.\ \ref{fig:test}). For $A_1 = (\{A,B,C\}, \{A+B \to C, C \to 2 A \}$,
    \begin{align*}
        \overline{\mathbb{S}} &= 
        \begin{bmatrix}
            -1  & \phantom{-}2 \\
            -1 & \phantom{-}0 \\
            \phantom{-}1 & -1  
        \end{bmatrix}\\
        \overline{\mathbb{S}}_C &= 
        \begin{bmatrix}
            -1  & \phantom{-}2 \\
            \phantom{-}1 & -1 
        \end{bmatrix}\\
         \overline{\mathbb{S}}_C^{-1} &= 
       \begin{bmatrix}
           1 & 2\\
           1 & 1
       \end{bmatrix}.
    \end{align*}
    The flow-productive cone is 
    \begin{align*}
        \mathcal{F}(A_1) &= \text{cone}\left( \begin{bmatrix}
            1\\
            1
        \end{bmatrix}, \begin{bmatrix}
            2\\
            1
        \end{bmatrix}  \right),
    \end{align*}
        and the species-productive cone is 
    \begin{align*}
        \mathcal{P}(A_1) &= \text{cone}\left( \begin{bmatrix}
            \phantom{-}1\\
            -1\\
            \phantom{-}0
        \end{bmatrix}, \begin{bmatrix}
            \phantom{-}0\\
            -2\\
            \phantom{-}1
        \end{bmatrix}  \right).
    \end{align*}
    Notice that the graph $\mathcal{G}^*$ (from Example \ref{eg:CRN-extension}) has a positive conservation law $\textbf{x} = [1,1,2]$. The partition-productive cone for the partition $T_1=\{\{B\},\{\},\{\{A,C\}\}\}$ is 
    \begin{align*}
        \mathcal{Q}(T_1) &= \text{cone}\left(  \begin{bmatrix}
            \phantom{-}0\\
            -1\\
            \phantom{-}0
        \end{bmatrix}, 
         \begin{bmatrix}
            1\\
            0\\
            0
        \end{bmatrix},
         \begin{bmatrix}
            0\\
            0\\
            1
        \end{bmatrix}
        \right) \perp \textbf{x}\\
        &= \text{cone}\left( \begin{bmatrix}
            \phantom{-}1\\
            -1\\
            \phantom{-}0
        \end{bmatrix}, \begin{bmatrix}
            \phantom{-}0\\
            -2\\
            \phantom{-}1
        \end{bmatrix}  \right)\\ 
        &=\mathcal{P}(A_1).
    \end{align*}
For this example, the species-productive and partition-productive cones are identical, which is an instance of a general result proved in Theorem \ref{theorem:partition-prod}.    
\end{example}

\section{Organizing an autocatalytic ecosystem}
\label{sec:organization}

In the previous section, we have defined and proved properties of minimal autocatalytic subnetworks (MASs), or minimal CRNs that contain one autocatalytic core. Even mildly complicated CRNs can exhibit an abundance of autocatalytic cores \cite{personalconv,peng2022hierarchical}. 
\begin{definition}
    We define an \textbf{autocatalytic ecosystem} to be a CRN containing one or more minimal autocatalytic subnetworks (MASs). 
\end{definition}
In this section, we investigate some mathematical properties of autocatalytic ecosystems and give computational algorithms to detect and organize them. In Sec.\ \ref{sec:results}, we prove some mathematical results about MASs. In Sec.\ \ref{sec:algorithms} and Appendix \ref{app:alg}, we provide algorithms to exhaustively enumerate the MASs and identify their species-productive cones. Finally, we discuss the polyhedral geometry of an autocatalytic ecosystem and introduce a visualization scheme in Sec.\ \ref{sec:visualization}. 

\subsection{Mathematical results}
\label{sec:results}
To understand the geometry of the different cones defined in the previous subsection, we prove some results about their behavior. First, in Proposition \ref{prop:food-set} we give the conditions under which two MASs with different partitions will have non-intersecting species-productive regions. Next, in Theorem \ref{theorem:partition-prod}, we explore conditions under which the species-productive cone is identical with the partition-productive cone for a MAS. Under such conditions, the partition itself contains the information of the species-productive regions of all the MASs in that equivalence class. Finally, in Proposition \ref{prop:null-flows}, we clarify the topological properties a CRN must posses in order for the species-productive cones of two MASs to intersect. In Sec.\ \ref{sec:visualization} we will use these results to construct a visualization scheme for the list of MASs in any CRN.

\begin{proposition}
\label{prop:food-set}
Two autocatalytic subnetworks with different food sets have disjoint species-productive regions if their non-core member species sets are empty.
\end{proposition}

\begin{proof}
We will show that if two autocatalytic subnetworks $A$ and $B$ have distinct food sets and their non-core member species sets are empty, then the interiors of their species-productive cones (species-productive regions) do not intersect, 
\[\text{int}(\mathcal{P}(A)) \cap \text{int}(\mathcal{P}(B)) = \vn.\]
Let $A_{F}/B_{F}\neq \vn$. Recall that that the species-productive cone of the subnetworks lies within their partition-productive cone $\mathcal{Q}(\text{FWMC}(A))$ and $\mathcal{Q}(\text{FWMC}(B))$. Let $\textbf{y}$ be the vector of length $\mathcal{S}$ such that
\begin{align*}
    \bigr[ \textbf{y} \bigr]_i &=
    \begin{cases}
   -1   \text{ if } i \in A_{F}/B_{F}\\
    \phantom{-}0 \text{ otherwise.}
    \end{cases}
\end{align*}
Then, by definition of the partition-productive cones, $\textbf{y} \cdot \textbf{v}  >0 $ and $\textbf{y} \cdot \textbf{w} \leq 0 $, where $\textbf{v}$ and $\textbf{w}$ are any vectors in  $\text{int}(\mathcal{Q}(\text{FWMC}(A)))$ and $\text{int}(\mathcal{Q}(\text{FWMC}(B)))$, respectively. Thus, $\textbf{y}$ defines a hyperplane separating the two species-productive regions, and the two regions do not intersect.
\end{proof}
\begin{remark}
If the non-core member species set is not empty, then the result is no longer true. For instance, see Example \ref{eg:4species-partitionproductive}.
\end{remark}

\begin{example}
    \label{eg:food-intersect}
    Using Examples \ref{eg:FWMC} and \ref{eg:all-cones}, the species-productive cones for $A_1$ and $A_2$ in the species set $\{A,B,C,D\}$ are
    \begin{align*}
        \mathcal{P}(A_1) &=
        \text{cone}\left( \begin{bmatrix}
            \phantom{-}1\\
            -1\\
            \phantom{-}0\\
            \phantom{-}0
        \end{bmatrix}, \begin{bmatrix}
            \phantom{-}0\\
            -2\\
            \phantom{-}1\\
            \phantom{-}0
        \end{bmatrix}  \right),\\
        \mathcal{P}(A_2) &=
        \text{cone}\left( \begin{bmatrix}
            -1\\
            \phantom{-}1\\
            \phantom{-}0\\
            \phantom{-}0
        \end{bmatrix}, \begin{bmatrix}
            -2\\
            \phantom{-}0\\
            \phantom{-}0\\
            \phantom{-}1
        \end{bmatrix}  \right).
    \end{align*}
    If we let $\textbf{y} = [0, -1, 0, 0] $, then $\textbf{y}\cdot\text{int}(\mathcal{P}(A_1)) >0$ and $\textbf{y}\cdot\text{int}(\mathcal{P}(A_2)) <0$. Thus $\textbf{y}$ defines a separating hyperplane, and the two species-productive regions are disjoint. 
\end{example}

\begin{lemma}
\label{lemma:core_identity}
The restriction of the species-productive cones of all MASs to their core (member) species is the non-negative orthant in the core species.
\end{lemma}
\begin{proof}
    Recall that the species-productive cone of a MAS is defined as 
    \begin{align*}
         \mathcal{P}(A) &= \text{cone}(\text{cols}(\overline{\mathbb{S}} \cdot \overline{\mathbb{S}}_C^{-1})).
    \end{align*}
    By definition of the matrix inverse, the restriction of the product to the core species is the identity matrix, \[\overline{\mathbb{S}} \cdot \overline{\mathbb{S}}_C^{-1}\big|_C = I_C.\]
    Thus, the species-productive cone restricted to the core species is the non-negative orthant.
\end{proof}

As remarked in Remark~\ref{remark:inflow}, a MAS cannot contain an inflow or outflow reaction. Moreover, for material CRNs, where the species are made of constituents that are conserved in a reaction, a MAS must also possess at least one conservation law. The following results are applicable for such MASs. (In Sec.\ \ref{sec:applications}, we develop the formalism for CRNs with non-negative integer conservation laws and find their application in the examples.) 
\begin{lemma}
\label{lemma:food}
    If a CRN has any positive conservation laws, the union of the food set and the non-core member set of any MAS must be non-empty.
\end{lemma}
\begin{proof}
    Let $A$ be a MAS, with associated stoichiometric matrix $\overline{\mathbb{S}}$, of a CRN with a positive conservation law given by the vector $\textbf{x}$. Assume for the sake of contradiction that $A$ does not have any element in the food set $A_F$ or the non-core member set $A_{M/C}$. For any flow $\textbf{v}$ in the productive region of $A$, we know that $\overline{\mathbb{S}} \cdot \textbf{v} \gg 0$. Also, since $\textbf{x}$ is a positive conservation law, $\textbf{x}\cdot \overline{\mathbb{S}} \cdot \textbf{v} \overset{!}{=} 0$. Since the sum of positive values cannot add up to zero, we have a contradiction. Thus, there must be at least one element in the food set or the non-core member set of a MAS with positive conservation laws. (See example \ref{eg:4species-partitionproductive}.)
\end{proof}

\begin{theorem}
\label{theorem:partition-prod}
    For a CRN with positive conservation laws, the species-productive cone of any MAS with exactly one more species than the core set is identical to its partition-productive cone.  
\end{theorem}

\begin{proof}
    Let $A$ be a MAS of a CRN with positive conservation laws $\{\textbf{x}_i\}$ with exactly one more species than the core set, denoted by $C = \{ C_1, \ldots, C_C \}$. From Lemma \ref{lemma:food}, we know that the extra species must be either in the food set or non-core member set of $A$. In either case, the species must be net consumed by the subnetwork to respect the positive conservation laws, and we denote it by $f$. \\
    Let us label the species set as $\{f, C_1, \ldots, C_C\}$. Recall that the partition-productive cone is defined as $\mathcal{Q}(\text{FWMC}(A)) = \text{cone}(\mathcal{B})\perp \{\textbf{x}_i\}$, where
    \begin{align*}
        \mathcal{B} &= \left\{ 
        \begin{bmatrix}
            -1 \\
            \phantom{-} 0 \\
            \phantom{-}\vdots \\
            \phantom{-}0
        \end{bmatrix} ,
        \begin{bmatrix}
            0 \\
            1 \\
            \vdots \\
            0
        \end{bmatrix} ,
        \ldots ,
                \begin{bmatrix}
            0 \\
            0 \\
            \vdots \\
            1
        \end{bmatrix} 
        \right\}.   
    \end{align*}
    Note that the additional basis vector for the non-core member species with positive entries is omitted since we know that it must be consumed in the species-productive region to respect the positive conservation laws. Moreover, using Lemma \ref{lemma:core_identity}, we know that the restriction of the species-productive cone to the core species is the non-negative orthant in the core species. In particular, for the species-productive cone $\mathcal{P}(A) = \text{cone}(\text{cols}(\overline{\mathbb{S}} \cdot \overline{\mathbb{S}}_C^{-1})$, we have 
   \begin{align*}
       \overline{\mathbb{S}} \cdot \overline{\mathbb{S}}_C^{-1}   
       &= 
                        \begin{bmatrix}
                            \textbf{f}\\
                            I_C
                        \end{bmatrix},
    \end{align*}
    where $\textbf{f}$ is a row vector $[\textbf{f}_1, \ldots, \textbf{f}_C]$ and $I_C$ is the identity matrix of size $C$. Since the CRN also has conservation laws, these coefficients must satisfy
    \begin{align*}
        \bigr[\textbf{x}_i\bigr]_f \cdot \textbf{f}_1 +  \bigr[\textbf{x}_i\bigr]_{C_1} \cdot 1 &= 0 \\
        \vdots \\
        \bigr[\textbf{x}_i\bigr]_f \cdot \textbf{f}_C +  \bigr[\textbf{x}_i\bigr]_{C_C} \cdot 1 &= 0, \\
    \end{align*}
    for every conservation law indexed by $i$. But these are simply the basis vectors of the partition-productive cone $\mathcal{B}$ when made orthogonal to the conservation laws. Thus the partition-productive cone is identical to the species-productive cone. 
\end{proof}

\begin{example}
    \label{eg:4species-partitionproductive}
    Consider two MASs (from Sec.\ \ref{sec:CCRN-1constituent} Fig.\ \ref{fig:cluster_n4}) 
    \begin{align*}
        A_1 &= (\{ \overline{1}, \overline{2}, \overline{3}, \overline{4}\},
        \{ \overline{1} + \overline{3} \to \overline{4}, \\
        & \phantom{=((} \overline{1} + \overline{4} \to \overline{2} + \overline{3},
         \overline{2}+ \overline{2} \to \overline{3} + \overline{1}\}  ),\\
             A_2 &= (\{ \overline{1}, \overline{2}, \overline{3}, \overline{4}\},
    \{ \overline{1} + \overline{3} \to \overline{4}, \\
    & \phantom{=((} \overline{1} + \overline{4} \to \overline{2} + \overline{3},
     \overline{2}+ \overline{2} \to \overline{4}\}  ).
    \end{align*}
    The partitions of $A_1$ and $A_2$ are $\{\{\},\{\},\{\overline{1},\{\overline{2},\overline{3},\overline{4}\}\}\}$ and $\{\{\overline{1}\},\{\},\{\{\overline{2},\overline{3},\overline{4}\}\}\}$, respectively. A simple calculation yields that their species-productive cones are in fact identical, and
    \begin{align*}
    \mathcal{P}(A_1) = \mathcal{P}(A_2) &=
         \text{cone}\left( \begin{bmatrix}
            -2\\
            \phantom{-}1\\
            \phantom{-}0\\
            \phantom{-}0
        \end{bmatrix}, \begin{bmatrix}
            -3\\
            \phantom{-}0\\
            \phantom{-}1\\
            \phantom{-}0
        \end{bmatrix}, \begin{bmatrix}
            -4\\
            \phantom{-}0\\
            \phantom{-}0\\
            \phantom{-}1
        \end{bmatrix}  \right).
    \end{align*}
\end{example}

\begin{definition}
For a CRN $\mathcal{G} = (\mathcal{S},\mathcal{R})$, the reversible CRN 
$\mathcal{G}_\text{rev}= (\mathcal{S},\mathcal{R}')$ is defined such that for every reaction $y_1\to y_2 \in \mathcal{R}$, the set $\mathcal{R}'$ contains both $y_1 \to y_2$ and $y_2 \to y_1$. 
\end{definition}
\begin{remark}
     Reversible CRNs are worth considering insofar as chemistry often assumes that all reactions are theoretically reversible (even if the rate constants of forward and reverse reactions differ greatly in magnitude).
\end{remark}

\begin{proposition}
\label{prop:null-flows}
    If the species-productive cones of two MASs have an intersection, the reversible CRN of their union has a null flow.
\end{proposition}
\begin{proof}
    Let the productive cones of two MASs $A$ and $B$, have a non-empty intersection, $\mathcal{P}(A)\cap \mathcal{P}(B)\neq \vn$. Let $A\cup B$ be the subnetwork obtained by taking the union of the two subnetworks, and let $\mathbb{S}_{A\cup B}$ be the associated stoichiometric matrix. Then, there must be non-identical flows $\textbf{v}_A$ and $\textbf{v}_B$ with support in $A$ and $B$, respectively, such that
    \[ \mathbb{S}_{A\cup B} \cdot \textbf{v}_A = \mathbb{S}_{A\cup B} \cdot \textbf{v}_B.\]
    This implies that $\mathbb{S}_{A\cup B} \cdot (\textbf{v}_A - \textbf{v}_B) = 0$, and thus the kernel of the stoichiometric matrix is non-empty. In general, the difference of their flows $(\textbf{v}_A - \textbf{v}_B)$ can have negative entries, however, it yields the interpretation of a null flow on a network if the union of the two subnetworks $A\cup B$ is made reversible.  
\end{proof}

\begin{remark}
    The converse statement, if the reversible CRN of the union of two MASs contains a null cycle then their species-productive cones must intersect, is not true. For example, consider the MASs (from Sec.\ \ref{sec:CCRN-1constituent} Fig.\ \ref{fig:cluster_n4})
    \begin{align*}
        A_1 &= (\{ \overline{1}, \overline{2}, \overline{3}\},
        \{ \overline{1} + \overline{2} \to \overline{3},  \overline{1} + \overline{3} \to \overline{2}+ \overline{2}\}),  \\
        A_2 &= (\{ \overline{1}, \overline{2}, \overline{3}\},
        \{ \overline{2} \to \overline{1} +  + \overline{1},  \overline{3} + \overline{1} \to \overline{2}+ \overline{2}\}).  
    \end{align*}
    Notice that $\text{FWMC}(A_1) = \{\{\overline{1}\},\{\},\{\{ \overline{2},\overline{3}\}\}\}$ and $\text{FWMC}(A_2) = \{\{\overline{3}\},\{\},\{\{ \overline{1},\overline{2}\}\}\}$. Thus, using Proposition \ref{prop:food-set}, the two species-productive cones do not intersect. However, the union of the two subnetworks yields 
    \begin{align*}
        A_1\cup A_2 &= (\{ \overline{1}, \overline{2}, \overline{3}\},  \{ \overline{1} + \overline{2} \to \overline{3},\\
        &\phantom{=((}
         \overline{1} + \overline{3} \to \overline{2}+ \overline{2}, \overline{2} \to \overline{1} +  \overline{1}\}).
    \end{align*}
    which has a null cycle (with null flow $[1,1,1]^T$).
\end{remark}

\begin{figure}[t]
\includegraphics[scale=0.5]{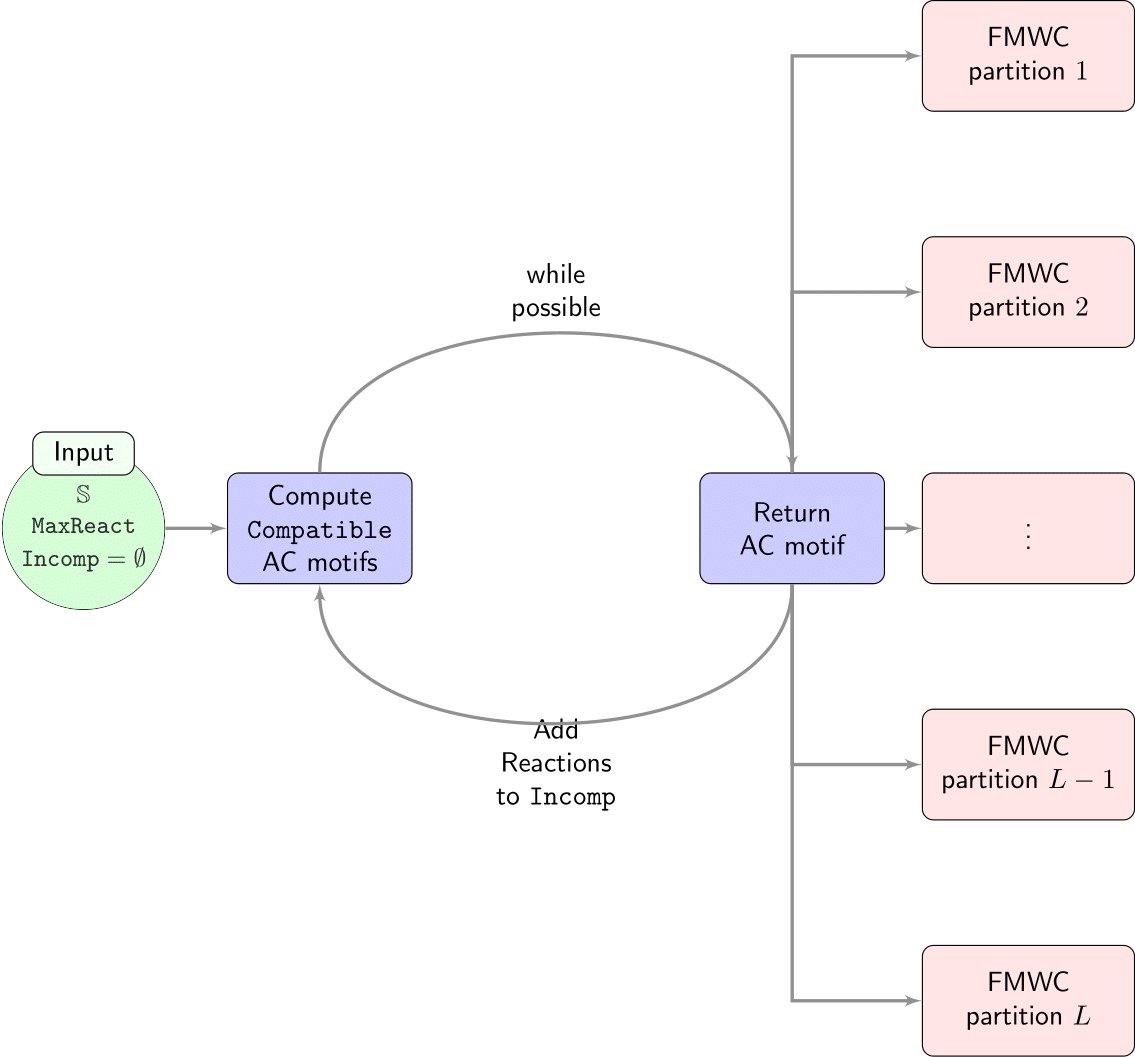}
\caption{Flowchart of our procedure to construct minimal autocatalytic subnetworks.\label{fig:flowchart}}
\end{figure}

\subsection{Algorithms}
\label{sec:algorithms}

In what follows we describe a mathematical optimization based approach to enumerate the whole list of exclusively autocatalytic cores in a CRN. The model that we propose uses the following variables:
\begin{align*}
y_i &= \begin{cases}
1 & \mbox{if species $i\in$ Core set,}\\
0 & \mbox{otherwise}
\end{cases}.
\end{align*}
for all $i \in \mathcal{S}$.

$\mathbf{v} = [v_1, \ldots, v_{|\mathcal{R}|}] \in [0,1]^{|\mathcal{R}|}_+$: flow inducing a MAS.

$z_r = \begin{cases}
1 & \mbox{if $v_r>0$},\\
0 & \mbox{otherwise}
 \end{cases}$,
for all $r \in \mathcal{R}$.

These variables are adequately combined by means of linear constraints to ensure the verification of the properties of Theorem \ref{theorem:properties_MAS}. Specifically:
\begin{itemize}
    \item The species in the Core set ($y_i=1$) are the productive species of the selected reactions:
    \begin{align}
     \sum_{r\in \mathcal{R}} (\mathbb{S})_{i}^{r} v_r \geq & \epsilon - \Delta_i (1-y_i), \forall i \in \mathcal{S}, \label{eq:semi_positivity}
     \end{align}
     Note that in case $y_i=1$, the linear inequality is activated ensuring that the $i$-th species is produced in net. Otherwise, the constraint is redundant.
     \item All species in the Core set are both consumed and produced by the reactions in the core:
     \begin{align}
         \sum_{r\in \mathcal{R}:\atop (\mathbb{S}^+)^r_i>0} z_r\geq & y_i, \quad \forall i \in \mathcal{S},
      \label{eq:exc_out}\\
          \sum_{r\in \mathcal{R}: \atop (\mathbb{S}^-)^r_i>0} z_r \geq & y_i, \quad     \forall i \in \mathcal{S},
    \label{eq:exc_in}
     \end{align}
     The two sets of inequalities above ensure that in case $y_i=1$, there must exist reactions in the core (those with $z_r=1$) for which species $i$ with $(\mathbb{S}^+)^r_i>0$ (for \eqref{eq:exc_out}) and $(\mathbb{S}^-)^r_i>0$ (for \eqref{eq:exc_in}).
     \item All reactions in the core both consume and produce core species:
\begin{align}
           \sum_{i\in \mathcal{S}: \atop (\mathbb{S}^+)^r_i} y_i \geq & z_r, \quad  \forall r \in \mathcal{R},
      \label{eq:cons_out}\\ 
    \sum_{i\in \mathcal{S}: \atop (\mathbb{S}^-)^r_i} y_i  \geq & z_r, \quad \forall r \in \mathcal{R}. 
    \label{eq:consum_in}
\end{align}
In case reaction $r$ is part of the core ($z_r=1$), there exists at least one produced core species (inequality \eqref{eq:cons_out}) and at least one consumed species (inequality \eqref{eq:consum_in}).
\item The non-selected reactions have zero-flow:
\begin{equation}\label{eq:support}
v_r \leq z_r, \forall r \in \mathcal{R}.
\end{equation}
In case reaction $r$ is not included in the core ($z_r=0)$, the flow component for this reaction is forced to be zero ($v_r=0$). Otherwise, the flow is unrestricted ($v_r\leq 1$).
\end{itemize}
Then, the master mixed integer programming (MILP) problem that allows us to generate the entire list of exclusively autocatalytic cores consists of minimizing the number of reactions in a network with all the above conditions:
\begin{align*}
    \min  & \sum_{r\in \mathcal{R}} z_r\\
    \mbox{s.t. } & \eqref{eq:semi_positivity}-\eqref{eq:support},\\
    & y_i \in \{0,1\},  \forall i \in \mathcal{S},\\
    & z_r \in \{0,1\}, \forall r \in \mathcal{R},\\
    & \mathbf{v} \in [0,1]^{|\mathcal{R}|}_+.
\end{align*}
The general procedure consists of sequentially solving the above optimization problem and adding to it new linear constraints to allow the generation of different cores. A flowchart of our procedure to construct the minimal exclusively autocatalytic cores is shown in Fig. \ref{fig:flowchart}. Define a list of \emph{sets of incompatible reactions}, \texttt{Incomp}, initialized to the empty set, that stores, at each iteration, the sets of reactions that are part of the already computed MASs, and avoids generating MASs containing those sets of reactions. Then, in the first step, our algorithm computes an autocatalytic subnetwork by selecting a set of core species and reactions taking part in a subnetwork with minimum number of reactions. The obtained subnetwork is stored, and its FWMC partition is computed. This set of reactions, say $\mathcal{R}^\prime$, is added to \texttt{Incomp}, and the procedure is repeated, by adding the following constraint to the MILP to assure that the subnetworks obtained in future iterations do not contain any of the sets of reactions in \texttt{Incomp}:
$$
\sum_{r\in \mathcal{R}^\prime} z_r \leq |\mathcal{R}^\prime|-1.
$$
The algorithm stops when there does not exist sets of reactions not \emph{incompatible} with those in \texttt{Incomp}. Minimality is assured by the minimization of the number of reactions in the subnetwork in each run and by the conditions imposed by the set \texttt{Incomp}. Note that the complete enumeration of all the possible combinations of reactions/species that may be included in a MAS, and the detection of a MAS is computationally prohibitive in practice. However, our procedure avoids this enumeration by solving an MILP at each iteration. Moreover, if one is interested on finding just a certain number, $k$, of MAS, it can be done by terminating the algorithm after $k$ iterations and giving as output the $k$ obtained MASs.

The identification of the FWMC partition for a given autocatalytic subnetwork results from checking both the signs of the restricted stoichiometric matrix, $\overline{\mathbb{S}}$ and the production vector $\overline{\mathbb{S}} v^*$ for the flow $v^*$ also obtained after each iteration. Specifically, for each species $i$ in the subnetwork: if all the components in $\overline{\mathbb{S}}_i$ are non positive (resp.\ non negative), $i$ is classified as a food (resp.\ waste) species; in case $\bar S_i$ has negative and positive entries, $i$ is classified as a member species; and if $\overline{\mathbb{S}}_i$ has negative and positive entries and $(\overline{\mathbb{S}} v^*)_i$ is strictly positive, then $i$ is a core species.  

Once the set of MASs is obtained, they are classified by means of their FWMC partitions. (Additional checks must be made to determine if there is more than one autocatalytic core in a MAS. In this case, the FWMC partition of the subnetwork is non-unique, and the MAS must be assigned to all the FWMC classes to which its autocatalytic cores belong.) Within each FWMC class, we check the \textit{pairwise intersection} of the species-productive cones. For each subnetwork $A_1, A_2$ in the same class, we first check whether the cones $\mathcal{P}(A_1)$ and $\mathcal{P}(A_2)$ intersect by finding a nonzero vector shared by both cones. In case the cones intersect, we check whether the cones are identical, one is contained in the other, or they partially intersect. This check is performed by computing the distances between the (normalized) generators of the cones. If all these distances are zero, the cones are identical; if one of the sets of distances is zero but the other is not, one of the cones is contained in the other; but if in both sets there are positive distances, the cones partially intersect. Using the same procedure, we also check the pairwise intersection of the partition-productive cones of each equivalence class. Further details on the described approaches are provided in Appendix \ref{app:check_intersection}.

\begin{figure*}
    \centering
    \includegraphics[width=0.95\textwidth]{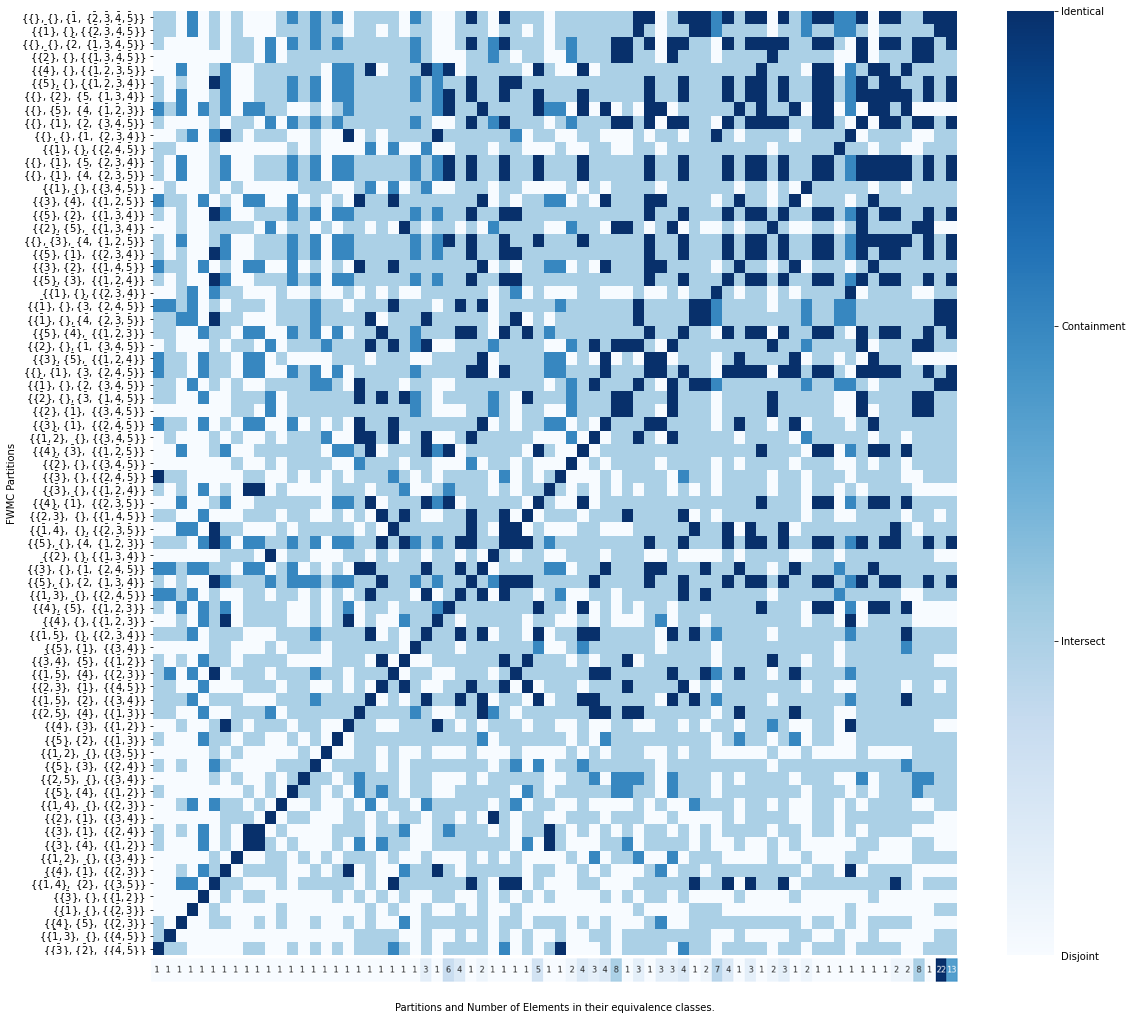}
    \caption{The intersection data of partition-productive cones for all partitions of the complete $L=5$ 1-constituent CCRN (described in Sec.\ \ref{sec:CCRN-1constituent}). }
    \label{fig:intersection_partition_5}
\end{figure*}

\subsection{Geometry and visualization}
\label{sec:visualization}

In the last subsection, we proposed an algorithm that, given a stoichiometric matrix of a CRN, outputs a list of the MASs it contains. In this section, we will explain how to take the list of MASs and visualize their combinatorics and geometry. Recall that for each MAS, we define a flow-productive cone in the flow space of the graph where each core member species is strictly produced, and a species-productive cone on the space of changes in concentration (population). Geometrically, the space of changes in chemical concentration (i.e., the velocity space of chemical concentrations) is the stoichiometric subspace of the CRN. Thus, the list of MASs can be seen as yielding a partial polyhedral decomposition of the stoichiometric space induced from the flows on the hypergraph (CRN). We remark that, for a more complete decomposition one must also consider the consumptive cones of the minimal drainable subnetworks. However, it is not necessary that the union of the autocatalytic and drainable subnetworks span the complete stoichiometric space (for example, see Sec.\ \ref{sec:L_3_example} Fig.\ \ref{fig:cluster_L3}).

In subsec.\ \ref{sec:definitions}, we showed that each subnetwork can be assigned its FWMC equivalence class(es). As explained in Proposition \ref{prop:food-set}, while there are cases where the species-productive cones of different equivalence classes can be shown to not intersect, in general the productive regions of different equivalence classes can intersect. To visualize the list of equivalence classes to which the MASs belong, we run the algorithm for finding an intersection between each pair of partition-productive cones and obtain a two-dimensional square matrix,  $C_{\text{pp}}$, of dimension equal to the number of equivalence classes. Let us denote the list of equivalence classes by $\mathcal{L}$. The entry of $C_{\text{pp}}$ in the $i^{\text{th}}$ row and $j^{\text{th}}$ columns is given by,
\begin{align*}
    [C_{\text{pp}}]_i^j & = 
    \begin{cases}
    \phantom{-}2 \text{ if } Q(\mathcal{L}_i) = Q(\mathcal{L}_j),\\
        \phantom{-}1 \text{ if } Q(\mathcal{L}_i) \subsetneq Q(\mathcal{L}_j),\\
        \phantom{-}0 \text{ if } \varnothing \neq Q(\mathcal{L}_i) \cap Q(\mathcal{L}_j) \subsetneq Q(\mathcal{L}_i), Q(\mathcal{L}_j),\\
        -1 \text{ if }  Q(\mathcal{L}_i) \cup Q(\mathcal{L}_j) = \varnothing.
    \end{cases}
\end{align*}
Notice that any asymmetry in entries across the diagonal indicates that only one of the partition-productive cones completely contain the other. This matrix can then be visualized as a heat map, for example see Fig.\ \ref{fig:intersection_partition_5}.

\begin{figure*}[t]
\centering
\begin{subfigure}{.5\textwidth}
  \centering
  \includegraphics[width=0.9\linewidth]{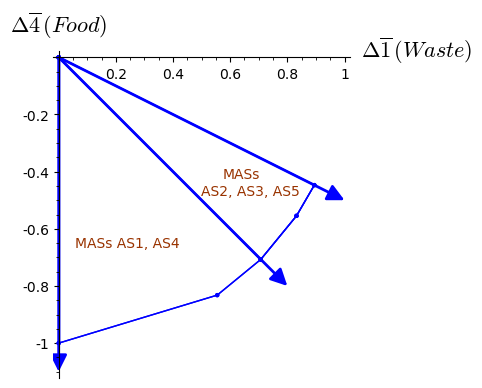}
  
\end{subfigure}%
\begin{subfigure}{.5\textwidth}
  \centering
  \includegraphics[width=.9\linewidth]{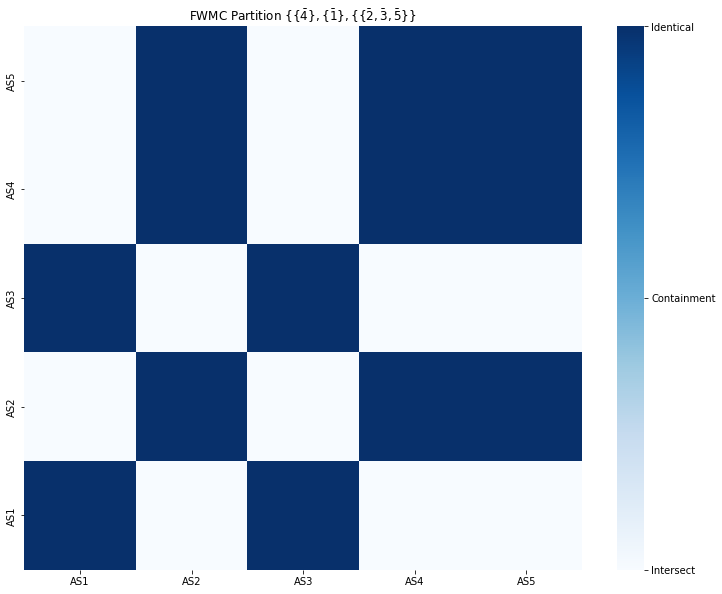}
  
\end{subfigure}
\caption{All MASs belonging to the $\{\{\overline{4}\},\{\overline{1}\},\{\{\overline{2},\overline{3},\overline{5}\}\}\}$ class in the complete $L=5$ 1-constituent CCRN are considered. The projections of their species-productive cones on the food and waste species are shown in the left panel (where the change in the concentration of species $\overline{n}$ is denoted as $\Delta \overline{n}$). The right panel displays the intersection information of MASs within a class analogous to Fig.\ \ref{fig:intersection_partition_5}. Note that there are no MASs in the class that are disjoint, and thus the `disjoint' label is omitted. }
\label{fig:cone_intersection}    
\end{figure*}

\begin{figure*}
    \includegraphics[width = \textwidth]{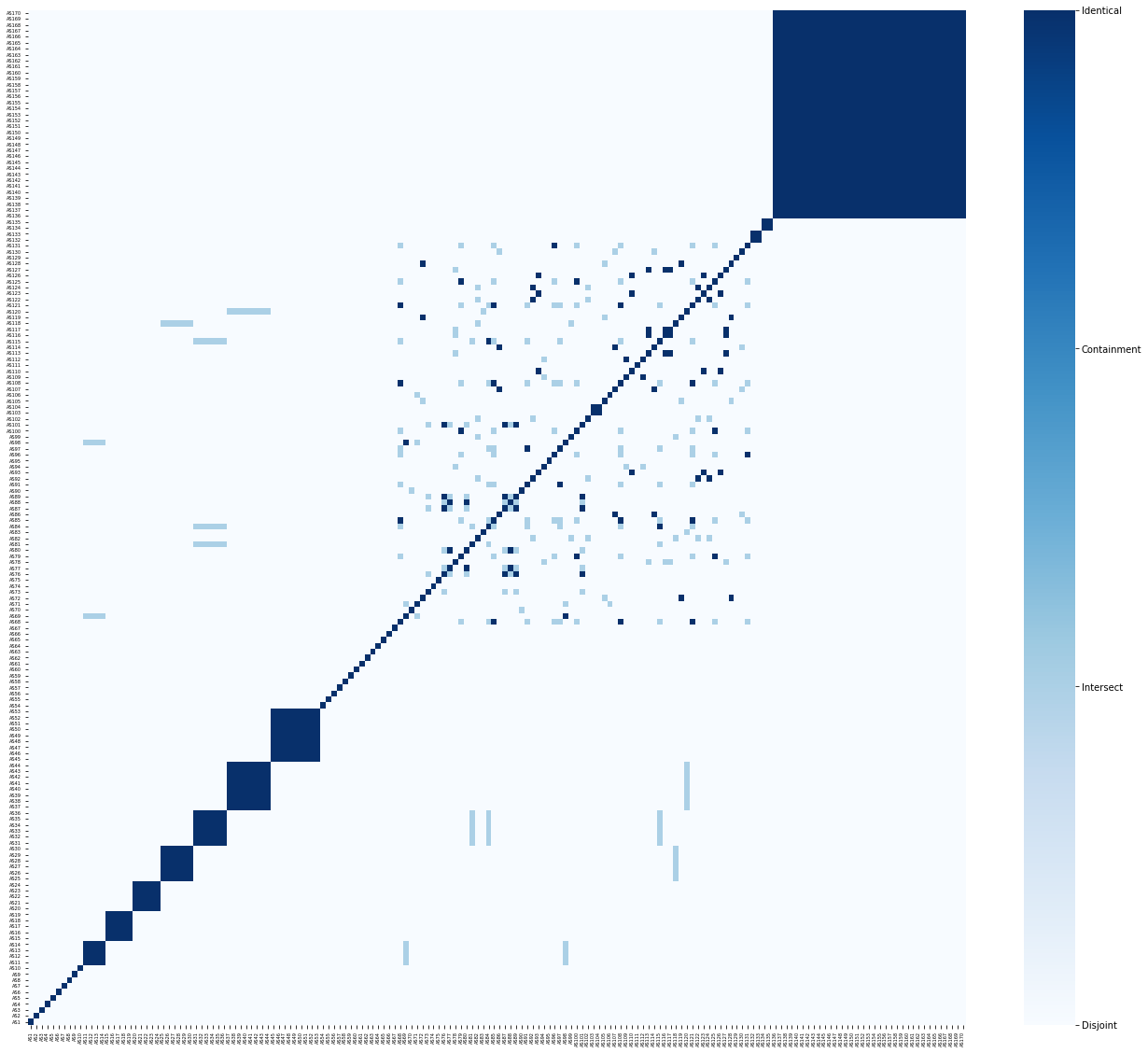}
    \caption{The pairwise intersection statistic for all the species-productive cones of the $L=5$ 1-constituent complete CCRN of order two. The reaction network and the list of MASs can be found in Appendix \ref{sec:MAS_list}. \label{fig:all_intersect}}
\end{figure*}

Each equivalence class can contain several MASs. The species-productive cones of the MASs in a class are not always identical. While they will share the same projection on the core species, the productive regions in the non-core species can be very different. For example, let us pick the equivalence class $\{\{\overline{4}\},\{\overline{1}\},\{\{\overline{2},\overline{3},\overline{5}\}\}\}$ from the list of classes shown in Fig.\ \ref{fig:intersection_partition_5}. From the figure, we know that it contains $5$ MASs. We plot the projection of the species-productive region in the non-core species in the top panel of Fig.\ \ref{fig:cone_intersection}. In higher dimensions when more non-core members are involved, this representation can get rather cumbersome. Thus, we employ a similar visualization as $C_\text{pp}$ to depict the intersection of MASs within an equivalence class. Whether or not there is an intersection between the species-productive cones can be ascertained using our algorithm outlined in the previous subsection. For an example of the resulting visualization for the same equivalence class considered above, see the bottom panel of Fig.\ \ref{fig:cone_intersection}. In the same manner, a visualization for the information of pairwise intersection of the productive cones of all MASs for a CRN can also be obtained, for example see Fig.\ \ref{fig:all_intersect}. 

\begin{example}
    Consider the complete $L=5$ 1-constituent CCRN of order two (introduced in Section \ref{sec:applications}). The list of all the FWMC classes, the number of MASs they contain, and their intersection information is summarized in Fig.\ \ref{fig:intersection_partition_5}. Consider the class $\{\{\overline{4}\},\{\overline{1}\},\{\{\overline{2},\overline{3},\overline{5}\}\}\}$. It contains the MASs:
    \begin{align*}
        \text{AS1} =\{&  \overline{2}+\overline{2}\to \overline{3} + \overline{1},\\
        &\overline{5}\to \overline{2} + \overline{3},\\
        &\overline{4}+\overline{3}\to \overline{2} + \overline{5}\},\\
        \text{AS2} =\{&  \overline{4}+\overline{2}\to \overline{1} + \overline{5},\\
        &\overline{5}\to \overline{2} + \overline{3},\\
        &\overline{4}+\overline{3}\to \overline{2} + \overline{5}\},
    \end{align*}
    \begin{align*}
        \text{AS3} =\{&  \overline{3}+\overline{3}\to \overline{5} + \overline{1},\\
        &\overline{5}\to \overline{2} + \overline{3},\\
        &\overline{4}+\overline{2}\to \overline{3} + \overline{3}\},\\ 
        \text{AS4} =\{&  \overline{3}+\overline{3}\to \overline{5} + \overline{1},\\
        &\overline{4}+\overline{2}\to \overline{5} + \overline{1},\\
        &\overline{5}\to \overline{2} + \overline{3}\},\\
        \text{AS5} =\{&  \overline{3}\to \overline{2} + \overline{1},\\
        &\overline{4}+\overline{2}\to \overline{5} + \overline{1},\\
        &\overline{5}\to \overline{2} + \overline{3}\}.        
    \end{align*}
    The projection of the species-productive cones for the above MASs and their intersection information is summarized in Fig.\ \ref{fig:cone_intersection}.
\end{example}

\begin{figure*}
    \centering
    \includegraphics[width = \textwidth]{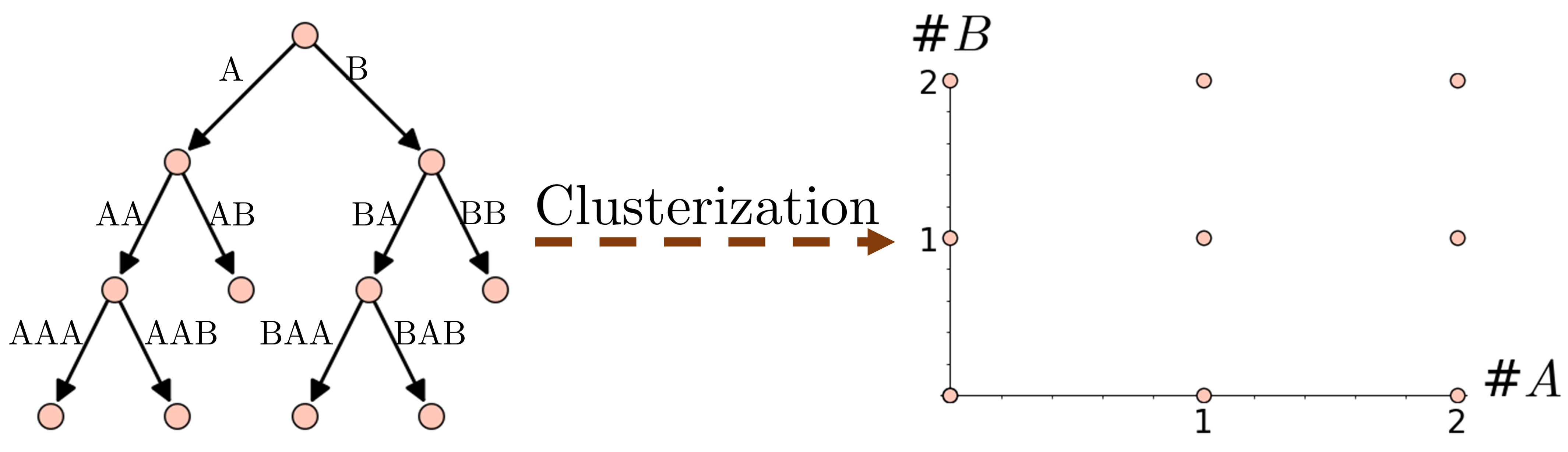}
    \caption{Consider a CRN of polymers whose species set is arranged as a binary tree in two letters, shown to the left. Then the \textit{cluster} resulting from each binary string is obtained by counting the number of occurrences of each alphabet. The \textit{clusterization} process projects from a tree structure over polymers to a lattice structure over clusters, shown to the right.}
    \label{fig:clusterization}
\end{figure*}
\section{Cluster chemical reaction networks (CCRN) framework}
\label{sec:applications}

All fully-balanced chemical reactions conserve the number of atoms of each type. Often, but not always, reaction models of polymerization and hydrolysis also conserve the number of each monomer. The existence of such conservation laws allows us to coarse-grain the underlying detailed CRNs, projecting away the bond structure of each molecule or polymer and only counting the number of atoms or monomers, respectively. Following terminology from Chapter 7 of \cite{kelly2011reversibility}, we will refer to this coarse-grained description of each species as a \textit{cluster} (see Fig.\ \ref{fig:clusterization}). This also allows us to coarse-grain a realistic CRN into a new CRN which has clusters as species and reactions between multisets of clusters induced from the reactions in the original CRN. We will call the resulting CRN, a \textit{cluster chemical reaction network} (CCRN) obtained from the original CRN.

There are three advantages of considering CCRN descriptions of CRNs. First, consider a CRN whose species set consists of all linear polymers formed of $D$ monomers up to length $N$. Then, the CCRN induced by the CRN would reduce the number of species from exponential ($D^N$) to polynomial ($N^D$) in the length of the polymer, reducing number of species and changing the connectivitity in the reaction graph from a tree to a lattice for many reaction models. Secondly, since each reaction in the original CRN induces a reaction in the CCRN with the same topology, the mapping would preserve the autocatalytic property. In other words, if a CRN is autocatalytic, its induced CCRN will also be autocatalytic. However, since it is a many-to-one map, multiple autocatalytic subnetworks of the CRN might map to the same autocatalytic subnetwork in its CCRN or the CCRN might contain MASs that do not correspond to any motif in the CRN (see Example \ref{eg:CCRN_polymer}). Thirdly, one can later re-introduce structure into the CCRN by adding more species with the same cluster counts (conserved quantities). In this way, the coarse-graining can be systematically refined to recover the original CRN by adding new dimensions, and allows a gradual complexification of the model.

\begin{example}
\label{eg:CCRN_polymer}
    Consider the CRN 
    \begin{align*}
        \mathcal{G} &= (\mathcal{S},\mathcal{R})\\
        \mathcal{S} &= \{ A, B, BAA, ABB, ABBA, BAAB, ABBABAAB\}\\
        \mathcal{R} &= \{ ABBA + BAA + B \to ABBABAAB,\\
        & \phantom{000.} BAAB + ABB + A \to ABBABAAB,\\
        & \phantom{0000} ABBABAAB \to ABBA + BAAB. \}
    \end{align*}
    The CCRN obtained by counting the number of $A$s and $B$s in each string and representing the resulting clusters as $\overline{n_A, n_B}$ is
    \begin{align*}
        \text{CCRN}(\mathcal{G}) &= (\mathcal{S}_C,\mathcal{R}_C)\\
        \mathcal{S}_C &= \{ \overline{1,0}, \overline{0,1}, \overline{2,1}, \overline{1,2}, \overline{2,2}, \overline{4,4}\}\\
        \mathcal{R}_C &= \{ \overline{2,2} + \overline{2,1} + \overline{0,1} \to \overline{4,4},\\
        & \phantom{000} \overline{2,2} + \overline{1,2} + \overline{1,0} \to \overline{4,4},\\
        & \phantom{000} \overline{4,4} \to \overline{2,2} + \overline{2,2}. \}
    \end{align*}
Notice that while the CRN has one minimal autocatalytic subnetwork (MAS), the CCRN has two MASs. This can be remedied by remembering that $BAAB$ and $ABBA$ are distinct by introducing a new species $\overline{2,2}^*$. Resulting in the CCRN 
    \begin{align*}
        \text{CCRN}(\mathcal{G})^* &= (\mathcal{S}_C^*,\mathcal{R}_C^*)\\
        \mathcal{S}_C^* &= \{ \overline{1,0}, \overline{0,1}, \overline{2,1}, \overline{1,2}, \overline{2,2},\overline{2,2}^*, \overline{4,4}\}\\
        \mathcal{R}_C^* &= \{ \overline{2,2} + \overline{2,1} + \overline{0,1} \to \overline{4,4},\\
        & \phantom{000} \overline{2,2}^* + \overline{1,2} + \overline{1,0} \to \overline{4,4},\\
        & \phantom{000} \overline{4,4} \to \overline{2,2} + \overline{2,2}^*. \}
    \end{align*}
    Note that $\text{CCRN}(\mathcal{G})^*$ has identical information to the original CRN and should, therefore, yield identical inferences to the original CRN. 
\end{example}

We formally define a CCRN in Sec.\ \ref{sec:CCRN-formalism}, and systematically explore the properties of CCRNs with one conserved quantity (type of atom or monomer) in Sec.\ \ref{sec:CCRN-complete}. We also provide the statistics of autocatalytic subnetworks found in such reaction networks and present a worked-out-example in Sec.\ \ref{sec:L_3_example}. We briefly discuss the computational challenges in scaling the algorithm for fully connected models in Sec.\ \ref{sec:comp_challenge} and introduce rule-generated CCRNs in Sec.\ \ref{sec:rule_gen}.

\subsection{Formalism}
\label{sec:CCRN-formalism}
A \textbf{cluster chemical reaction network (CCRN)} is a CRN with the additional structure that, upon excluding the inflow and outflow reactions (see Remark \ref{remark:inflow}), the network has at least one nonnegative integer conservation law (see Sec.\ \ref{sec:conservation}). Each conservation law $\textbf{x}$ yields a conserved quantity $\textbf{x}\cdot \textbf{n}$ for a population in state $\textbf{n}$. Each conservation law corresponds to a type of \textbf{constituent} that is conserved, and the magnitude of the conserved quantity denotes the amount of that constituent in the population state. We denote the number of distinct constituents by $D$ and label them as $A_1, \ldots, A_D$. The \textit{\textbf{species}} of a CCRN, termed \textbf{clusters}, are multisets of constituents and denoted by an overline (notation chosen to be consistent with \cite{liu2018mathematical}) over a vector of nonnegative integers representing the number of each constituent in the cluster. 
\begin{notation}
\[\overline{\textbf{n}} := \overline{n_1,\ldots,n_D}\]
    will be used to denote a cluster comprising $n_1, \ldots, n_D$ of constituents $A_1, \ldots, A_D$, respectively. 
\end{notation}

Denoting the population state with exactly one $\overline{\textbf{n}}$ type particle as $1_{\overline{\textbf{n}}}$, if the conservation laws are labelled $\textbf{x}_d$ where $d$ takes values from $1$ to $D$, then
\[\textbf{x}_d \cdot 1_{\overline{\textbf{n}}} = n_d.\]
A cluster thus corresponds to the vector of its conserved quantities. In chemistry, the compositional formula of a chemical is an example of its cluster representation. For example, the cluster of water or $H_2 O$, in a basis of constituents $\{H,O\}$ will be represented as $\overline{2,1}$. Here the first and second conservation laws counts the number of $H$ and $O$ atoms, respectively. In case of multiple clusters with identical vectors of conserved quantities, we distinguish them with an asterisk $^*$. For example, in Table \ref{table:cluster-paradigm-rules}, $\overline{1}$ and $\overline{1}^*$ refer to a monomer and an activated monomer, respectively.

\begin{definition}
    We define the \textbf{length} of a cluster to be the sum of its conserved quantities, \[\ell(\textbf{n}) := \sum_{i=1}^D n_i.\]   
    The length of a cluster is a scalar quantity.
\end{definition}

In a CCRN, \textit{\textbf{complexes}} are multisets of clusters and are denoted by \[c^\alpha := \sum_\textbf{n} c^\alpha_\textbf{n} \phantom{\cdot} \overline{\textbf{n}}.\] As an abuse of notation, we denote the stoichiometry of complex $c^\alpha$ also by $c^\alpha$, where now it is the column vector with entries $c^\alpha_\textbf{n}$. 

\begin{definition}
The \textbf{size per constituent} of a complex is the vector sum of conserved quantities of each type, and is denoted
\[|c^\alpha| = \sum_\textbf{n} c^\alpha_\textbf{n} \phantom{\cdot} \textbf{n}.\] 
\end{definition}

\begin{definition}
 The \textbf{width} of a complex is the total number of clusters in the multiset, and denoted by \[w(c^\alpha):= \sum_\textbf{n} c^\alpha_\textbf{n}.\]
\end{definition}

In a CCRN, every \textit{\textbf{reaction}} must be such that the sizes per constituent of the source and target complexes are identical. Thus, a reaction $c^\alpha \to c^\beta$ is allowed only if \[|c^\alpha| = \sum_\textbf{n} c^\alpha_\textbf{n} \phantom{\cdot} \textbf{n}= \sum_\textbf{n} c^\beta_\textbf{n} \phantom{\cdot} \textbf{n} = |c^\beta|.\] This restriction is required in order for the constituents to be conserved quantities, as 
\[\sum_\textbf{n} \textbf{n}(c^\beta - c^\alpha)_\textbf{n} = 0.\] 

\begin{definition}
Let $\mathcal{H}_D = (\mathcal{S}, \mathcal{R})$ denote a CCRN with $D$ constituents. Then for any finite CCRN, 
\begin{align*}
    \mathcal{S} &\subset \{\overline{\textbf{n}} | \textbf{n} \in \mathbb{Z}_{\geq 0}^D  \},\\
    \mathcal{R} &\subseteq \{c^\alpha \to c^\beta | |c^\alpha| = |c^\beta|  \}.
\end{align*}
For a CCRN $\mathcal{H}_D$: 
    \begin{itemize}
        \item The length of the CCRN is defined to be the maximum length of its clusters, and denoted  \[\ell(\mathcal{H}_D) = \max_{\textbf{n} \in \mathcal{S}} \ell(\textbf{n}).\] 
        \item The \textbf{order} of the CCRN is defined to be the maximum width of its complexes, and denoted by \[o(\mathcal{H}_D) = \max_{\mathcal{R}} w(c^\alpha).\] 
    \end{itemize}
\end{definition}

\begin{definition}
We define a \textbf{complete CCRN} of length $L$ and order $w$ to consist of all species up to length $L$ and all possible reactions that are allowed by the conservation laws with complexes of width up to $w$.    
\end{definition}
 \begin{remark}
 Since any reaction consisting of more than two reactants or products can be written as chains of second-order reactions, for our analysis \textit{we will restrict to CCRNs of order two}. Notice that this restriction ensures that any exclusively autocatalytic subnetwork will also be stoichiometrically autocatalytic because the reactants and products are necessarily distinct.    
 \end{remark}

\begin{table}[h]
\begin{tabular}{c|ccccc|ccccc}
   Max.\ length & \multicolumn{5}{c|}{CCRN properties}  & \multicolumn{5}{c}{$\#$ reactions in a MAS}  \\
   $L$ & $|\mathcal{C}|$ & $|\mathcal{R}|$ & $\ell$ & $\imath$ & $\delta$ & 2 & 3 & 4 & 5 &6  \\
   \hline 
    3 & 6 & 6 & 3 & 3 & 1 & 2 & 0 & 0 & 0 &0\\
    4 & 11 & 14 & 5 & 8 & 3 & 9 & 11 & 0 & 0 &0 \\
    5 & 17 & 26 & 7 & 16 & 6 & 24 & 98 & 48 & 0&0\\
    6 & 24 & 44  & 9 & 29 & 10 & 48 & 461 & 768 & 331&0\\
    7 & 32 & 68 & 11 & 47 & 15 & 85 & 1549& 6028& 5673& 1709
\end{tabular}
\caption{\label{table:1cluster} The complete 1-constituent CCRN of length $L$ and order two is considered for $L$ from $3-7$. The topological properties of the CCRN, such as number of linkage classes $\ell$ and deficiency $\delta$, are shown in the left half (for calculations, see Appendix \ref{app:kernel}). The number of MASs for a given number of reactions is shown in the right half (for a visual representation, see Fig.\ \ref{fig:1_const_MASs_data}).}
\end{table}

\subsection{Complete CCRN}
\label{sec:CCRN-complete}

\begin{figure}
    \centering
    \includegraphics[width=0.45\textwidth]{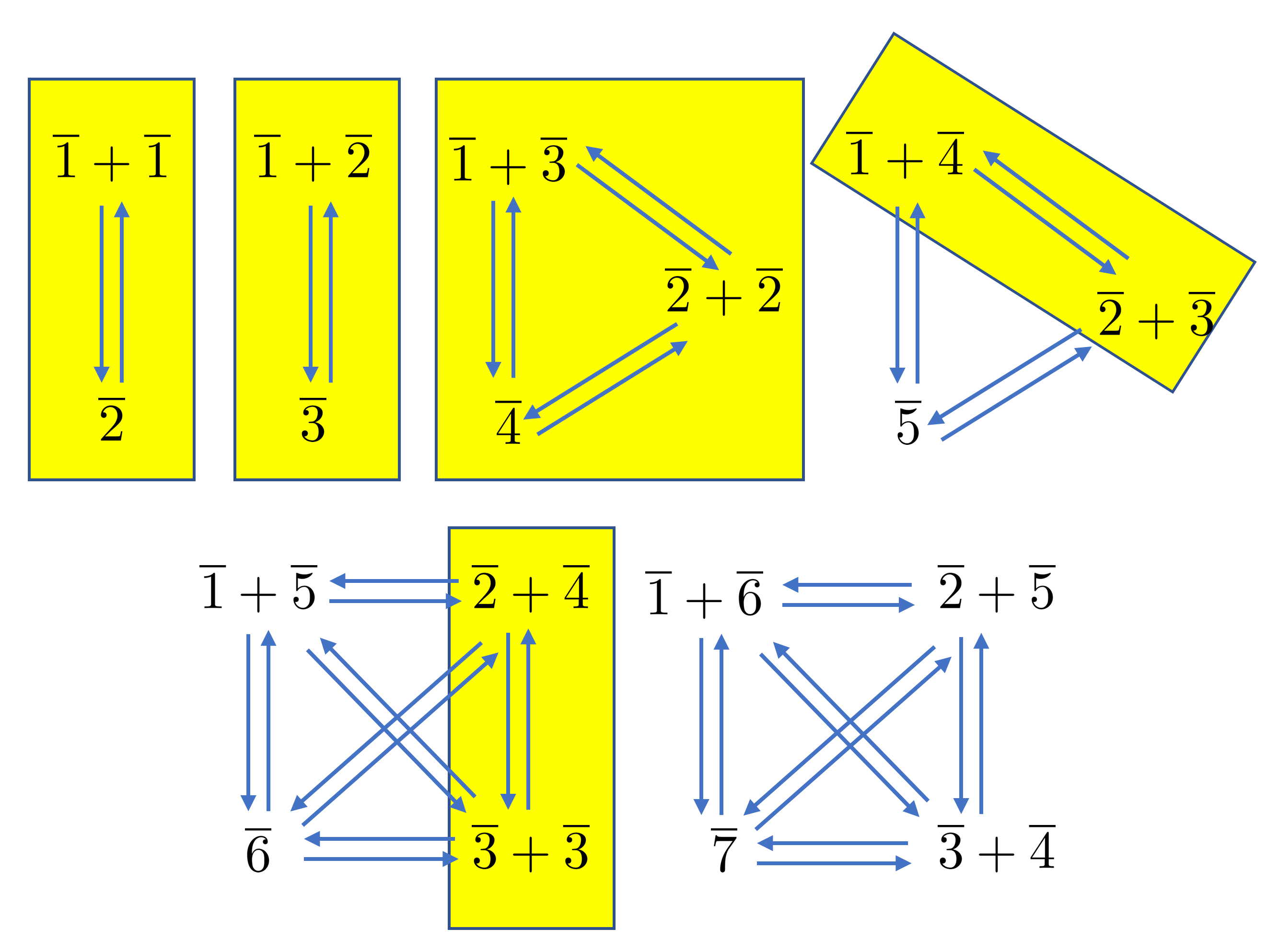}
    \caption{The first few reactions in the complete 1-constituent CCRN of order two. The reactions containing clusters only up to length four ($L=4$) are shown in yellow boxes. }
    \label{fig:CCRN_1d}
\end{figure}

\begin{figure}
    \centering
    \includegraphics[width = 0.5\textwidth]{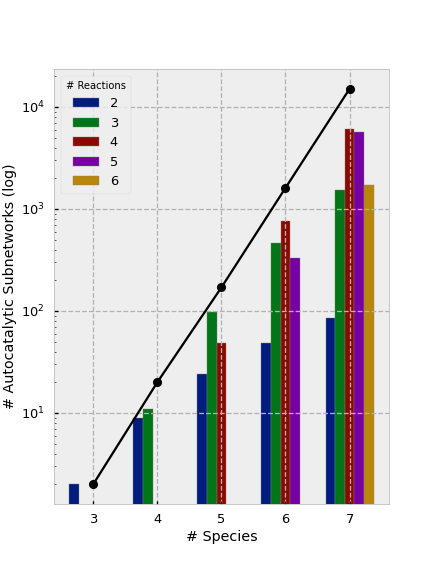}
    \caption{Number of minimal autocatalytic subnetworks (MASs) varying with $L$ for the complete 1-constituent CCRN of order two. For each $L$, the distribution of MAS with the number of reactions is also shown. Notice that total MASs for each $L$, plotted as black dots, increases exponentially with $L$.  }
    \label{fig:1_const_MASs_data}
\end{figure}

\subsubsection{1-constituent CCRN}
\label{sec:CCRN-1constituent}
Let $\mathcal{H}_1^L = (\mathcal{S},\mathcal{C},\mathcal{R})$ denote a complete 1-constituent CCRN of length $L$ and order two. 1-constituent CCRN of length $L$ means that the species set consists of clusters up to length $L$, denoted $\mathcal{S} = \{ \overline{1}, \overline{2}, \ldots, \overline{L}\}$. By order two, we mean that all complexes are at most of width two, and the set of complexes is denoted as $\mathcal{C} = \{ \overline{a} \}\cup\{ \overline{a} + \overline{b}\}$ for $\overline{a},\overline{b} \in \mathcal{S}$ and $b\leq a \leq L$. By \textit{a complete CCRN}, we mean that the set of reactions contains all the possible reactions that are allowed by the conservation law, i.e.\ 
\begin{align*}
   \mathcal{R} =& \{ \ce{$\overline{a} + \overline{b}$ <=> $\overline{c} + \overline{d}$} | a+b = c+d \} \cup \\
&   \{ \ce{$\overline{a} $ <=> $\overline{c} + \overline{d}$} | a = c+d \}. 
\end{align*}
The first few linkage classes of the complete 1-constituent CCRN of order two are shown in Fig.\ \ref{fig:CCRN_1d} and the complete CCRN for $L=4$ is highlighted in yellow. The information about the reaction networks and the MASs that they contain for the complete 1-constituent CCRNs of $L$ ranging from $3$ to $7$ is collected in Table \ref{table:1cluster} and Fig.\ \ref{fig:1_const_MASs_data}.

\begin{figure*}[t]
  \centering
  \includegraphics[width=\linewidth]{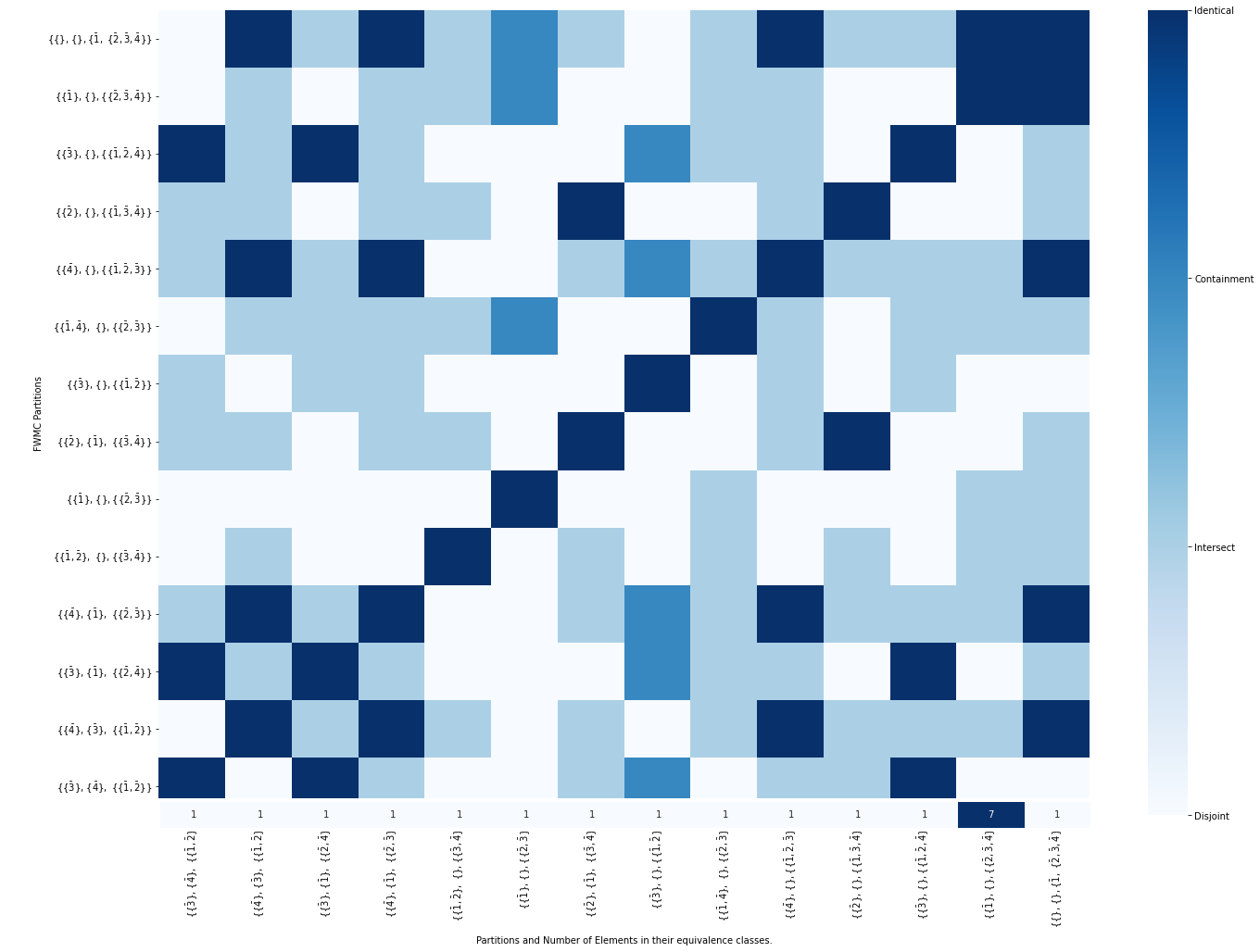}
\caption{The list of the FWMC classes with the information of their intersection for the complete 1-constituent CCRN with $L=4$ is shown.}
    \label{fig:cluster_n4_2}        
\end{figure*}

Consider the complete 1-constituent CCRN with $L=4$ of order two, the reaction network for which is shown in Fig.\ \ref{fig:CCRN_1d}. Notice that this subnetwork has $5$ linkage classes, $11$ complexes, and $14$ reactions. An application of the algorithm for enumerating all the MASs of the CCRN discussed in Sec.\ \ref{sec:algorithms} yields a list of $20$ subnetworks, as shown in Fig.\ \ref{fig:cluster_n4}. In the list, there are $9$ MASs with two reactions and $11$ MASs with three reactions. Since the CCRN has $4$ species, and $1$ conservation law, the stoichiometric subspace is of dimension $3$. Moreover, due to the conservation law, from Lemma \ref{lemma:food} each MAS must possess at least one non-core species. This means that there cannot be any $4$ reaction MAS as it would require at least $5$ distinct species, which would be a contradiction. 

Upon obtaining a list of MASs for a given CRN, we would like to understand how their species-productive and partition-productive cones intersect. Observe that each MAS in the $L=4$ CCRN considered above has a unique autocatalytic core. For the CCRN, using Theorem \ref{theorem:partition-prod}, any MASs with the same FWMC partitions have identical species-productive cones and it is identical to the partition-productive cone. Thus, we do not show any plot for the species-productive cone intersection data. Moreover, any two partitions consisting of the same core of $3$ species must have the same partition-productive cone, as the stoichiometric subspace is of dimension $3$. For example, the partition-productive cones of $\{\{\overline{1}\},\{\},\{\{\overline{2},\overline{3},\overline{4}\}\}\}$ and $\{\{\},\{\},\{\overline{1},\{\overline{2},\overline{3},\overline{4}\}\}\}$ are identical. The intersections of partition-productive cones for different FWMC partitions for the complete 1-constituent
CCRN with $L = 4$ are shown in the Fig.\ \ref{fig:cluster_n4_2}.  

Unlike the $L=4$ case, for the complete 1-constituent CCRN with $L=5$ of order two, there can be equivalent MASs with identical FWMC-partitions whose species-productive cones do not intersect. Moreover, it is not a priori clear whether the partition-productive cones of different FWMC partitions will intersect. This is why we employed our visualization scheme in Fig.\ \ref{fig:intersection_partition_5}. As described in Sec.\ \ref{sec:visualization}, for a particular equivalence class we also show the intersection information of the species-productive cones in Fig.\ \ref{fig:cone_intersection}.

\subsubsection{L=3 1-constituent CCRN}
\label{sec:L_3_example}
\begin{figure}
    \centering
    \includegraphics[width=0.4\textwidth]{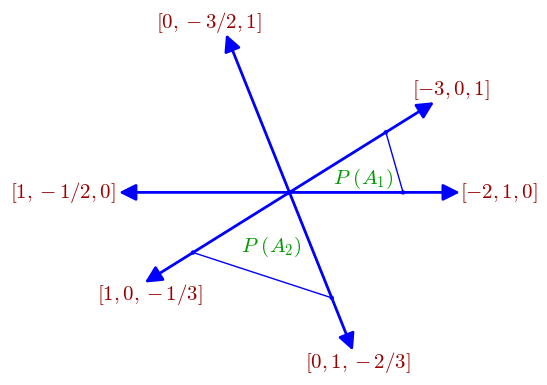}
    \caption{The 2-dimensional stoichiometric subspace along with the autocatalytic productive cones for the $L=3$ complete 1-constituent CCRN are shown. The $6$ rays emanating from the origin, labelled by their directions in the $3-$dimensional space, form the edges of productive cones where one species is consumed and another is produced. For instance, the ray $[-3,0,1]$ corresponds to the direction in the stoichiometric space where $\overline{1}$ is consumed, $\overline{3}$ is produced, and $\overline{2}$ remains constant.}
    \label{fig:cluster_L3}
\end{figure}

Consider the complete 1-constituent CCRN with $L=4$ of order two, given by 
\begin{align*}
    \mathcal{H}_1^3 &= (\mathcal{S},\mathcal{R})\\
    \mathcal{S} &= \{ \overline{1},\overline{2},\overline{3}\}\\
    \mathcal{R} &= \{ \ce{$\overline{1} + \overline{1}$ <=>[k_11][k_2] $\overline{2}$},\\
    & \phantom{--} \ce{$\overline{1} + \overline{2}$ <=>[k_12][k_3] $\overline{3}$},\\
    & \phantom{--} \ce{$\overline{1} + \overline{3}$ <=>[k_13][k_22] $\overline{2}+ \overline{2}$}\}.
\end{align*}
This network has two MASs, namely
\begin{align*}
    A_1 =& \{ \overline{1} + \overline{2} \to \overline{3},\\
         & \phantom{0} \overline{1} + \overline{3} \to \overline{2} + \overline{2}\},\\
     A_2 =&  \phantom{...} \{ \phantom{...} \overline{2} \to \overline{1} + \overline{1},\\
     & \phantom{0} \overline{3} + \overline{1} \to \overline{2} + \overline{2}\}.
\end{align*}

Since the CCRN has one conservation law, the stoichiometric subspace is of dimension $2$ and perpendicular to $[1,2,3]^T$. The two-dimensional stoichiometric subspace is shown in Fig.\ \ref{fig:cluster_L3}. The $6$ rays emanating from the origin, labelled by their directions in the $3-$dimensional space, form the edges of productive cones where one species is consumed and another is produced. The species-productive cones of the two autocatalytic cycles are also labelled. Note that the species-productive regions is the complete cone within the bounding rays, and the finite boundary is drawn simply to enhance visualization. For example, the species-productive cone of the MAS $A_1$ is bounded by the rays $[-2,1,0]$ and $[-3,0,1]$. Notice that the FWMC partition of $A_1$ is $\{\{\overline{1}\},\{\},\{\{\overline{2},\overline{3}\}\}\}$, consistent with the species-productive region that strictly consumes species $\overline{1}$ and strictly produces species $\overline{2}$ and $\overline{3}$.

We also want to make a few remarks about details that are not visible in Fig.\ \ref{fig:cluster_L3}. Firstly, for every MAS, there is a minimal drainable subnetwork obtained by reversing the reaction edges. The species-productive cones of these drainable networks will be the negatives of their autocatalytic counterparts. Secondly, notice that there is no MAS with the partition $\{\{\overline{2}\},\{\},\{\{\overline{1},\overline{3}\}\}\}$. Had there been one, as would have been the case had we allowed reactions of order $3$, then its species-productive cone would be bound by the rays $[0,-3/2,1]$ and $[1,-1/2,0]$.

\begin{figure}
    \centering
    \includegraphics[width=0.4\textwidth]{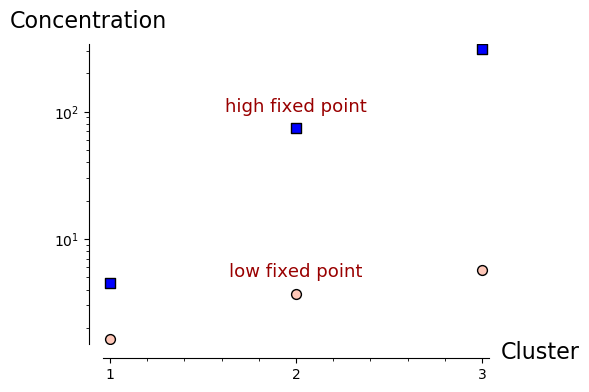}
    \caption{Multiple fixed points for the complete 1-constituent CCRN with $L=3$ and order two obtained using Feinberg's deficiency one algorithm (see Appendix \ref{app:deficiency-1}). The parameters for the algorithm are chosen to be $\mu_{\overline{1}}=1,\mu_{\overline{2}}=3,\mu_{\overline{3}}=4=\eta$, and the rate constants for the model are $k_{11} = 1, k_2 = 1.001, k_{12} =1.1 , k_{3}=1.011, k_{13} = 0.146, k_{22} = 0.027$. }
    \label{fig:cluster_L3_fp}
\end{figure}

Notice from Table \ref{table:1cluster} that the complete CCRN with $L=3$ has a deficiency of one. This means that under mass-action kinetics, it has the potential to exhibit multistability. To investigate this, we used Feinberg's deficiency one algorithm \cite{feinberg1988chemical} in Appendix \ref{app:deficiency-1} and obtained rate constants for which the CCRN has multiple steady states. An example of multiple steady states and the rate constants using which they are obtained are reported in Fig.\ \ref{fig:cluster_L3_fp}.

\subsubsection{Number theoretic result for 1-constituent CCRN of order two}
\label{sec:num_theor}
Since CCRNs are defined by number conservation laws, finding autocatalytic subnetworks in any CCRN corresponds to finding number-valued solutions of systems of equations. In this subsection we will show that if there exists a two-reaction MAS in a partition of a 1-constituent CCRN, then it must be unique.

\begin{lemma}
    Let $A$ be a $2$-reactions MAS in a 1-constituent CCRN of order two. Then, there does not exist any other MASs in the CCRN with the same FWMC partition as $A$.
\end{lemma}
\begin{proof}
    Let $A_C = \{\overline{i}, \overline{j}\}$ be the core species in $A$ (assume without loss of generality that $i>j$). 
    The possible 2-reaction MASs with that set of core species are given by reactions of the form
    \begin{align*}
        \begin{cases}
            \overline{i} + \overline{\ell_1} \to& \overline{j} + \overline{\ell_2},\\
            \overline{j} + \overline{\ell_3} \to& 2\overline{i}.
        \end{cases}\\
        \begin{cases}
            \overline{i} + \overline{m_1} \to & 2\overline{j}, \\
            \overline{j} + \overline{m_2} \to & \overline{i} + \overline{m_3}.
        \end{cases}
    \end{align*}
    Note that each non-core species can appear with the value $0$, indicating the absence of a reactant. However, due to Lemma \ref{lemma:food}, the conservation law will require that at least one food species is non zero. Observe, also, that in both types of potential MASs there is a reaction which is uniquely determined by $\bar i$ and $\bar j$ (being then $\bar \ell_3$ and $\bar m_1$ also unique for this choice of the core set). Concretely $\bar \ell_3 = \overline{2i-j}$ and $\bar m_1 = \overline{2j-i}$. Note that these reactions are different since it is assumed that $i\neq j$. Thus, the above reactions reduce to:
        \begin{align*}
        M_1: \begin{cases}
            \overline{i} + \overline{j+\ell-i} \to& \overline{j} + \overline{\ell},\\
            \overline{j} + \overline{2i-j} \to& 2\overline{i}.
        \end{cases}\\
        M_2: \begin{cases}
            \overline{i} + \overline{2j-i} \to & 2\overline{j}, \\
            \overline{j} + \overline{i+m-j} \to & \overline{i} + \overline{m}.
        \end{cases}
    \end{align*}
     Now we will do a case-by-case analysis to show that each FWMC-partition will contain at most one solution of the systems of equations. 

     Note that the only possible partition of the species induced by the reactions in $M_1$ is:
     $FWMC(M_1) = 
     \{\bar j+\bar\ell-\bar i, \overline{2i-j}\}, \{\bar\ell\}, \{\{\bar i, \bar j\}\}
     $
     whereas the partition for $M_2$ is:
     $
     FWMC(M_2)=      \{\overline{2j-i}, \bar i+\bar m-\bar j, \}, \{\bar m\}, \{\{\bar i, \bar j\}\}
     $
     From both types of partition, the only possibility for which both coincide is that $\bar l = \bar m$ and $\bar i = \bar j$ contradicting the assumption that $i>j$.

\end{proof}
Given two different species $\overline{i}$, $\overline{j}$ with $i>j\geq 1$, the above result also provides a way to construct all the 2-reactions MASs with those species being the core species. The explicit reactions taking part of the MASs are in the form:
\begin{itemize}
    \item In case $i \leq \frac{L}{2}$:
$$
MAS_1 = \left\{\begin{array}{rrl}
     R_1: &       \overline{i} + \overline{j+\ell-i} \to& \overline{j} + \overline{\ell},\\
     R_2: &       \overline{j} + \overline{2i-j} \to& 2\overline{i}.
\end{array}\right.
$$
for all $\ell \in [i-j,L] \cap \mathbb{Z}$.
\item In case $\frac{i}{2} \leq j \leq \frac{L}{2}$:
$$
MAS_2 = \left\{\begin{array}{rrl}
      R_1:  &      \overline{i} + \overline{2j-i} \to & 2\overline{j}, \\
     R_2:  &       \overline{j} + \overline{i+m-j} \to & \overline{i} + \overline{m}.
\end{array}\right.
$$
for all $m \in [0,L+j-i] \cap \mathbb{Z}$.
\end{itemize}

\subsubsection{Computational challenges in scaling}
\label{sec:comp_challenge}
Consider the complete 2-constituent CCRN of order two, defined as
\begin{align*}
\mathcal{H}_2 =& (\mathcal{S},\mathcal{R})  \\
\mathcal{S} = & \{ \overline{a,b} | a,b \in \mathbb{Z}_{\geq 0} \}\\
\mathcal{R} =& \{ \overline{a,b} + \overline{c,d} \to \overline{e,f} + \overline{g,h} |\\
              & \phantom{--}    a+c = e+g \text{ and } b+d = f+h \}.
\end{align*}
We wish to remark that for $\mathcal{H}_2$ and other CRNs with a large number of reactions, generating the whole set of autocatalytic subnetworks is computationally challenging. As the number of reaction increases, the dimension of the space where the MASs are to be found also increases, and even finding a  single MAS in the network implies solving a difficult optimization problem that might be computationally costly. Even if finding the MASs is polynomial-time solvable (which does not seem to be the case), the complexity of the enumeration algorithm may turn into exponential (for e.g., see \cite{garey1979computers}), since the number of subnetworks increases considerably with the number of species and reactions. Thus for practical reasons, it may be more feasible to consider sparser rule-generated networks, as explained in the next subsection and shown in Fig.\ \ref{table:cluster-paradigm-rules}.


\subsection{Rule generated CCRNs}
\label{sec:rule_gen}
\begin{table}[]
    \centering
    \begin{tabular}{cc}
    \hline\\
 Monomer activation &           \ce{$\overline{1}$ <=>  $\overline{1}^*$}    \\
Catalyzed activation &    \ce{$\overline{C} + \overline{1}$ <=>  $\overline{1}^*+\overline{C}$}       \\ \\
    \hline\\
    Polymerization & \ce{$\overline{1}+\overline{1}^*$ <=>  $\overline{2}$}\\
                    & \ce{$\overline{2}+\overline{1}^*$ <=>  $\overline{3}$}\\
                    & \vdots \\
    1-constituent   & \ce{$\overline{j}+\overline{1}^*$ <=>  $\overline{j+1}$}\\ \\
  2-constituent     & 
        $\begin{cases}
            \ce{$\overline{j,k}+\overline{1,0}^*$ <=>  $\overline{j+1,k}$}\\
            \ce{$\overline{j,k}+\overline{0,1}^*$ <=>  $\overline{j,k+1}$} 
        \end{cases}$
  \\ 
  \vdots  & \vdots \\ \\
  \hline \\
  \begin{tabular}{@{}c@{}}Templated \\ polymerization\end{tabular}  & 
  $\begin{cases}
      \ce{$\overline{j,k}+ \overline{k-1,j} +\overline{1,0}^*$ <=>  $\overline{j,k}+ \overline{k,j}$}\\
      \ce{$\overline{j,k}+ \overline{k,j-1} +\overline{0,1}^*$ <=>  $\overline{j,k}+ \overline{k,j}$}
  \end{cases}$\\ \\
  \hline \\
  Fusion & \ce{$\overline{j} + \overline{k}$ <=>  $\overline{j+k}$ }\\
  Fission & \ce{$\overline{j}$ <=>  $\overline{j-k} + \overline{k}$ } \\
  \\
  \hline
    \end{tabular}
    \caption{Several examples of rules for generating a sparse CCRN relevant to biochemistry. Here, in accordance with the definition of a CCRN, the variables $j,k$ must be chosen such that the resulting clusters have a non-negative integer quantity of each constituent. }
    \label{table:cluster-paradigm-rules}
\end{table}

A naive computational modelling of artificial chemistry \cite{banzhaf2015artificial} often runs into issues of memory, storage and processing due to a combinatorial explosion of chemical species and reactions involved in the CRN. For instance, if one uses string chemistry \cite{moyer2020stoichiometric} (or polymer sequence chemistry) to model polymers of length up to $N$ with $D$ distinct monomers ($D$ constituents), a straightforward calculation shows that one is required to track $\sum_{i=1}^N D^i > D^N$ species. Since for any physically relevant model $N\gg D$, so it is clear that the computation will soon become intractable as one increases the length of the string (polymer) $N$. Even a simple graph-grammar for generating artificial chemistry, for e.g.\ with the application of \cite{andersen2016software}, can generate a reaction network that is computationally intractable. Moreover, even in the computationally tractable regime, an enumeration of autocatalytic subnetworks or motifs will generically yield an extremely long list. In such a scenario, it might be unclear how to simplify the model without stripping away the essential details or adding artifacts.

We argue that the \textit{Cluster framework} can help alleviate some of the computational issues. Any rule generated CRN, in string chemistry or built using graph-grammar methods must induce rules on the CCRN. Since the induced CCRN can have an exponentially reduced species set, it offers a way to constrain the model to a more computationally tractable regime. Since autocatalytic networks are preserved under this coarsening (with some caveats mentioned in the introduction to this section), one can use the CCRN to obtain a more manageable list of MASs. One can then gradually complexify the CCRN by adding more species with the same conserved quantities until the desired behavior of the CRN is captured by the CCRN.

For example, consider polymer chemistry with two monomers $A$ and $B$. For an arbitrary polymer $\omega$, the addition of an $A$ or $B$ yields $\omega \cdot A$ or $\omega \cdot B$, respectively. In the cluster framework, if we map $A$, $B$, and $\omega$ to $\overline{1,0}$, $\overline{0,1}$, and $\overline{j,k}$, respectively, the polymerization reactions in the CCRN become
\begin{align*}
    \ce{$\overline{j,k}+\overline{1,0}$ <=>  $\overline{j+1,k}$},\\
        \ce{$\overline{j,k}+\overline{0,1}$ <=>  $\overline{j,k+1}$}.
\end{align*}
In realistic chemistry, however, a monomer addition to a polymer is only done by an activated monomer and not an unactivated monomer. Thus, we may want to distinguish activated and unactivated monomers in our model, for which we can add extra species $\overline{1,0}^*$ and $\overline{0,1}^*$. The resulting polymerization reactions and other examples of rule generated CCRNs are shown in Table \ref{table:cluster-paradigm-rules}.

\section{Discussion and future research}
\label{sec:discussion}

In this work, we began by presenting a linear-algebraic representation of CRNs in terms of an input-output matrix pair. We introduced different notions of autocatalysis existing in literature as nested classes of conditions on the CRN. In particular, the results and analysis in this work is for the class of \textit{exclusive autocatalysis} \cite{andersen2021defining}, which is more encompassing of true cases than the notion of autocatalysis used by Blokhuis et al.\ in \cite{blokhuis2020universal}. We defined minimal autocatalytic subnetworks (MASs), proved properties of their stoichiometric matrices, and explained their induced polyhedral geometry. Next, we proposed mathematical optimization based algorithms to exhaustively compute all the MASs in a CRN and create their visualizations. Finally, we introduced the notion of cluster CRNs (CCRNs) as a useful coarse-graining induced by general CRNs, and employed our techniques on CCRNs obtained from one constituent. Notably, we show that the list of MASs for maximally connected 1-constituent CCRNs with length up to $L$ increases exponentially with $L$.

The mathematical theory developed here facilitated the development of effective algorithms for identifying MASs in general CRNs. Nevertheless, our optimization-based approach requires solving a series of binary mathematical programming problems with an increasing number of linear constraints. This approach is computationally costly for large CRNs. However, there are reasons to hope that further research on these optimization problems, in particular, its polyhedral properties, will allow us to strengthen the formulations and solve them more efficiently. Exploiting the algebraic properties behind conservation laws for sequences of integers holds particular promise in this regard.
We have also introduced a simple coarse graining of CRNs as CCRNs. By adjusting the number of dimensions included in a CCRN it is possible to trade-off computational tractability against fidelity to the complete CRN and to accurately capture all autocatalytic motifs. This flexibility is a great strength of the CCRN model. Additionally, although we here only considered clusters with nonnegative constituents, it also should be possible to consider clusters defined on the complete integer lattice. Such a reaction network could be used to model nuclear reactions where the positive conserved quantities correspond to positive charges or subatomic particles and the negative conserved quantities would refer to negative charges or antiparticles.

In our explorations we only considered fully connected CCRNs, but some reactions that are consistent with mass conservation are , nonetheless, impossible in real chemistry. In future work it would be interesting, similar to \cite{anderson2022prevalence,garcia2023chemically},  to consider randomly generated CCRNs in an Erdos-Renyi framework and compute a bound on the number of autocatalytic cycles that persist as a function of the probability of a reaction’s being allowed in the network. An additional, and perhaps even more promising approach to representing real chemistry would be to expand our understanding of the kinetics of fully-connected CCRNs when rate constants differ over many orders of magnitude. Such an effort might be aided by understanding the effects of adding thermodynamic parameters such as free-energy assignments to different clusters. 

Overall, the mathematical framework and computational tools developed here have potential applications for any kind of CRN that has multiple MASs: an autocatalytic ecosystem. This is of particular significance for better connecting biology  and chemistry at multiple scales of analysis. As discussed by \cite{baum2023ecology}, a cell’s ability to grow and reproduce rests on the fact that its metabolism is an autocatalytic ecosystem. Even ignoring autocatalytic motifs that include genetic mechanisms (i.e., those whose members are proteins or nucleic acids), metabolic networks contain many autocatalytic motifs \cite{peng2022hierarchical}. The tools developed here will facilitate detecting MASs and their ecological interactions within metabolic networks, which has the potential to help elucidate diverse biochemical processes. 

Studies of natural ecological communities, from coral reefs to rainforests, could also be informed by studies of autocatalytic CRNs. An organism (a cell or multicellular system) can be coarse-grained as a single autocatalytic motif composed of numerous lower-level cooperating MASs. Here, the chemical complex that corresponds to the organism is a member of an autocatalytic motif, which sits in a species-productive cone due to the input of food derived from the physical environment (e.g., light and carbon dioxide in the case of photosynthetic organisms) and/or other from other organisms. However, although it ought to be possible to apply the tools developed here to enrich community ecosystem, in practice the lack of stoichiometric data for ecology (e.g., how many rabbits react with one wolf to make two wolves?) make this challenging at the current time.

Even if we currently lack the data sets needed to mathematically analyze metabolisms or other biological autocatalytic ecosystems, the work here has important implications for studying origins of life. Biological autocatalysis is not discretely different from that found in non-biological CRNs. The continuity between the two implies that the origin of life can be viewed as a process of expansion and complexification of autocatalytic ecosystems via the progressive activation (i.e., transitioning to positive flux) of more and more MASs over time. In future work, we hope to introduce kinetics to CCRNs and expand the insights gained here to identify conditions that permit and promote the open-ended exploration of the attractor space associated with CCRNs in an autocatalytic ecosystem.


\section*{Acknowledgements}
We want to thank Connor Simpson for help with combinatorics and programming, Alex Plum, Tymofii Sokolskyi, and Zhen Peng for biological discussions, and Asvin G., Dave Auckly, Diego Rojas, Gautam Neelakantham, Jeff Linderoth, and Polly Yu for mathematical discussions. We additionally want to thank Vladimir Sotirov for a careful reading of the manuscript and numerous suggestions. Finally, we thank an anonymous reviewer for devoting time and effort to significantly improve the work. 

\section*{Declarations}

 \subparagraph{Competing interests} The authors declare that they have no conflict of interest.

\subparagraph{Authors' contributions} Conceptualization: Praful Gagrani, Victor Blanco, Eric Smith, David Baum; Formal analysis and investigation: Praful Gagrani; Computational analysis and investigation: Victor Blanco; Writing - original draft preparation: Praful Gagrani, Victor Blanco; Writing - review and editing: Praful Gagrani, Victor Blanco, David Baum, Eric Smith; Funding acquisition: David Baum, Eric Smith, Victor Blanco. 
 
\subparagraph{Funding} 
 This work was partially funded by the National Science Foundation, Division of Environmental Biology (Grant No: DEB-2218817). V. Blanco acknowledges the financial support by the Spanish Ministerio de Ciencia e Innovaci\'on AEI/FEDER grant number PID2020-114594GBC21 and IMAG-Maria de Maeztu grant CEX2020-001105-M /AEI /10.13039/501100011033, Junta de Andalucía projects P18-FR-1422/2369, AT 21\_00032, FEDERUS-1256951, B-FQM-322-UGR20, and UE-NextGenerationEU (ayudas de movilidad para la recualificaci\'on del profesorado universitario).
 
\subparagraph{Availability of data and materials}
The Python codes for detecting minimal autocatalytic subnetworks are made available at \url{https://github.com/vblancoOR/autocatatalyticsubnetworks} \cite{Blanco_Python_Codes_for_2023}.

\appendix
    
\section{Deficiency theory and multistability in CRNs}
\label{app:deficiency}
In \cite{feinberg2019foundations}, a CRN is defined by the triple $\{\mathcal{S},\mathcal{C},\mathcal{R}\}$, where $\mathcal{S}, \mathcal{C},$ and $\mathcal{R}$ are the sets of species, complexes, and reactions respectively. Recall from Sec.\ \ref{sec:hypergraphs} that a CRN is a hypergraph with hypervertices in the set $\mathcal{C}$ and hyperedges in $\mathcal{R}$. The stoichiometric matrix $\mathbb{S}$ is defined such that (see Eq. \ref{eq:stoich_mat})
\begin{align*}
    \text{cols}(\mathbb{S}) &= \{y'-y|y\to y' \in \mathcal{R}\}. \nonumber
\end{align*}

Alternatively, $\mathbb{S}$ can be written as the product $Y \cdot \mathbb{M}$, where the columns of $Y$ are the stoichiometries of the hypervertices 
\begin{align*}
    \text{cols}(Y) = \{ \phantom{.} y \phantom{.} | y \in \mathcal{C}\}
\end{align*}
and $\mathbb{M}$ is the oriented vertex-edge incidence matrix of the hypergraph (see \cite{balandin1940structural,feinberg2019foundations})
\begin{align*}
    \text{cols}(\mathbb{M}) = \{e_{y'} - e_y |y\to y' \in \mathcal{R}\},
\end{align*}
where $e_y$ is the vector with $1$ at hypervertex $y$ and $0$ otherwise. Note that $Y$ and $\mathbb{M}$ are of sizes $|\mathcal{S}|\times|\mathcal{C}|$ and $|\mathcal{C}|\times|\mathcal{R}|$, respectively, yielding $\mathbb{S}$ of size $|\mathcal{S}|\times|\mathcal{R}|$. 

Since $\mathbb{S}$ is a linear operator, using the rank-nullity theorem we get
\begin{align}
    \text{dim}(\text{domain}(\mathbb{S})) &= \text{dim}(\text{im}(\mathbb{S})) + \text{dim}(\text{ker}(\mathbb{S})). 
    \label{eq:rank_null_stoic}
\end{align}
Let $E, V, s$ denote the number of reactions (hyperedges), complexes (hypervertices) and the dimension of the stoichiometric subspace $S$, respectively. Then, by definition,
\begin{align*}
    \text{dim}(\text{domain}(\mathbb{S})) &= E,\\
    \text{dim}(\text{im}(\mathbb{S})) &= s.
\end{align*}
Using $\mathbb{S} = Y \cdot \mathbb{M}$, we have
\begin{align}
    \text{dim}(\text{ker}(\mathbb{S})) &= \text{dim}(\text{ker}(\mathbb{M})) + \text{dim}(\text{ker}(Y) \cap \text{im}(\mathbb{M})). \label{eq:ker_S}
\end{align}
Notice that the complexes and reactions, without referring to the underlying species, form a graph, the vertex-edge incidence matrix of which is given by $\mathbb{M}$. Using properties of the vertex-edge incidence matrix, if the number of independent loops in the graph are $\imath$, then
\[ \text{dim}(\text{ker}(\mathbb{M})) = \imath. \]
Denoting $\text{dim}(\text{ker}(Y) \cap \text{im}(\mathbb{M}))$ by $\delta$, also referred to as the \textit{deficiency} of the CRN, substituting the above and Eq. \ref{eq:ker_S} in Eq. \ref{eq:rank_null_stoic}, we get
\begin{align}
    \delta &= E - \imath - s. \label{eq:delta_1}
\end{align}

Furthermore, let us denote the number of connected components or \textit{linkage classes} of the complex-reaction graph by $\ell$. Notice that $\ell$ and $\imath$ are the zeroth and first Betti numbers of a graph, respectively \cite{hatcher2005algebraic}. Using the formula for the Euler characteristic of a graph, we have the relation
\begin{align}
    V - E & = \ell - \imath, \label{eq:euler}
\end{align}
where $V$ is the number of vertices in the graph. Substituting the above in Eq. \ref{eq:delta_1}, we obtain the more widely used expression for the deficiency
\begin{align}
    \delta & = V - \ell - s. \label{eq:delta_2}
\end{align}
For direct derivations of Eq. \ref{eq:delta_2} without resorting to a counting of independent loops, see \cite{smith2017flows, feinberg2019foundations}.

Intuitively, deficiency is a topological property of the reaction network that informs about the steady state dynamics. In particular, a non-zero deficiency is a necessary, but not sufficient, condition for the dynamics to exhibit \textit{multistability} or multiple steady states. In certain cases, such as for deficiency one networks \cite{feinberg1988chemical}, there exist algorithms that can help classify whether or not multiple steady states can exist for a given CRN and identify the rate constants such that under mass-action kinetics it exhibits multistability. Once a model exhibiting multistability is finalized, one can also use algorithms from \cite{gagrani2023action,dykman1994large} to estimate the most-likely path the system will take, along with its probability, to escape from one steady state to another.

\section{Kernel of the stoichiometric matrix for the complete 1-constituent CCRN}
\label{app:kernel}
\begin{figure}
    \centering
    \includegraphics[width=0.5\textwidth]{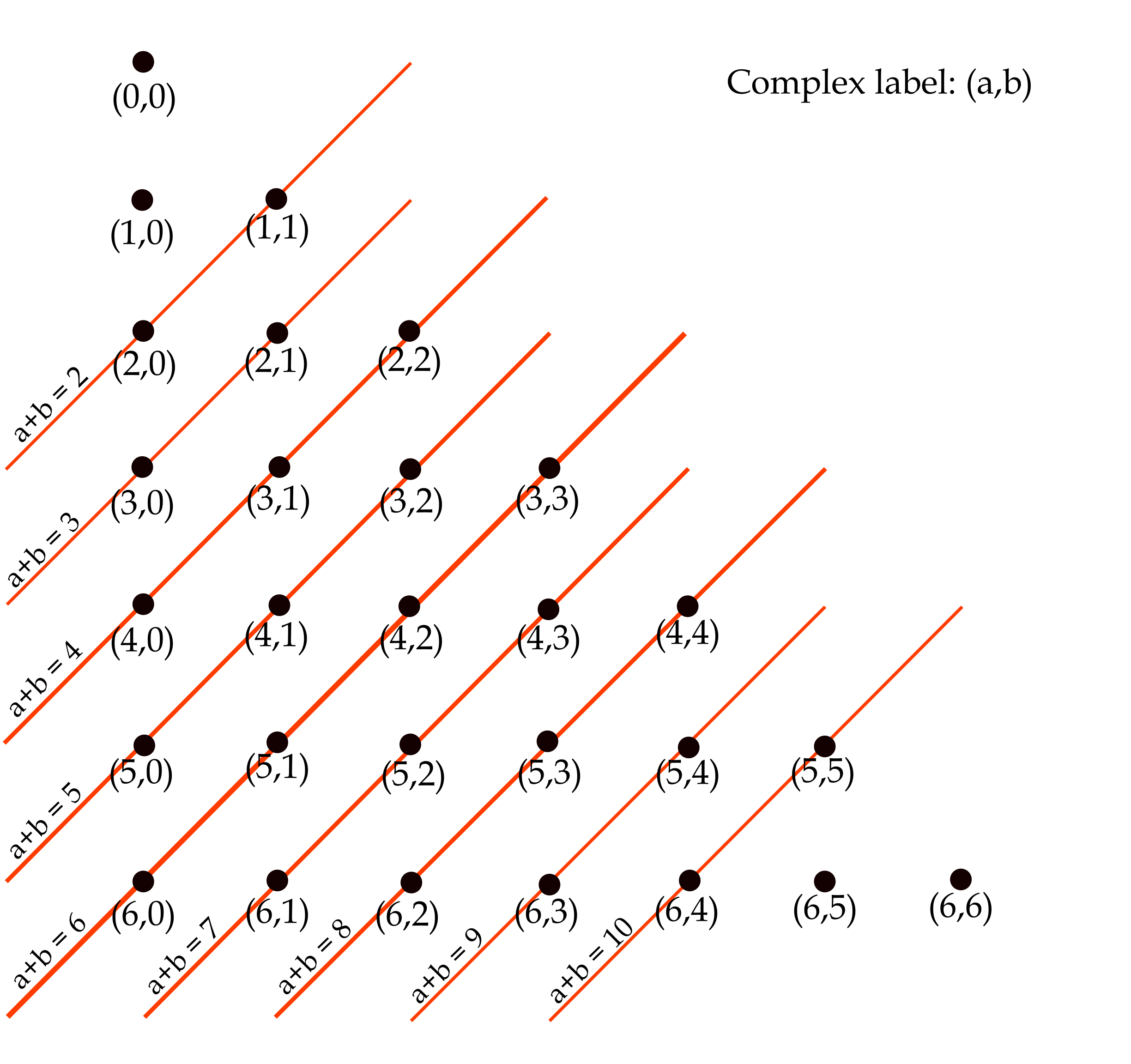}
    \caption{Representation of the complete 1-constituent CCRN with $L=6$. Complex label $(a,b)$ on a lattice point corresponds to the complex $\overline{a} + \overline{b}$. All the complexes connected by a red line correspond to a fully connected graph in the reaction network. For example, the red line connecting $\{(5,0),(4,1),(3,2)\}$, labelled $a+b=5$, corresponds to the reactions $\{\overline{5} \iff \overline{4} + \overline{1} \iff \overline{3} + \overline{2} \iff \overline{5}\}$ in the CCRN. }
    \label{fig:1-cluster-complexes}
\end{figure}

Let $\mathcal{H}_1^L = (\mathcal{S},\mathcal{C},\mathcal{R})$ denote a complete 1-constituent CCRN of length $L$ and order two. 1-constituent CCRN of length $L$ means that the species set consists of clusters up to length $L$, denoted as $\mathcal{S} = \{ \overline{1}, \overline{2}, \ldots, \overline{L}\}$ (for a detailed definition, see Sec.\ \ref{sec:CCRN-1constituent}.) For a diagrammatic representation of the reaction network, see Fig.\ \ref{fig:1-cluster-complexes}. Here, following Appendix \ref{app:deficiency}, we will calculate the kernel of the stoichiometric matrix for a general length $L$.

First, we will count the number of complexes in a complete 1-constituent CCRN of length $L$. Let us denote the set of complexes of size $N$ as $\mathcal{C}^N_L$. Recall that the size per constituent of a complex is defined as the sum of the conserved quantities in each constituent. For $2\leq N\leq L$, 
\[
\mathcal{C}^N_L=\{\overline{N},\overline{N-1}+\overline{1},\ldots, \overline{\floor{N/2}{}} +\overline{\ceil{N/2}{}}\}.
\] 
For $L < N < 2L -2$, 
\[
\mathcal{C}^N_L = \{ \overline{N-L} + \overline{L}, \ldots, \overline{\floor{N/2}{}} +\overline{\ceil{N/2}{}}\}.
\]
Thus, the number of complexes as a function of $N$ and $L$ are 
\begin{align*}
    |\mathcal{C}^N_L| &= 
    \begin{cases}
        \floor{N/2} + 1 \text{ for } 2\leq N\leq L,\\
        L + 1 - \ceil{N/2}  \text{ for } L < N \leq 2L-2.
    \end{cases}
\end{align*}
The total number of complexes for a CCRN of length $L$ is thus 
\[|\mathcal{C}_L| = \sum_N |\mathcal{C}^N_L| = \frac{(L+1)(L+2)}{2}-4.
\]
This can be found by either summing the individual contributions or by observing from Fig. \ref{fig:1-cluster-complexes} that the complexes occupy the lower triangle including the diagonal for a lattice with $L+1$ rows and columns excluding $4$ points. 

Let us denote the set of reactions of complexes of size $N$ by $\mathcal{R}^N_L$. Since the reaction network of the complete CCRN is obtained by connecting all allowed reactions, for complexes of size $N$, the number of reactions is given by 
\begin{align*}
    |\mathcal{R}^N_L| &=
    \begin{cases}
        2{ \floor{N/2} + 1 \choose 2} \text{ for } 2\leq N\leq L,\\
        2{L + 1 - \ceil{N/2}  \choose 2} \text{ for } L < N \leq 2L-2,        
    \end{cases}
\end{align*}
where the factor of $2$ is multiplied since all reactions are reversible.
Since, each $N$ contributes exactly one linkage class (or connected component), the total number of linkage classes $\ell$ for the complete CCRN is $2L - 3$.

Recall from Appendix \ref{app:deficiency}, the kernel of the stoichiometric matrix is the sum of the independent loops $\imath$ and the deficiency $\delta$, where using Equations \ref{eq:euler} and \ref{eq:delta_2} we have
\begin{align}
    \imath &= |\mathcal{R}_L| - |\mathcal{C}_L| + \ell\\ \nonumber 
    &= \sum _{N=2}^{2L-2}|\mathcal{R}^N_L| - \frac{L^2-5L-2}{4} \nonumber \\
    \delta &= |\mathcal{C}_L| - \ell - s \\\nonumber
        &= \frac{(L+1)(L+2)}{2}-4 - (2L - 3) - (L-1) \\
        &= \frac{(L-1)(L-2)}{2}. \nonumber 
\end{align}

The intersection $\text{Im}(\mathbb{M}) \cap \text{Ker}(Y)$, the dimension of which is the deficiency $\delta$, counts the null flows of the CCRN which are not simple loops in the network. While an explicit basis can be obtained using subnetworks of the form
\begin{align*}
    \overline{m} \to \overline{m-k} + \overline{k}\\
    \overline{n} + \overline{k} \to \overline{n+k} \\
    \overline{n+k} + \overline{m-k} \to \overline{n} + \overline{m},
\end{align*}
we will not be pursuing it since we do not need it for any result in this work. 

\section{Deficiency-one algorithm for L=3 1-constituent CCRN}
\label{app:deficiency-1}
The complete $L=3$ 1-constituent CCRN is the simplest complete CCRN, and as we will see, it possesses non-trivial dynamical properties. In particular, from Table \ref{table:1cluster}, it can be seen that the deficiency $\delta$ for the CCRN is one. This hints at the possibility that the CRN, when taken with mass-action kinetics, exhibits multistability. In this section, using Feinberg's deficiency-one algorithm from \cite{feinberg1988chemical,feinberg2019foundations}, we will confirm that the network can indeed exhibit multiple steady states, and find a relation between the rate constants and concentrations where that is the case. The algorithm works by converting the CRN into a system of linear inequalities, and if they have a solution, they can be used to obtain a relation between the rate constants and multiple steady states. 

The $L=3$ 1-constituent CCRN, denoted by $\mathcal{G} = (\mathcal{S},\mathcal{C},\mathcal{R})$, is given by the species set $\mathcal{S} = \{ \overline{1},\overline{2},\overline{3}\}$, complex set $\mathcal{C} = \{\overline{1} + \overline{1},\overline{2},\overline{1} + \overline{2},\overline{3},\overline{1} + \overline{3},\overline{2} + \overline{2}\} $ and reaction set
\begin{align*}
    \mathcal{R} &= \{ \ce{$\overline{1} + \overline{1}$ <=>[k_11][k_2] $\overline{2}$},\\
    & \phantom{--} \ce{$\overline{1} + \overline{2}$ <=>[k_12][k_3] $\overline{3}$},\\
    & \phantom{--} \ce{$\overline{1} + \overline{3}$ <=>[k_13][k_22] $\overline{2}+ \overline{2}$}\}.
\end{align*}
Notice that the above reaction network is a regular \cite{feinberg1988chemical} deficiency-one reaction network, and is thus a valid candidate for the deficiency-one algorithm. To proceed with our calculation, we will follow closely the example in Chapter 17 of \cite{feinberg2019foundations} and Sec.\ 5 in \cite{feinberg1988chemical}.

The confluence vector $g\in \mathbb{R}^\mathcal{C}$ for our network can be calculated to be (up to sign)
\begin{align*}
    g_{\overline{2}} &= -1, &     g_{\overline{1} + \overline{1}} &= 1,\\
     g_{\overline{1} + \overline{2}} &= -1, &    g_{\overline{3}} &= 1,\\
    g_{\overline{1} + \overline{3}} &= -1, &     g_{\overline{2}+\overline{2}} &= 1.\\
\end{align*}
For our construction, we will use the following upper-middle-lower partition
\begin{align*}
    U := & \{\overline{1} + \overline{3},\overline{2} + \overline{2} \},\\
    M := & \{\overline{1} + \overline{2},\overline{3} \},\\
    L := & \{  \overline{1} + \overline{1},\overline{2}\}.
\end{align*}
The partition induces the following system of inequalities on the chemical potential,
\begin{align*}
    2 \mu_{\overline{2}} > \mu_{\overline{1}} + \mu_{\overline{3}} > \mu_{\overline{3}} = \mu_{\overline{1}}+\mu_{\overline{2}} > \mu_{\overline{2}} > 2 \mu_{\overline{1}}.
\end{align*}
The inequalities are satisfied by 
\begin{align*}
     \mu_{\overline{1}} & > 0,\\
      \mu_{\overline{2}} & > 2  \mu_{\overline{1}},\\
    \mu_{\overline{3}} &= \mu_{\overline{1}}+\mu_{\overline{2}},
\end{align*}
which clearly have a solution, for e.g.\ $ \mu_{\overline{1}} = 1,  \mu_{\overline{2}} = 3,  \mu_{\overline{3}} = 4$. Thus, the CCRN can exhibit multiple steady states. 

To find the rate constants at which the system exhibits multistability, we use subsection 5.3 in \cite{feinberg1988chemical}. Following the reference, let us denote the mass-action kinetics rate vector by $\kappa_{y\to y'} := k_{y\to y'} x^y $, where $x$ is the concentration vector and $x^y:= \prod_{i\in \mathcal{S}} x_i^{y_i}$. Choosing the monotonically increasing function to be the exponential $\phi(x) = e^x$, and choosing $\eta$ so that $\mu_{\overline{1}} + \mu_{\overline{3}} < \eta < \mu_{\overline{2}} $ we get 
\begin{align*}
    \kappa_{\overline{1} + \overline{3} \to \overline{2} + \overline{2}}
    &= 
    1 \cdot \frac{e^{2\mu_{\overline{2}}}- e^{\eta}}{e^{2\mu_{\overline{2}}} - e^{\mu_{\overline{1}}}e^{\mu_{\overline{3}}}}  = k_{13} x_{\overline{1}}  x_{\overline{3}},
    \\
        \kappa_{ \overline{2} + \overline{2} \to \overline{1} + \overline{3} }
    &= 
    1 \cdot \frac{e^{\mu_{\overline{1}+\overline{3}}}- e^{\eta}}{e^{2\mu_{\overline{2}}} - e^{\mu_{\overline{1}}}e^{\mu_{\overline{3}}}} 
    =
    k_{22} x_{\overline{2}}^2 ,\\
    \kappa_{\overline{1} + \overline{2} \to \overline{3}}
    &= 
     k_{12} x_{\overline{1}}  x_{\overline{2}},\\
        \kappa_{ \overline{3} \to \overline{1} + \overline{2} }
    &= 
     k_{12} x_{\overline{1}}  x_{\overline{2}} - 1  = k_{3} x_{\overline{3}} ,\\
    \kappa_{\overline{2} \to \overline{1} + \overline{1}}
    &= 
    1 \cdot \frac{e^{2\mu_{\overline{1}}}- e^{\eta}}{e^{2\mu_{\overline{1}}} - e^{\mu_{\overline{2}}}} = k_{2}  x_{\overline{2}},\\
        \kappa_{ \overline{1} + \overline{1} \to \overline{2} }
    &= 
    1 \cdot \frac{e^{\mu_{\overline{2}}}- e^{\eta}}{e^{2\mu_{\overline{1}}} - e^{\mu_{\overline{2}}}} = k_{11} x_{\overline{1}}^2. 
\end{align*}
From \cite{feinberg1988chemical}, it is also known that if $x := (x_{\overline{1}},x_{\overline{2}},x_{\overline{3}})$, then so is $x^* := (x_{\overline{1}} e^{\mu_{\overline{1}}}, x_{\overline{2}} e^{\mu_{\overline{2}}}, x_{\overline{3}} e^{\mu_{\overline{3}}})$, thus yielding multiple steady states for our system. 

\section{Details of the algorithm for detecting stoichiometrically autocatalytic motifs}\label{app:alg}

In Section \ref{sec:algorithms}, we give the main ideas behind our algorithm to construct exclusively autocatalytic subnetworks using the input and output stoichiometric matrices. In what follows we detail the algorithm that we propose to enumerate the entire list of minimal \textit{stoichiometrically} autocatalytic subnetworks of a CRN, and that we applied to the CCRNs in our computational experiments.

\begin{notation}
    In this section, we abbreviate \textit{minimal stoichiometrically autocatalytic subnetwork} as MAS.
\end{notation}

The goal of the algorithm that we propose is to determine sets of core species and reactions conforming to the requirements of a MAS by solving, sequentially, a series of nested mathematical optimization problems.
Apart from the variables $y$, $z$ and $\mathbf{v}$ already described in Section \ref{sec:algorithms}, we use the following set of variables in our Mixed-Integer Linear Programming (MILP) problem:

\begin{align*}
q_i &= \begin{cases}
1 & \mbox{if species $i$ is a non-core present species}\\
& \mbox{in the MAS,} \\
0 & \mbox{otherwise}
 \end{cases},
\end{align*}
and $\Delta_i$ is a constant to be defined later
for all $i \in \mathcal{S}$.



The {\sc Master} optimization problem to construct MASs in this case is the following:
\begin{align}
\min & \dsum_{r\in \mathcal{R}} z_r\label{c:0}\\
& v_r \leq z_r, \forall r \in \mathcal{R}, \label{c:1}\\
& \dsum_{r \in \mathcal{R}} z_r \geq 2, \label{c:11}\\
& y_i + q_i \leq 1, \forall i \in \mathcal{S}, \label{c:2}\\
& y_i + q_i\geq z_r, \forall r \in \mathcal{R}, i \in \mathcal{S} \text{ with } \mathbb{S}^r_i>0 \neq 0,\label{c:3}\\
& \dsum_{i \in \mathcal{S}:\atop \mathbb{S}^r_i>0} y_i \geq z_r, \forall r \in \mathcal{R}, \label{c:5}\\
& \dsum_{i \in \mathcal{S}:\atop \mathbb{S}^r_i<0} y_i \geq z_r, \forall r \in \mathcal{R}, \label{c:6}\\
 & \dsum_{r \in \mathcal{R}:\atop |\mathbb{S}^r_i|>0} z_r \geq q_i, \forall i \in \mathcal{S}, \label{c:7}\\
 & \dsum_{r \in \mathcal{R}:\atop \mathbb{S}^r_i>0} z_r \geq y_i, \forall i \in \mathcal{S}, \label{c:8}\\
 & \dsum_{r \in \mathcal{R}:\atop \mathbb{S}^r_i<0} z_r \geq y_i, \forall i \in \mathcal{S}, \label{c:9}\\
& \dsum_{r \in \mathcal{R}} \mathbb{S}_i^r v_r \geq \varepsilon - \Delta_i (1-y_i), \forall i \in \mathcal{S}, \label{c:12}\\
& y_i, q_i\in \{0,1\}, \forall i \in \mathcal{S}, \label{c:13}\\
& v_r \in [0,1], z_r \in \{0,1\}\forall r \in \mathcal{R}.
\end{align}
There, the objective function accounts for minimizing the number of reactions in the MAS. This objective assures, that each of the generated autocatalytic subnetworks is support-inclusion minimal. Constraints \eqref{c:1} allow the correct representation  of variable $z_r$ (if $v_r>0$, then $z_r$ is forced to take value one, if $v_r=0$, by the minimization criterion, $z_r$ will also take value $0$). Constraint \eqref{c:11} avoids solutions with a single reaction. Constraints \eqref{c:2} assure that species are either core species or present non-core species but not both (it is also possible that they do not participate in the MAS). In case $y_i+q_i=1$, species $i$ is involved in the MASs either as food, member, core or waste species. Otherwise, species $i$ is not a part of the MAS. Constraints \eqref{c:3} assure that in case a reaction $r$ is selected to be part of the MAS ($z_j=1$), the reactants and product species of reaction $r$ must be present (either as core species or present non-core species). Since $q_i$ and $y_i$ are binary variables, the sum of them being greater than one (but because of the previous constraint the equality holds) implies that either $i$ is a core species ($y_i=1$), or non-core present species ($q_i=1$). Otherwise, the constraint is redundant.  Constraints \eqref{c:5} and \eqref{c:6} ensure that in case a reaction is in the MAS, there must be a core species as a reactant and another as a product in the restricted stoichiometric matrix. Observe that in case $z_r=1$, the sum of the binary variables $y_i$ with $i$ being reactant/product of $r$ must be at least one, implying that at least one of the reactant/product species must be activated as core species. In case $z_r=0$, the constraint is redundant.  Similarly, Constraints \eqref{c:7} enforces that in case a species is selected to be a present non-core species ($q_i=1$), then there must be a reaction where it is present.
\eqref{c:12} is the productivity condition (see Remark \ref{remark:productivity_implies}). There, in case $y_i=1$ ($i$ is a core species in the motif), the right-hand side in the constraint becomes $1$, implying that $\dsum_{r \in \mathcal{R}} \mathbb{S}^r_i v_r \geq \varepsilon >0$ (here $\varepsilon$ is a given positive but small enough constant to assure that the production is positive). 

Constants $\Delta_i$, for $i\in \mathcal{S}$, are assumed to be big enough assuring that in case $i$ is a non-core (present or not) species ($y_i=0$), the \textit{production} of species $i$ with flow vector $\mathbf{v}$ must always be greater than $\varepsilon - \Delta_i$ (making the constraint redundant). Specifically, for those non-core species whose production is negative (as food species), the absolute value production is upper bounded by $\Delta_i$. We take $\Delta_i = -\sum_{r\in \mathcal{R}:\atop \mathbb{S}^r_i<0} \mathbb{S}^r_i$, for all $i\in \mathcal{S}$.

The above model belongs to the class of discrete optimization problems, which are known to be computationally challenging~\cite{papadimitriou1981complexity}, but at the same time tremendously useful in determining solutions of complex decision problems in various fields~\cite{taha2014integer}. Although the problem of detecting a MAS is known to be a NP-complete \cite{andersen2012maximizing}, there are several available softwares with implemented routines to obtain solutions of integer programming problems (such as Gurobi, CPLEX, ...) in reasonable CPU time for medium size instances. 

In order to speed up the solution approach, the above master problem can be strengthened by adding valid inequalities that reduce the search of solutions. For instance, as shown in Theorem \ref{theorem:core_inverse}, since the number of reactions in an MAS coincides with the number of core species that are part of it, one can add this (redundant) constraint to the model:
$$
\dsum_{i \in \mathcal{S}} y_i = \dsum_{i \in \mathcal{R}} z_j
$$
Furthermore, in a reversible CRN, denoting by 
\begin{align*}
    \mathrm{Rev} = \{&(r,r') \in \mathcal{R}\times \mathcal{R}: \\
    &\mbox{reaction $r$ is the reverse of reaction $r'$}\},
\end{align*} 
one must also impose the following constraints that avoid using a reaction and its reverse in the same subnetwork:
$$
z_r + z_{r'} \leq 1, \forall (r, r')\in \mathrm{Rev}.
$$
That is, if $r$ and $r'$ are reverse reaction of each other, the $z$-variables of those reactions cannot be both one.

The solution of the problem above is a MAS with minimum number of reactions. Once the problem is solved, with optimal values in the variables $(\bar y, \bar q, \bar z, \overline{\mathbf{v}})$, the set of core species is obtained as:
\[{\rm Core} =\{i\in \mathcal{S}: \bar y_i=1\},\]
and the set of reactions involved in the MAS is:
$$
\mathcal{R}'= \{r \in \mathcal{R}: \bar z_r=1\}.
$$

Among the set of present non-core species, ${\rm Q}=\{i \in N: \bar q_i=1\}$, one can classify them based on the sign of the entries of the restricted stoichiometric matrix:
\begin{align*}
{\rm Food} &=\{i\in {\rm Q}: \mathbb{S}^r_i\leq 0, \forall r \in \mathcal{R}'\}, \\
{\rm Waste} &=\{i\in {\rm Q}: \mathbb{S}^r_i\geq 0, \forall r \in \mathcal{R}'\},
\\
{\rm Member} &=({\rm Q} \cup {\rm Core})\backslash ({\rm Food} \cup {\rm Waste})
\end{align*}

The algorithm to generate the whole set of MASs is described in Algorithm \ref{alg:ac}. There, at each iteration, $\ell$, the above Master problem is solved. Once a solution is obtained, say $(y^{\ell},   q^{\ell}, z^{\ell}, \textbf{v}^{\ell})$, the tuple of reactions that are part of the MAS is considered to restrict the next MASs. In particular, it is disallowed for the next MAS to use the same reactions tuples, which assures the minimality of the generated MASs. This restriction is incorporated in the model by means of Constraint \eqref{incomp}. The process is repeated until no more autocatalytic subnetwork can be found respecting the pool of incompatibilities that have been added during the procedure. With this algorithm the MASs are generated while being sorted by number of reactions involved in the subnetwork since each time the problem is solved the number of reactions is minimized.

 \begin{algorithm}[H]
  \SetAlgoLined
 \KwData{$\mathcal{S}$, $\mathcal{R}$, $\mathbb{S}$, $\ell=0$}

 \While{{\sc Feasible}}{
  Solve {\sc Master}: $\textsc{Mas}^{\ell} := (y^{\ell},   q^{\ell}, z^{\ell}, \textbf{v}^{\ell})$.
  
  \If{{\sc Master} is feasible}{
   Save the solution and add to {\sc Master} the constraint:
   \begin{equation}\label{incomp}
\dsum_{r \in \mathcal{R}:\atop z_r=1} z^{\ell}_j \leq |\{r\in \mathcal{R}: z_r^{\ell} =1\}|-1
\end{equation}
$\ell +1 \leftarrow \ell$.}
 }
  \KwResult{Set of MASs of a CRN: $\textsc{Mas}^{1}, \ldots, \textsc{Mas}^{\ell}$.}
 \caption{Optimization-based approach to enumerate MASs.\label{alg:ac}}
\end{algorithm}
 
One of the advantages of using mathematical optimization-based approaches to determine \textit{special} sets from an underlying set (as MASs from a CRN) is its flexibility to be adapted to different features. For instance, one can easily filter the obtained MASs to have at most a number of reactions, {\sc MaxReactions}, involved by adding the following constraint to the model:
$$
\dsum_{r\in \mathcal{R}} z_r \leq \text{\sc MaxReactions}.
$$
Analogously, one can also generate MASs where just a given set of species takes part in the core species set, $\mathcal{S}_0 \subseteq \mathcal{S}$, by adding:
$$
\dsum_{i\in \mathcal{S}_0} y_i =0.
$$

\begin{figure}[h]
\includegraphics[scale=0.5]{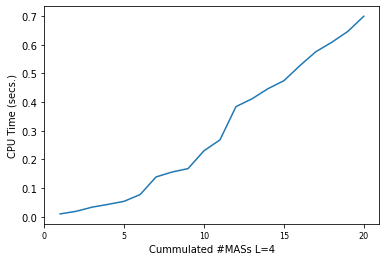}\\\includegraphics[scale=0.5]{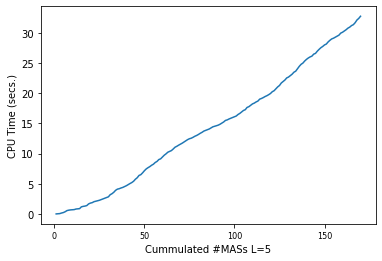}\\\includegraphics[scale=0.5]{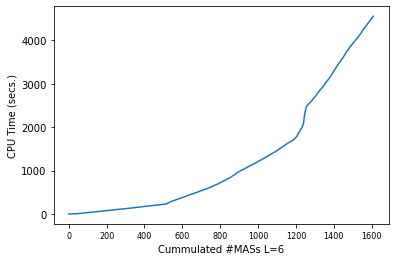}
\caption{CPU time vs. \# MASs for 1-constituent CCRN.\label{times}}
\end{figure}

As already mentioned, computing the entire list of MASs in a CRN is computationally challenging. Each iteration in the proposed procedure requires solving a mixed integer linear optimization problem with the constraints in the initial Master problem, but also those in the shape of \eqref{incomp} that are added each time a MASs is found. Thus, the difficulty of solving this optimization problem increases with the number of MASs for a given CRN. In Figure \eqref{times} we show the performance of the cumulative CPU time required to generate the MASs for 1-constituent CCRNS with $L=4, 5,$ and $6$. As can be observed from the plots, the consumed CPU time increases linearly in the number of MASs, with a slope strictly greater than one.

\begin{table}[h]
\begin{tabular}{rc}
Reaction id & Reaction\\\hline
R1/R2 & \ce{$2{\bar 1}$ <=> $ {\bar 2}$}\\
R3/R4 & \ce{${\bar 1} +{\bar 2}$ <=> $ {\bar 3}$}\\
R5/R6 & \ce{${\bar 1} +{\bar 3}$ <=> $ {\bar 4}$}\\
R7/R8 & \ce{${\bar 1} +{\bar 4}$ <=> $ {\bar 2} + {\bar 3}$}\\
R9/R10 & \ce{${\bar 1} +{\bar 3}$ <=> $ 2{\bar 2}$}\\
R11/R12 & \ce{$2{\bar 2}$ <=> $ {\bar 4}$}\\
R13/R14 & \ce{${\bar 2} +{\bar 4}$ <=> $ 2{\bar 3}$}\\\hline
\end{tabular}
\caption{List of reactions for a CCRN with 1 constinuent and $L=4$.\label{t:react4}}
\end{table}

\subsection{Checking intersection of cones}
\label{app:check_intersection}
Once the whole set of MASs has been generated, $\mathcal{M} = \{\textsc{Mas}_1, \ldots, \textsc{Mas}_K\}$, we have also developed an optimization-based methodology to check the pairwise intersection of the different polyhedral flow-productive cones and partition-productive cones induced from the set $\mathcal{M}$. Consider two of these cones in $\mathbb{R}^d_+$, say $C_1(v_1, \ldots, v_s)$ and $C_2(w_1, \ldots, w_r)$, described by means of its generating vectors (see Section \ref{sec:cones}). We first check if they the interior of the cones intersect or not by finding a non negative vector that belongs to the intersection of the closed cones maximizing the sum of the entries of that vector with respect to each of the generators of the cones, i.e.:
\begin{alignat*}{4}
\rho^*(C_1,C_2) := & \max &&\sum_{l=1}^s \lambda_l + \sum_{l=1}^r \mu_l\\
& \mbox{s.t.} && x = \dsum_{l=1}^s \lambda_l v_l,\\
& && x = \dsum_{l=1}^r \mu_l w_l,\\
&&& x \in \mathbb{R}^d_+,\\
&&& \lambda_l \in [0,1], \forall l=1, \ldots, s,\\
&&& \mu_l \in [0,1], \forall l=1, \ldots, r.
\end{alignat*}
If the optimal solution of the above linear programming problem is $\rho^*(C_1, C_2)=0$, then, the cones do not intersect (otherwise there should be another vector with greatest coefficients in the representation). If $\rho^*(C_1,C_2)>0$, then, the cones intersect. 

In case the cones intersect, different situations are possible: the cones are identical, one of the cones is strictly contained in the other, or the cones partially intersect. To detect each of the above possibilities, we compute the distances between each of the  normalized generating vectors of one of the cones to the generating vectors in the other cones:
$$
\phi^1_l = D(v_l, \{w_1, \ldots, w_r\}) = \min_{j=1, \ldots, r} \{\|v_l-w_j\|\}
$$
for $l=1, \ldots, s,$
and
$$
\phi^2_l = D(w_l, \{v_1, \ldots, v_s\}) = \min_{j=1, \ldots, s} \{\|w_l-v_j\|\}
$$
for $l=1, \ldots, r$.

With these values, in case the cones intersect:
\begin{itemize}
\item If $\sum_{l=1}^s \phi^1_l = \sum_{l=1}^r \phi^2_l =0$, then for each of the generating vectors in both cones, there is a normalized generating vector of the other cone at distance zero from it. Thus, both cones are identical.
\item If $\sum_{l=1}^s \phi^1_l = 0$, but $ \sum_{l=1}^r \phi^2_l >0$, then, $C_1 \subset C_2$ since for all the generating vectors in $C_1$, there exist a generating vector in $C_2$ at distance zero from it.
\item If $\sum_{l=1}^s \phi^1_l >0$, but $ \sum_{l=1}^r \phi^2_l =0$, then, $C_2 \subset C_1$ by the same reasoning as above.
\item If $\sum_{l=1}^s \phi^1_l >0$ and $ \sum_{l=1}^r \phi^2_l >0$, then, the cones intersect but only partially.
\end{itemize}

\section{Minimal Autocatalytic Subnetworks for CCRN with $L=4$ and $L=5$}
\label{sec:MAS_list}

Tables \ref{t:react4} and \ref{t:react5} show the set of feasible reactions for 1-constituent CCRNs with $L=4$ and $L=5$, respectively. For those CCRNs we report in Tables \ref{t:mas4}, \ref{t:mas5a}, and \ref{t:mas5b} the list of MASs obtained with our approach. Each row in those tables represent a MAS. In the first column of those tables (id) indicates the identifier of the MAS. The second column (Reactions (Flow)) provides the list of reactions taking part of the MAS. In parenthesis we provide a feasible flow for autocatalysis. In the third column (FWMC Partition), we report the FWMC partition for such a MAS. Finally, in the fourth column (Production) we show the production vector for the feasible flow in the second column, for each of the core species in the MAS.

\begin{table}[h]
\begin{tabular}{rc}
Reaction id & Reaction\\\hline
R1/R2 & \ce{$2 {\bar 1}$ <=> $ {\bar 2}$}\\
R3/R4 & \ce{${\bar 1} +{\bar 2}$ <=> $ {\bar 3}$}\\
R5/R6 & \ce{${\bar 1} +{\bar 3}$ <=> $ {\bar 4}$}\\
R7/R8 & \ce{${\bar 1} +{\bar 4}$ <=> $ {\bar 5}$}\\
R9/R10 & \ce{${\bar 1} +{\bar 5}$ <=> $ 2{\bar 3}$}\\
R11/R12 & \ce{${\bar 1} +{\bar 5}$ <=> $ {\bar 2} + {\bar 4}$}\\
R13/R14 & \ce{${\bar 1} +{\bar 4}$ <=> $ {\bar 2} + {\bar 3}$}\\
R15/R16 & \ce{${\bar 1} +{\bar 3}$ <=> $ 2{\bar 2}$}\\
R17/R18 & \ce{$2 {\bar 2}$ <=> $ {\bar 4}$}\\
R19/R20 & \ce{${\bar 2} +{\bar 3}$ <=> $ {\bar 5}$}\\
R21/R22 & \ce{${\bar 2} +{\bar 5}$ <=> $ {\bar 3} + {\bar 4}$}\\
R23/R24 & \ce{${\bar 2} +{\bar 4}$ <=> $ 2{\bar 3}$}\\
R25/R26 & \ce{${\bar 3} +{\bar 5}$ <=> $ 2{\bar 4}$}\\
\end{tabular}
\caption{List of reactions for a CCRN with 1 constinuent and $L=5$.\label{t:react5}}
\end{table}

\begin{table}[h]
\begin{adjustbox}{width=0.42\textwidth}
\begin{tabular}{ccccc}
MAS id & Reactions (Flow) & FWMC Partition &  Production\\\hline
1 & $\{ {\rm R2} (1),  {\rm R9} (1)\}$   & $\{\{\bar 3\}, \{\},\{\{\bar 1,\bar 2\}\}$  & $(1,  1)$\\
2 & $\{ {\rm R9} (1),  {\rm R13} (1)\}$   & $\{\{\bar 1,\bar 4\}, \{\},\{\{\bar 2,\bar 3\}\}$  & $(1,  1)$\\
3 & $\{ {\rm R4} (3),  {\rm R13} (2)\}$   & $\{\{\bar 4\}, \{\bar 1\}, \{\{\bar 2,\bar 3\}\}$  & $(1,  1)$\\
4 & $\{ {\rm R3} (3),  {\rm R9} (2)\}$   & $\{\{\bar 1\}, \{\},\{\{\bar 2,\bar 3\}\}$  & $(1,  1)$\\
5 & $\{ {\rm R8} (3),  {\rm R13} (2)\}$   & $\{\{\bar 2\}, \{\bar 1\}, \{\{\bar 3,\bar 4\}\}$  & $(1,  1)$\\
6 & $\{ {\rm R5} (3),  {\rm R13} (2)\}$   & $\{\{\bar 1,\bar 2\}, \{\},\{\{\bar 3,\bar 4\}\}$  & $(1,  1)$\\
7 & $\{ {\rm R8} (3),  {\rm R9} (2)\}$   & $\{\{\bar 3\}, \{\bar 4\}, \{\{\bar 1,\bar 2\}\}$  & $(1,  1)$\\
8 & $\{ {\rm R2} (2),  {\rm R7} (3)\}$   & $\{\{\bar 4\}, \{\bar 3\}, \{\{\bar 1,\bar 2\}\}$  & $(1,  1)$\\
9 & $\{ {\rm R8} (3),  {\rm R12} (2)\}$   & $\{\{\bar 3\}, \{\bar 1\}, \{\{\bar 2,\bar 4\}\}$  & $(1,  1)$\\
10 & $\{ {\rm R4} (2),  {\rm R7} (2),  {\rm R10} (1)\}$   & $\{\{\bar 4\}, \{\},\{\{\bar 1,\bar 2, \bar 3\}\}$  & $(1,  2,  1)$\\
11 & $\{ {\rm R3} (2),  {\rm R6} (1),  {\rm R8} (2)\}$   & $\{\{\bar 2\}, \{\},\{\{\bar 1,\bar 3, \bar 4\}\}$  & $(1,  1,  1)$\\
12 & $\{ {\rm R2} (2),  {\rm R5} (3),  {\rm R12} (2)\}$   & $\{\{\bar 3\}, \{\},\{\{\bar 1,\bar 2, \bar 4\}\}$  & $(1,  2,  1)$\\
13 & $\{ {\rm R3} (2),  {\rm R5} (4),  {\rm R7} (3)\}$   & $\{\{\bar 1\}, \{\},\{\{\bar 2,\bar 3, \bar 4\}\}$  & $(1,  1,  1)$\\
14 & $\{ {\rm R5} (4),  {\rm R7} (5),  {\rm R11} (2)\}$   & $\{\{\bar 1\}, \{\},\{\{\bar 2,\bar 3, \bar 4\}\}$  & $(1,  1,  1)$\\
15 & $\{ {\rm R3} (5),  {\rm R5} (4),  {\rm R12} (3)\}$   & $\{\{\bar 1\}, \{\},\{\{\bar 2,\bar 3, \bar 4\}\}$  & $(1,  1,  1)$\\
16 & $\{ {\rm R3} (6),  {\rm R7} (3),  {\rm R14} (4)\}$   & $\{\{\bar 1\}, \{\},\{\{\bar 2,\bar 3, \bar 4\}\}$  & $(1,  1,  1)$\\
17 & $\{ {\rm R5} (6),  {\rm R7} (5),  {\rm R10} (2)\}$   & $\{\{\}, \{\},\{\bar 1, \{\bar 2,\bar 3, \bar 4\}\}$  & $(1,  1,  1)$\\
18 & $\{ {\rm R7} (5),  {\rm R9} (4),  {\rm R11} (6)\}$   & $\{\{\bar 1\}, \{\},\{\{\bar 2,\bar 3, \bar 4\}\}$  & $(1,  1,  1)$\\
19 & $\{ {\rm R3} (9),  {\rm R12} (3),  {\rm R14} (4)\}$   & $\{\{\bar 1\}, \{\},\{\{\bar 2,\bar 3, \bar 4\}\}$  & $(1,  1,  1)$\\
20 & $\{ {\rm R7} (9),  {\rm R11} (6),  {\rm R14} (4)\}$   & $\{\{\bar 1\}, \{\},\{\{\bar 2,\bar 3, \bar 4\}\}$  & $(1,  1,  1)$
 \end{tabular}
 \end{adjustbox}
\caption{List of MASs for 1-constituent CCRN $L=4$.\label{t:mas4}}
\end{table}

\begin{table}[h]
\begin{adjustbox}{width=0.4\textwidth}{
\begin{tabular}{ccccc}
MAS id & Reactions (Flow) & FWMC Partition &  Production\\\hline
1 & $\{ {\rm R2} (1),  {\rm R15} (1)\}$   & $\{\{\bar 3\}, \{\},\{\{\bar 1,\bar 2\}\}$  & $(1,  1)$\\
2 & $\{ {\rm R23} (1),  {\rm R25} (1)\}$   & $\{\{\bar 2,\bar 5\}, \{\},\{\{\bar 3,\bar 4\}\}$  & $(1,  1)$\\
3 & $\{ {\rm R15} (1),  {\rm R23} (1)\}$   & $\{\{\bar 1,\bar 4\}, \{\},\{\{\bar 2,\bar 3\}\}$  & $(1,  1)$\\
4 & $\{ {\rm R13} (3),  {\rm R25} (2)\}$   & $\{\{\bar 1,\bar 5\}, \{\bar 2\}, \{\{\bar 3,\bar 4\}\}$  & $(1,  1)$\\
5 & $\{ {\rm R22} (3),  {\rm R23} (2)\}$   & $\{\{\bar 4\}, \{\bar 5\}, \{\{\bar 2,\bar 3\}\}$  & $(1,  1)$\\
6 & $\{ {\rm R2} (2),  {\rm R13} (3)\}$   & $\{\{\bar 4\}, \{\bar 3\}, \{\{\bar 1,\bar 2\}\}$  & $(1,  1)$\\
7 & $\{ {\rm R12} (3),  {\rm R25} (2)\}$   & $\{\{\bar 2,\bar 3\}, \{\bar 1\}, \{\{\bar 4,\bar 5\}\}$  & $(1,  1)$\\
8 & $\{ {\rm R7} (3),  {\rm R25} (2)\}$   & $\{\{\bar 1,\bar 3\}, \{\},\{\{\bar 4,\bar 5\}\}$  & $(1,  1)$\\
9 & $\{ {\rm R9} (2),  {\rm R22} (3)\}$   & $\{\{\bar 1,\bar 4\}, \{\bar 2\}, \{\{\bar 3,\bar 5\}\}$  & $(1,  1)$\\
10 & $\{ {\rm R6} (3),  {\rm R25} (2)\}$   & $\{\{\bar 5\}, \{\bar 1\}, \{\{\bar 3,\bar 4\}\}$  & $(1,  1)$\\
11 & $\{ {\rm R14} (3),  {\rm R15} (2)\}$   & $\{\{\bar 3\}, \{\bar 4\}, \{\{\bar 1,\bar 2\}\}$  & $(1,  1)$\\
12 & $\{ {\rm R5} (3),  {\rm R23} (2)\}$   & $\{\{\bar 1,\bar 2\}, \{\},\{\{\bar 3,\bar 4\}\}$  & $(1,  1)$\\
13 & $\{ {\rm R14} (3),  {\rm R23} (2)\}$   & $\{\{\bar 2\}, \{\bar 1\}, \{\{\bar 3,\bar 4\}\}$  & $(1,  1)$\\
14 & $\{ {\rm R15} (2),  {\rm R21} (3)\}$   & $\{\{\bar 1,\bar 5\}, \{\bar 4\}, \{\{\bar 2,\bar 3\}\}$  & $(1,  1)$\\
15 & $\{ {\rm R22} (3),  {\rm R25} (2)\}$   & $\{\{\bar 3\}, \{\bar 2\}, \{\{\bar 4,\bar 5\}\}$  & $(1,  1)$\\
16 & $\{ {\rm R12} (3),  {\rm R15} (2)\}$   & $\{\{\bar 3,\bar 4\}, \{\bar 5\}, \{\{\bar 1,\bar 2\}\}$  & $(1,  1)$\\
17 & $\{ {\rm R14} (3),  {\rm R18} (2)\}$   & $\{\{\bar 3\}, \{\bar 1\}, \{\{\bar 2,\bar 4\}\}$  & $(1,  1)$\\
18 & $\{ {\rm R2} (2),  {\rm R11} (3)\}$   & $\{\{\bar 5\}, \{\bar 4\}, \{\{\bar 1,\bar 2\}\}$  & $(1,  1)$\\
19 & $\{ {\rm R4} (3),  {\rm R9} (2)\}$   & $\{\{\bar 5\}, \{\bar 2\}, \{\{\bar 1,\bar 3\}\}$  & $(1,  1)$\\
20 & $\{ {\rm R9} (2),  {\rm R14} (3)\}$   & $\{\{\bar 2,\bar 5\}, \{\bar 4\}, \{\{\bar 1,\bar 3\}\}$  & $(1,  1)$\\
21 & $\{ {\rm R18} (2),  {\rm R21} (3)\}$   & $\{\{\bar 5\}, \{\bar 3\}, \{\{\bar 2,\bar 4\}\}$  & $(1,  1)$\\
22 & $\{ {\rm R3} (3),  {\rm R15} (2)\}$   & $\{\{\bar 1\}, \{\},\{\{\bar 2,\bar 3\}\}$  & $(1,  1)$\\
23 & $\{ {\rm R4} (3),  {\rm R23} (2)\}$   & $\{\{\bar 4\}, \{\bar 1\}, \{\{\bar 2,\bar 3\}\}$  & $(1,  1)$\\
24 & $\{ {\rm R9} (2),  {\rm R19} (3)\}$   & $\{\{\bar 1,\bar 2\}, \{\},\{\{\bar 3,\bar 5\}\}$  & $(1,  1)$\\
25 & $\{ {\rm R11} (1),  {\rm R14} (2),  {\rm R22} (2)\}$   & $\{\{\bar 3\}, \{\},\{\bar 1, \{\bar 2,\bar 4, \bar 5\}\}$  & $(1,  1,  1)$\\
26 & $\{ {\rm R13} (2),  {\rm R21} (2),  {\rm R24} (1)\}$   & $\{\{\bar 1,\bar 5\}, \{\},\{\{\bar 2,\bar 3, \bar 4\}\}$  & $(1,  2,  1)$\\
27 & $\{ {\rm R11} (2),  {\rm R19} (2),  {\rm R22} (1)\}$   & $\{\{\bar 1,\bar 3\}, \{\},\{\{\bar 2,\bar 4, \bar 5\}\}$  & $(1,  1,  1)$\\
28 & $\{ {\rm R3} (2),  {\rm R6} (1),  {\rm R14} (2)\}$   & $\{\{\bar 2\}, \{\},\{\{\bar 1,\bar 3, \bar 4\}\}$  & $(1,  1,  1)$\\
29 & $\{ {\rm R12} (2),  {\rm R20} (2),  {\rm R22} (1)\}$   & $\{\{\bar 4\}, \{\bar 1\}, \{\{\bar 2,\bar 3, \bar 5\}\}$  & $(1,  1,  1)$\\
30 & $\{ {\rm R4} (2),  {\rm R13} (1),  {\rm R21} (2)\}$   & $\{\{\bar 5\}, \{\},\{\bar 1, \{\bar 2,\bar 3, \bar 4\}\}$  & $(1,  1,  1)$\\
31 & $\{ {\rm R6} (2),  {\rm R11} (2),  {\rm R14} (1)\}$   & $\{\{\bar 5\}, \{\},\{\bar 2, \{\bar 1,\bar 3, \bar 4\}\}$  & $(1,  1,  1)$\\
32 & $\{ {\rm R3} (2),  {\rm R20} (1),  {\rm R22} (2)\}$   & $\{\{\bar 1,\bar 4\}, \{\},\{\{\bar 2,\bar 3, \bar 5\}\}$  & $(1,  1,  1)$\\
33 & $\{ {\rm R4} (1),  {\rm R12} (2),  {\rm R13} (2)\}$   & $\{\{\bar 4\}, \{\bar 5\}, \{\{\bar 1,\bar 2, \bar 3\}\}$  & $(1,  1,  1)$\\
34 & $\{ {\rm R4} (2),  {\rm R13} (2),  {\rm R16} (1)\}$   & $\{\{\bar 4\}, \{\},\{\{\bar 1,\bar 2, \bar 3\}\}$  & $(1,  2,  1)$\\
35 & $\{ {\rm R5} (2),  {\rm R8} (1),  {\rm R12} (2)\}$   & $\{\{\bar 2,\bar 3\}, \{\},\{\{\bar 1,\bar 4, \bar 5\}\}$  & $(1,  1,  1)$\\
36 & $\{ {\rm R5} (2),  {\rm R13} (2),  {\rm R21} (1)\}$   & $\{\{\bar 1,\bar 5\}, \{\},\{\{\bar 2,\bar 3, \bar 4\}\}$  & $(1,  1,  1)$\\
37 & $\{ {\rm R11} (2),  {\rm R18} (1),  {\rm R19} (3)\}$   & $\{\{\bar 1,\bar 3\}, \{\},\{\{\bar 2,\bar 4, \bar 5\}\}$  & $(1,  1,  1)$\\
38 & $\{ {\rm R19} (3),  {\rm R21} (2),  {\rm R23} (1)\}$   & $\{\{\bar 2\}, \{\},\{\{\bar 3,\bar 4, \bar 5\}\}$  & $(1,  1,  1)$\\
39 & $\{ {\rm R3} (3),  {\rm R18} (2),  {\rm R25} (2)\}$   & $\{\{\bar 1,\bar 5\}, \{\},\{\{\bar 2,\bar 3, \bar 4\}\}$  & $(1,  1,  2)$\\
40 & $\{ {\rm R7} (2),  {\rm R19} (2),  {\rm R21} (3)\}$   & $\{\{\bar 1,\bar 2\}, \{\},\{\{\bar 3,\bar 4, \bar 5\}\}$  & $(1,  1,  1)$\\
41 & $\{ {\rm R8} (2),  {\rm R14} (2),  {\rm R22} (3)\}$   & $\{\{\bar 3\}, \{\bar 1\}, \{\{\bar 2,\bar 4, \bar 5\}\}$  & $(1,  1,  1)$\\
42 & $\{ {\rm R12} (2),  {\rm R19} (2),  {\rm R21} (3)\}$   & $\{\{\bar 2\}, \{\bar 1\}, \{\{\bar 3,\bar 4, \bar 5\}\}$  & $(1,  1,  1)$\\
43 & $\{ {\rm R11} (2),  {\rm R17} (2),  {\rm R22} (3)\}$   & $\{\{\bar 1,\bar 3\}, \{\},\{\{\bar 2,\bar 4, \bar 5\}\}$  & $(1,  1,  1)$\\
44 & $\{ {\rm R5} (2),  {\rm R12} (3),  {\rm R21} (2)\}$   & $\{\{\bar 2\}, \{\},\{\bar 3, \{\bar 1,\bar 4, \bar 5\}\}$  & $(1,  1,  1)$\\
45 & $\{ {\rm R18} (2),  {\rm R19} (3),  {\rm R25} (2)\}$   & $\{\{\bar 3\}, \{\},\{\{\bar 2,\bar 4, \bar 5\}\}$  & $(1,  2,  1)$\\
46 & $\{ {\rm R12} (3),  {\rm R13} (2),  {\rm R20} (2)\}$   & $\{\{\bar 4\}, \{\bar 3\}, \{\{\bar 1,\bar 2, \bar 5\}\}$  & $(1,  1,  1)$\\
47 & $\{ {\rm R19} (3),  {\rm R21} (3),  {\rm R26} (1)\}$   & $\{\{\bar 2\}, \{\},\{\{\bar 3,\bar 4, \bar 5\}\}$  & $(1,  1,  1)$\\
48 & $\{ {\rm R7} (2),  {\rm R11} (3),  {\rm R19} (2)\}$   & $\{\{\bar 1,\bar 3\}, \{\},\{\{\bar 2,\bar 4, \bar 5\}\}$  & $(1,  1,  1)$\\
49 & $\{ {\rm R2} (2),  {\rm R5} (3),  {\rm R18} (2)\}$   & $\{\{\bar 3\}, \{\},\{\{\bar 1,\bar 2, \bar 4\}\}$  & $(1,  2,  1)$\\
50 & $\{ {\rm R16} (2),  {\rm R20} (2),  {\rm R22} (3)\}$   & $\{\{\bar 4\}, \{\bar 1\}, \{\{\bar 2,\bar 3, \bar 5\}\}$  & $(1,  1,  1)$\\
51 & $\{ {\rm R4} (2),  {\rm R6} (3),  {\rm R11} (4)\}$   & $\{\{\bar 5\}, \{\bar 2\}, \{\{\bar 1,\bar 3, \bar 4\}\}$  & $(1,  1,  1)$\\
52 & $\{ {\rm R14} (4),  {\rm R20} (2),  {\rm R22} (3)\}$   & $\{\{\}, \{\bar 1\}, \{\bar 3, \{\bar 2,\bar 4, \bar 5\}\}$  & $(1,  1,  1)$\\
53 & $\{ {\rm R4} (2),  {\rm R12} (4),  {\rm R20} (3)\}$   & $\{\{\bar 4\}, \{\bar 1\}, \{\{\bar 2,\bar 3, \bar 5\}\}$  & $(1,  1,  1)$\\
54 & $\{ {\rm R7} (4),  {\rm R11} (3),  {\rm R14} (2)\}$   & $\{\{\bar 3\}, \{\},\{\bar 1, \{\bar 2,\bar 4, \bar 5\}\}$  & $(1,  1,  1)$\\
55 & $\{ {\rm R11} (3),  {\rm R13} (2),  {\rm R19} (4)\}$   & $\{\{\bar 1\}, \{\},\{\bar 3, \{\bar 2,\bar 4, \bar 5\}\}$  & $(1,  1,  1)$\\
56 & $\{ {\rm R13} (2),  {\rm R19} (4),  {\rm R21} (3)\}$   & $\{\{\bar 1\}, \{\},\{\bar 2, \{\bar 3,\bar 4, \bar 5\}\}$  & $(1,  1,  1)$\\
57 & $\{ {\rm R12} (4),  {\rm R14} (2),  {\rm R21} (3)\}$   & $\{\{\bar 2\}, \{\bar 1\}, \{\{\bar 3,\bar 4, \bar 5\}\}$  & $(1,  1,  1)$\\
58 & $\{ {\rm R6} (4),  {\rm R11} (3),  {\rm R21} (2)\}$   & $\{\{\bar 5\}, \{\bar 3\}, \{\{\bar 1,\bar 2, \bar 4\}\}$  & $(1,  1,  1)$\\
59 & $\{ {\rm R8} (4),  {\rm R19} (2),  {\rm R22} (3)\}$   & $\{\{\bar 3\}, \{\bar 1\}, \{\{\bar 2,\bar 4, \bar 5\}\}$  & $(1,  1,  1)$\\
60 & $\{ {\rm R7} (4),  {\rm R14} (2),  {\rm R21} (3)\}$   & $\{\{\bar 2\}, \{\},\{\bar 1, \{\bar 3,\bar 4, \bar 5\}\}$  & $(1,  1,  1)$\\
61 & $\{ {\rm R4} (4),  {\rm R6} (2),  {\rm R21} (3)\}$   & $\{\{\bar 5\}, \{\bar 1\}, \{\{\bar 2,\bar 3, \bar 4\}\}$  & $(1,  1,  1)$\\
62 & $\{ {\rm R5} (2),  {\rm R7} (4),  {\rm R21} (3)\}$   & $\{\{\bar 1,\bar 2\}, \{\},\{\{\bar 3,\bar 4, \bar 5\}\}$  & $(1,  1,  1)$\\
63 & $\{ {\rm R3} (4),  {\rm R11} (2),  {\rm R22} (3)\}$   & $\{\{\bar 1\}, \{\},\{\bar 4, \{\bar 2,\bar 3, \bar 5\}\}$  & $(1,  1,  1)$\\
64 & $\{ {\rm R5} (2),  {\rm R14} (3),  {\rm R22} (4)\}$   & $\{\{\bar 3\}, \{\bar 5\}, \{\{\bar 1,\bar 2, \bar 4\}\}$  & $(1,  1,  1)$\\
65 & $\{ {\rm R4} (3),  {\rm R11} (2),  {\rm R21} (4)\}$   & $\{\{\bar 5\}, \{\bar 4\}, \{\{\bar 1,\bar 2, \bar 3\}\}$  & $(1,  1,  1)$\\
66 & $\{ {\rm R3} (2),  {\rm R5} (4),  {\rm R13} (3)\}$   & $\{\{\bar 1\}, \{\},\{\{\bar 2,\bar 3, \bar 4\}\}$  & $(1,  1,  1)$\\
67 & $\{ {\rm R8} (2),  {\rm R12} (3),  {\rm R13} (4)\}$   & $\{\{\}, \{\bar 3\}, \{\bar 4, \{\bar 1,\bar 2, \bar 5\}\}$  & $(1,  1,  1)$\\
68 & $\{ {\rm R6} (2),  {\rm R19} (4),  {\rm R21} (3)\}$   & $\{\{\bar 2\}, \{\bar 1\}, \{\{\bar 3,\bar 4, \bar 5\}\}$  & $(1,  1,  1)$\\
69 & $\{ {\rm R3} (4),  {\rm R12} (2),  {\rm R14} (3)\}$   & $\{\{\bar 2\}, \{\bar 5\}, \{\{\bar 1,\bar 3, \bar 4\}\}$  & $(1,  1,  1)$\\
70 & $\{ {\rm R5} (2),  {\rm R8} (3),  {\rm R22} (4)\}$   & $\{\{\bar 3\}, \{\bar 2\}, \{\{\bar 1,\bar 4, \bar 5\}\}$  & $(1,  1,  1)$\\
71 & $\{ {\rm R7} (2),  {\rm R12} (3),  {\rm R20} (4)\}$   & $\{\{\bar 4\}, \{\bar 3\}, \{\{\bar 1,\bar 2, \bar 5\}\}$  & $(1,  1,  1)$\\
72 & $\{ {\rm R12} (4),  {\rm R13} (3),  {\rm R22} (2)\}$   & $\{\{\bar 4\}, \{\bar 5\}, \{\{\bar 1,\bar 2, \bar 3\}\}$  & $(1,  1,  1)$\\
73 & $\{ {\rm R3} (4),  {\rm R14} (5),  {\rm R26} (2)\}$   & $\{\{\bar 2\}, \{\bar 5\}, \{\{\bar 1,\bar 3, \bar 4\}\}$  & $(1,  1,  1)$\\
74 & $\{ {\rm R11} (5),  {\rm R19} (4),  {\rm R26} (2)\}$   & $\{\{\bar 1\}, \{\},\{\bar 3, \{\bar 2,\bar 4, \bar 5\}\}$  & $(1,  1,  1)$\\
75 & $\{ {\rm R8} (4),  {\rm R17} (2),  {\rm R22} (5)\}$   & $\{\{\bar 3\}, \{\bar 1\}, \{\{\bar 2,\bar 4, \bar 5\}\}$  & $(1,  1,  1)$\\
76 & $\{ {\rm R10} (2),  {\rm R12} (4),  {\rm R13} (5)\}$   & $\{\{\bar 4\}, \{\bar 5\}, \{\{\bar 1,\bar 2, \bar 3\}\}$  & $(1,  1,  1)$\\
77 & $\{ {\rm R10} (2),  {\rm R12} (4),  {\rm R20} (5)\}$   & $\{\{\bar 4\}, \{\bar 1\}, \{\{\bar 2,\bar 3, \bar 5\}\}$  & $(1,  1,  1)$
\end{tabular}}
\end{adjustbox}
\caption{List of MASs for 1-constituent CCRN $L=5$.\label{t:mas5a}}
\end{table}

\begin{table}
\begin{adjustbox}{width=0.4\textwidth}{
\begin{tabular}{ccccc}
MAS id & Reactions (Flow) & FWMC Partition &  Production\\\hline
78 & $\{ {\rm R1} (2),  {\rm R12} (5),  {\rm R20} (4)\}$   & $\{\{\bar 4\}, \{\bar 3\}, \{\{\bar 1,\bar 2, \bar 5\}\}$  & $(1,  1,  1)$\\
79 & $\{ {\rm R7} (4),  {\rm R10} (2),  {\rm R21} (5)\}$   & $\{\{\bar 2\}, \{\},\{\bar 1, \{\bar 3,\bar 4, \bar 5\}\}$  & $(1,  1,  1)$\\
80 & $\{ {\rm R1} (2),  {\rm R14} (5),  {\rm R22} (4)\}$   & $\{\{\bar 3\}, \{\bar 5\}, \{\{\bar 1,\bar 2, \bar 4\}\}$  & $(1,  1,  1)$\\
81 & $\{ {\rm R5} (4),  {\rm R13} (5),  {\rm R17} (2)\}$   & $\{\{\bar 1\}, \{\},\{\{\bar 2,\bar 3, \bar 4\}\}$  & $(1,  1,  1)$\\
82 & $\{ {\rm R6} (4),  {\rm R11} (5),  {\rm R16} (2)\}$   & $\{\{\bar 5\}, \{\bar 3\}, \{\{\bar 1,\bar 2, \bar 4\}\}$  & $(1,  1,  1)$\\
83 & $\{ {\rm R6} (5),  {\rm R11} (4),  {\rm R24} (2)\}$   & $\{\{\bar 5\}, \{\bar 2\}, \{\{\bar 1,\bar 3, \bar 4\}\}$  & $(1,  1,  1)$\\
84 & $\{ {\rm R10} (2),  {\rm R12} (4),  {\rm R21} (5)\}$   & $\{\{\bar 2\}, \{\bar 1\}, \{\{\bar 3,\bar 4, \bar 5\}\}$  & $(1,  1,  1)$\\
85 & $\{ {\rm R2} (3),  {\rm R9} (5),  {\rm R24} (4)\}$   & $\{\{\bar 5\}, \{\bar 4\}, \{\{\bar 1,\bar 2, \bar 3\}\}$  & $(1,  1,  2)$\\
86 & $\{ {\rm R5} (5),  {\rm R9} (3),  {\rm R12} (4)\}$   & $\{\{\bar 2\}, \{\},\{\bar 1, \{\bar 3,\bar 4, \bar 5\}\}$  & $(1,  1,  1)$\\
87 & $\{ {\rm R2} (3),  {\rm R5} (5),  {\rm R22} (4)\}$   & $\{\{\bar 3\}, \{\bar 5\}, \{\{\bar 1,\bar 2, \bar 4\}\}$  & $(1,  1,  1)$\\
88 & $\{ {\rm R2} (3),  {\rm R7} (5),  {\rm R20} (4)\}$   & $\{\{\bar 4\}, \{\bar 3\}, \{\{\bar 1,\bar 2, \bar 5\}\}$  & $(1,  1,  1)$\\
89 & $\{ {\rm R8} (4),  {\rm R15} (3),  {\rm R19} (5)\}$   & $\{\{\bar 3\}, \{\bar 4\}, \{\{\bar 1,\bar 2, \bar 5\}\}$  & $(1,  1,  1)$\\
90 & $\{ {\rm R11} (4),  {\rm R19} (5),  {\rm R23} (3)\}$   & $\{\{\bar 1\}, \{\},\{\bar 2, \{\bar 3,\bar 4, \bar 5\}\}$  & $(1,  1,  1)$\\
91 & $\{ {\rm R5} (5),  {\rm R7} (4),  {\rm R9} (3)\}$   & $\{\{\bar 1\}, \{\},\{\{\bar 3,\bar 4, \bar 5\}\}$  & $(1,  1,  1)$\\
92 & $\{ {\rm R3} (5),  {\rm R5} (4),  {\rm R18} (3)\}$   & $\{\{\bar 1\}, \{\},\{\{\bar 2,\bar 3, \bar 4\}\}$  & $(1,  1,  1)$\\
93 & $\{ {\rm R16} (4),  {\rm R18} (5),  {\rm R25} (3)\}$   & $\{\{\bar 5\}, \{\bar 1\}, \{\{\bar 2,\bar 3, \bar 4\}\}$  & $(2,  1,  1)$\\
94 & $\{ {\rm R8} (4),  {\rm R19} (5),  {\rm R23} (3)\}$   & $\{\{\bar 2\}, \{\bar 1\}, \{\{\bar 3,\bar 4, \bar 5\}\}$  & $(1,  1,  1)$\\
95 & $\{ {\rm R8} (4),  {\rm R18} (3),  {\rm R19} (5)\}$   & $\{\{\bar 3\}, \{\bar 1\}, \{\{\bar 2,\bar 4, \bar 5\}\}$  & $(1,  1,  1)$\\
96 & $\{ {\rm R6} (6),  {\rm R11} (5),  {\rm R17} (2)\}$   & $\{\{\bar 5\}, \{\bar 3\}, \{\{\bar 1,\bar 2, \bar 4\}\}$  & $(1,  1,  1)$\\
97 & $\{ {\rm R7} (6),  {\rm R11} (5),  {\rm R17} (2)\}$   & $\{\{\bar 1\}, \{\},\{\{\bar 2,\bar 4, \bar 5\}\}$  & $(1,  1,  1)$\\
98 & $\{ {\rm R13} (3),  {\rm R16} (4),  {\rm R22} (6)\}$   & $\{\{\bar 4\}, \{\bar 5\}, \{\{\bar 1,\bar 2, \bar 3\}\}$  & $(1,  1,  1)$\\
99 & $\{ {\rm R4} (6),  {\rm R21} (5),  {\rm R26} (2)\}$   & $\{\{\}, \{\bar 1\}, \{\bar 5, \{\bar 2,\bar 3, \bar 4\}\}$  & $(1,  1,  1)$\\
100 & $\{ {\rm R1} (2),  {\rm R4} (5),  {\rm R21} (6)\}$   & $\{\{\bar 5\}, \{\bar 4\}, \{\{\bar 1,\bar 2, \bar 3\}\}$  & $(1,  1,  1)$\\
101 & $\{ {\rm R3} (6),  {\rm R13} (3),  {\rm R24} (4)\}$   & $\{\{\bar 1\}, \{\},\{\{\bar 2,\bar 3, \bar 4\}\}$  & $(1,  1,  1)$\\
102 & $\{ {\rm R5} (6),  {\rm R13} (5),  {\rm R16} (2)\}$   & $\{\{\}, \{\},\{\bar 1, \{\bar 2,\bar 3, \bar 4\}\}$  & $(1,  1,  1)$\\
103 & $\{ {\rm R12} (6),  {\rm R21} (5),  {\rm R24} (2)\}$   & $\{\{\}, \{\bar 1\}, \{\bar 2, \{\bar 3,\bar 4, \bar 5\}\}$  & $(1,  1,  1)$\\
104 & $\{ {\rm R6} (5),  {\rm R10} (2),  {\rm R11} (6)\}$   & $\{\{\}, \{\bar 2\}, \{\bar 5, \{\bar 1,\bar 3, \bar 4\}\}$  & $(1,  1,  1)$\\
105 & $\{ {\rm R12} (6),  {\rm R20} (5),  {\rm R24} (2)\}$   & $\{\{\}, \{\bar 1\}, \{\bar 4, \{\bar 2,\bar 3, \bar 5\}\}$  & $(1,  1,  1)$\\
106 & $\{ {\rm R5} (6),  {\rm R21} (3),  {\rm R26} (4)\}$   & $\{\{\bar 1,\bar 2\}, \{\},\{\{\bar 3,\bar 4, \bar 5\}\}$  & $(1,  1,  1)$\\
107 & $\{ {\rm R12} (6),  {\rm R13} (5),  {\rm R24} (2)\}$   & $\{\{\}, \{\bar 5\}, \{\bar 4, \{\bar 1,\bar 2, \bar 3\}\}$  & $(1,  1,  1)$\\
108 & $\{ {\rm R4} (3),  {\rm R11} (6),  {\rm R16} (4)\}$   & $\{\{\bar 5\}, \{\bar 4\}, \{\{\bar 1,\bar 2, \bar 3\}\}$  & $(1,  1,  1)$\\
109 & $\{ {\rm R6} (6),  {\rm R21} (3),  {\rm R24} (4)\}$   & $\{\{\bar 5\}, \{\bar 1\}, \{\{\bar 2,\bar 3, \bar 4\}\}$  & $(1,  1,  1)$\\
110 & $\{ {\rm R7} (6),  {\rm R21} (5),  {\rm R24} (2)\}$   & $\{\{\bar 1\}, \{\},\{\bar 2, \{\bar 3,\bar 4, \bar 5\}\}$  & $(1,  1,  1)$\\
111 & $\{ {\rm R14} (6),  {\rm R21} (3),  {\rm R26} (4)\}$   & $\{\{\bar 2\}, \{\bar 1\}, \{\{\bar 3,\bar 4, \bar 5\}\}$  & $(1,  1,  1)$\\
112 & $\{ {\rm R11} (5),  {\rm R16} (6),  {\rm R18} (4)\}$   & $\{\{\bar 5\}, \{\bar 3\}, \{\{\bar 1,\bar 2, \bar 4\}\}$  & $(1,  1,  1)$\\
113 & $\{ {\rm R10} (6),  {\rm R21} (5),  {\rm R23} (4)\}$   & $\{\{\bar 2\}, \{\bar 1\}, \{\{\bar 3,\bar 4, \bar 5\}\}$  & $(1,  1,  1)$\\
114 & $\{ {\rm R6} (5),  {\rm R9} (4),  {\rm R24} (6)\}$   & $\{\{\bar 5\}, \{\bar 2\}, \{\{\bar 1,\bar 3, \bar 4\}\}$  & $(1,  1,  1)$\\
115 & $\{ {\rm R13} (5),  {\rm R15} (4),  {\rm R17} (6)\}$   & $\{\{\bar 1\}, \{\},\{\{\bar 2,\bar 3, \bar 4\}\}$  & $(1,  1,  1)$\\
116 & $\{ {\rm R10} (6),  {\rm R13} (5),  {\rm R23} (4)\}$   & $\{\{\bar 4\}, \{\bar 5\}, \{\{\bar 1,\bar 2, \bar 3\}\}$  & $(1,  1,  1)$\\
117 & $\{ {\rm R10} (6),  {\rm R20} (5),  {\rm R23} (4)\}$   & $\{\{\bar 4\}, \{\bar 1\}, \{\{\bar 2,\bar 3, \bar 5\}\}$  & $(1,  1,  1)$\\
118 & $\{ {\rm R5} (9),  {\rm R9} (3),  {\rm R26} (4)\}$   & $\{\{\bar 1\}, \{\},\{\{\bar 3,\bar 4, \bar 5\}\}$  & $(1,  1,  1)$\\
119 & $\{ {\rm R3} (9),  {\rm R18} (3),  {\rm R24} (4)\}$   & $\{\{\bar 1\}, \{\},\{\{\bar 2,\bar 3, \bar 4\}\}$  & $(1,  1,  1)$\\
120 & $\{ {\rm R10} (6),  {\rm R13} (9),  {\rm R16} (4)\}$   & $\{\{\bar 4\}, \{\bar 5\}, \{\{\bar 1,\bar 2, \bar 3\}\}$  & $(1,  1,  1)$\\
121 & $\{ {\rm R10} (6),  {\rm R21} (9),  {\rm R26} (4)\}$   & $\{\{\bar 2\}, \{\bar 1\}, \{\{\bar 3,\bar 4, \bar 5\}\}$  & $(1,  1,  1)$\\
122 & $\{ {\rm R13} (9),  {\rm R17} (6),  {\rm R24} (4)\}$   & $\{\{\bar 1\}, \{\},\{\{\bar 2,\bar 3, \bar 4\}\}$  & $(1,  1,  1)$\\
123 & $\{ {\rm R2} (1),  {\rm R9} (1),  {\rm R18} (1),  {\rm R25} (1)\}$   & $\{\{\bar 5\}, \{\},\{\{\bar 1,\bar 2, \bar 3, \bar 4\}\}$  & $(1,  1,  1,  1)$\\
124 & $\{ {\rm R3} (3),  {\rm R8} (3),  {\rm R12} (2),  {\rm R19} (2)\}$   & $\{\{\bar 2\}, \{\},\{\{\bar 1,\bar 3, \bar 4, \bar 5\}\}$  & $(2,  1,  1,  1)$\\
125 & $\{ {\rm R9} (2),  {\rm R16} (3),  {\rm R18} (2),  {\rm R24} (3)\}$   & $\{\{\bar 5\}, \{\},\{\{\bar 1,\bar 2, \bar 3, \bar 4\}\}$  & $(1,  1,  1,  1)$\\
126 & $\{ {\rm R4} (3),  {\rm R7} (3),  {\rm R16} (2),  {\rm R20} (2)\}$   & $\{\{\bar 4\}, \{\},\{\{\bar 1,\bar 2, \bar 3, \bar 5\}\}$  & $(2,  1,  1,  1)$\\
127 & $\{ {\rm R3} (5),  {\rm R8} (3),  {\rm R10} (2),  {\rm R12} (2)\}$   & $\{\{\bar 2\}, \{\},\{\{\bar 1,\bar 3, \bar 4, \bar 5\}\}$  & $(2,  1,  1,  1)$\\
128 & $\{ {\rm R3} (3),  {\rm R6} (2),  {\rm R8} (3),  {\rm R19} (4)\}$   & $\{\{\bar 2\}, \{\},\{\{\bar 1,\bar 3, \bar 4, \bar 5\}\}$  & $(2,  1,  1,  1)$\\
129 & $\{ {\rm R7} (3),  {\rm R10} (3),  {\rm R16} (2),  {\rm R20} (5)\}$   & $\{\{\bar 4\}, \{\},\{\{\bar 1,\bar 2, \bar 3, \bar 5\}\}$  & $(2,  1,  1,  1)$\\
130 & $\{ {\rm R3} (4),  {\rm R6} (2),  {\rm R10} (4),  {\rm R21} (3)\}$   & $\{\{\bar 2\}, \{\},\{\{\bar 1,\bar 3, \bar 4, \bar 5\}\}$  & $(2,  1,  1,  1)$\\
131 & $\{ {\rm R3} (5),  {\rm R8} (3),  {\rm R10} (4),  {\rm R23} (2)\}$   & $\{\{\bar 2\}, \{\},\{\{\bar 1,\bar 3, \bar 4, \bar 5\}\}$  & $(2,  1,  1,  1)$\\
132 & $\{ {\rm R3} (3),  {\rm R5} (2),  {\rm R7} (5),  {\rm R11} (4)\}$   & $\{\{\bar 1\}, \{\},\{\{\bar 2,\bar 3, \bar 4, \bar 5\}\}$  & $(1,  1,  1,  1)$\\
133 & $\{ {\rm R3} (3),  {\rm R8} (5),  {\rm R19} (4),  {\rm R26} (2)\}$   & $\{\{\bar 2\}, \{\},\{\{\bar 1,\bar 3, \bar 4, \bar 5\}\}$  & $(2,  1,  1,  1)$\\
134 & $\{ {\rm R3} (5),  {\rm R8} (3),  {\rm R12} (4),  {\rm R24} (2)\}$   & $\{\{\}, \{\},\{\bar 2, \{\bar 1,\bar 3, \bar 4, \bar 5\}\}$  & $(2,  1,  1,  1)$\\
135 & $\{ {\rm R3} (5),  {\rm R7} (5),  {\rm R11} (4),  {\rm R24} (2)\}$   & $\{\{\bar 1\}, \{\},\{\{\bar 2,\bar 3, \bar 4, \bar 5\}\}$  & $(1,  1,  1,  1)$\\
136 & $\{ {\rm R3} (7),  {\rm R6} (2),  {\rm R8} (3),  {\rm R10} (4)\}$   & $\{\{\bar 2\}, \{\},\{\{\bar 1,\bar 3, \bar 4, \bar 5\}\}$  & $(2,  1,  1,  1)$\\
137 & $\{ {\rm R3} (3),  {\rm R5} (6),  {\rm R7} (5),  {\rm R20} (4)\}$   & $\{\{\bar 1\}, \{\},\{\{\bar 2,\bar 3, \bar 4, \bar 5\}\}$  & $(1,  1,  1,  1)$\\
138 & $\{ {\rm R3} (5),  {\rm R7} (5),  {\rm R10} (2),  {\rm R11} (6)\}$   & $\{\{\}, \{\},\{\bar 1, \{\bar 2,\bar 3, \bar 4, \bar 5\}\}$  & $(1,  1,  1,  1)$\\
139 & $\{ {\rm R3} (7),  {\rm R8} (5),  {\rm R10} (4),  {\rm R26} (2)\}$   & $\{\{\bar 2\}, \{\},\{\{\bar 1,\bar 3, \bar 4, \bar 5\}\}$  & $(2,  1,  1,  1)$\\
140 & $\{ {\rm R3} (3),  {\rm R5} (7),  {\rm R11} (4),  {\rm R26} (5)\}$   & $\{\{\bar 1\}, \{\},\{\{\bar 2,\bar 3, \bar 4, \bar 5\}\}$  & $(1,  1,  1,  1)$\\
141 & $\{ {\rm R5} (2),  {\rm R7} (8),  {\rm R11} (7),  {\rm R16} (3)\}$   & $\{\{\}, \{\},\{\bar 1, \{\bar 2,\bar 3, \bar 4, \bar 5\}\}$  & $(1,  1,  1,  1)$\\
142 & $\{ {\rm R3} (5),  {\rm R7} (5),  {\rm R9} (4),  {\rm R24} (6)\}$   & $\{\{\bar 1\}, \{\},\{\{\bar 2,\bar 3, \bar 4, \bar 5\}\}$  & $(1,  1,  1,  1)$\\
143 & $\{ {\rm R7} (5),  {\rm R9} (4),  {\rm R15} (7),  {\rm R17} (6)\}$   & $\{\{\bar 1\}, \{\},\{\{\bar 2,\bar 3, \bar 4, \bar 5\}\}$  & $(2,  1,  1,  1)$\\
144 & $\{ {\rm R3} (3),  {\rm R5} (11),  {\rm R20} (4),  {\rm R26} (5)\}$   & $\{\{\bar 1\}, \{\},\{\{\bar 2,\bar 3, \bar 4, \bar 5\}\}$  & $(1,  1,  1,  1)$\\
145 & $\{ {\rm R5} (6),  {\rm R7} (8),  {\rm R17} (3),  {\rm R20} (7)\}$   & $\{\{\bar 1\}, \{\},\{\{\bar 2,\bar 3, \bar 4, \bar 5\}\}$  & $(1,  1,  1,  1)$\\
146 & $\{ {\rm R3} (9),  {\rm R7} (5),  {\rm R20} (4),  {\rm R24} (6)\}$   & $\{\{\bar 1\}, \{\},\{\{\bar 2,\bar 3, \bar 4, \bar 5\}\}$  & $(1,  1,  1,  1)$\\
147 & $\{ {\rm R5} (7),  {\rm R11} (7),  {\rm R17} (3),  {\rm R26} (8)\}$   & $\{\{\bar 1\}, \{\},\{\{\bar 2,\bar 3, \bar 4, \bar 5\}\}$  & $(1,  1,  1,  1)$\\
148 & $\{ {\rm R3} (10),  {\rm R11} (4),  {\rm R24} (7),  {\rm R26} (5)\}$   & $\{\{\bar 1\}, \{\},\{\{\bar 2,\bar 3, \bar 4, \bar 5\}\}$  & $(1,  1,  1,  1)$\\
149 & $\{ {\rm R7} (10),  {\rm R11} (9),  {\rm R16} (5),  {\rm R24} (2)\}$   & $\{\{\}, \{\},\{\bar 1, \{\bar 2,\bar 3, \bar 4, \bar 5\}\}$  & $(1,  1,  1,  1)$\\
150 & $\{ {\rm R7} (9),  {\rm R9} (8),  {\rm R17} (3),  {\rm R24} (7)\}$   & $\{\{\bar 1\}, \{\},\{\{\bar 2,\bar 3, \bar 4, \bar 5\}\}$  & $(1,  2,  1,  1)$\\
151 & $\{ {\rm R5} (9),  {\rm R7} (8),  {\rm R16} (3),  {\rm R20} (7)\}$   & $\{\{\}, \{\},\{\bar 1, \{\bar 2,\bar 3, \bar 4, \bar 5\}\}$  & $(1,  1,  1,  1)$\\
152 & $\{ {\rm R7} (10),  {\rm R10} (2),  {\rm R11} (11),  {\rm R16} (5)\}$   & $\{\{\}, \{\},\{\bar 1, \{\bar 2,\bar 3, \bar 4, \bar 5\}\}$  & $(1,  1,  1,  1)$\\
153 & $\{ {\rm R5} (10),  {\rm R11} (7),  {\rm R16} (3),  {\rm R26} (8)\}$   & $\{\{\}, \{\},\{\bar 1, \{\bar 2,\bar 3, \bar 4, \bar 5\}\}$  & $(1,  1,  1,  1)$\\
154 & $\{ {\rm R3} (10),  {\rm R10} (7),  {\rm R11} (6),  {\rm R13} (5)\}$   & $\{\{\}, \{\},\{\bar 1, \{\bar 2,\bar 3, \bar 4, \bar 5\}\}$  & $(1,  1,  1,  1)$\\
155 & $\{ {\rm R3} (10),  {\rm R9} (4),  {\rm R24} (11),  {\rm R26} (5)\}$   & $\{\{\bar 1\}, \{\},\{\{\bar 2,\bar 3, \bar 4, \bar 5\}\}$  & $(1,  1,  1,  1)$\\
156 & $\{ {\rm R7} (8),  {\rm R15} (6),  {\rm R17} (9),  {\rm R20} (7)\}$   & $\{\{\bar 1\}, \{\},\{\{\bar 2,\bar 3, \bar 4, \bar 5\}\}$  & $(1,  1,  1,  1)$\\
157 & $\{ {\rm R9} (4),  {\rm R15} (12),  {\rm R17} (11),  {\rm R26} (5)\}$   & $\{\{\bar 1\}, \{\},\{\{\bar 2,\bar 3, \bar 4, \bar 5\}\}$  & $(2,  1,  1,  1)$\\
158 & $\{ {\rm R11} (7),  {\rm R15} (7),  {\rm R17} (10),  {\rm R26} (8)\}$   & $\{\{\bar 1\}, \{\},\{\{\bar 2,\bar 3, \bar 4, \bar 5\}\}$  & $(1,  1,  1,  1)$\\
159 & $\{ {\rm R5} (14),  {\rm R17} (3),  {\rm R20} (7),  {\rm R26} (8)\}$   & $\{\{\bar 1\}, \{\},\{\{\bar 2,\bar 3, \bar 4, \bar 5\}\}$  & $(1,  1,  1,  1)$\\
160 & $\{ {\rm R3} (15),  {\rm R10} (7),  {\rm R11} (6),  {\rm R18} (5)\}$   & $\{\{\}, \{\},\{\bar 1, \{\bar 2,\bar 3, \bar 4, \bar 5\}\}$  & $(1,  1,  1,  1)$\\
161 & $\{ {\rm R3} (10),  {\rm R10} (7),  {\rm R11} (11),  {\rm R26} (5)\}$   & $\{\{\}, \{\},\{\bar 1, \{\bar 2,\bar 3, \bar 4, \bar 5\}\}$  & $(1,  1,  1,  1)$\\
162 & $\{ {\rm R3} (14),  {\rm R20} (4),  {\rm R24} (11),  {\rm R26} (5)\}$   & $\{\{\bar 1\}, \{\},\{\{\bar 2,\bar 3, \bar 4, \bar 5\}\}$  & $(1,  1,  1,  1)$\\
163 & $\{ {\rm R7} (10),  {\rm R9} (9),  {\rm R16} (5),  {\rm R24} (11)\}$   & $\{\{\}, \{\},\{\bar 1, \{\bar 2,\bar 3, \bar 4, \bar 5\}\}$  & $(1,  1,  1,  1)$\\
164 & $\{ {\rm R5} (17),  {\rm R16} (3),  {\rm R20} (7),  {\rm R26} (8)\}$   & $\{\{\}, \{\},\{\bar 1, \{\bar 2,\bar 3, \bar 4, \bar 5\}\}$  & $(1,  1,  1,  1)$\\
165 & $\{ {\rm R10} (7),  {\rm R11} (6),  {\rm R13} (15),  {\rm R17} (10)\}$   & $\{\{\}, \{\},\{\bar 1, \{\bar 2,\bar 3, \bar 4, \bar 5\}\}$  & $(1,  1,  1,  1)$\\
166 & $\{ {\rm R7} (14),  {\rm R17} (9),  {\rm R20} (13),  {\rm R24} (6)\}$   & $\{\{\bar 1\}, \{\},\{\{\bar 2,\bar 3, \bar 4, \bar 5\}\}$  & $(1,  1,  1,  1)$\\
167 & $\{ {\rm R11} (14),  {\rm R17} (10),  {\rm R24} (7),  {\rm R26} (15)\}$   & $\{\{\bar 1\}, \{\},\{\{\bar 2,\bar 3, \bar 4, \bar 5\}\}$  & $(1,  1,  1,  1)$\\
168 & $\{ {\rm R15} (14),  {\rm R17} (17),  {\rm R20} (7),  {\rm R26} (8)\}$   & $\{\{\bar 1\}, \{\},\{\{\bar 2,\bar 3, \bar 4, \bar 5\}\}$  & $(1,  1,  1,  1)$\\
169 & $\{ {\rm R10} (7),  {\rm R11} (21),  {\rm R17} (10),  {\rm R26} (15)\}$   & $\{\{\}, \{\},\{\bar 1, \{\bar 2,\bar 3, \bar 4, \bar 5\}\}$  & $(1,  1,  1,  1)$\\
170 & $\{ {\rm R9} (14),  {\rm R17} (10),  {\rm R24} (21),  {\rm R26} (15)\}$   & $\{\{\bar 1\}, \{\},\{\{\bar 2,\bar 3, \bar 4, \bar 5\}\}$  & $(1,  1,  1,  1)$
 \end{tabular}
} 
\end{adjustbox}
\caption{List of MASs for 1-constituent CCRN $L=5$ (cntd.).\label{t:mas5b}}
\end{table}

\bibliographystyle{unsrt} 
\bibliography{bibliography}
\end{document}